\documentclass[aip,reprint,author-year,nofootinbib]{revtex4-1}

\usepackage{graphicx}
\usepackage{amsmath,amssymb}             
\usepackage{stmaryrd}					
\usepackage{color}
\usepackage[dvipsnames]{xcolor}
\usepackage{color,hyperref}
\definecolor{darkblue}{rgb}{0.0,0.0,0.4}
\definecolor{red}{rgb}{0.7,0.0,0.0}
\definecolor{green}{rgb}{0.0,0.5,0.0}
\hypersetup{colorlinks,breaklinks,
            linkcolor=darkblue,urlcolor=darkblue,
            anchorcolor=darkblue,citecolor=RoyalBlue}

\def\apj{Astrophys. J.}

\def\apjs{Astrophys. J. Supp.}

\def\mnras{Mon. Not. R. Astron. Soc.}

\def\prd{Phys. Rev. D}

\def\physrep{Physics Reports}

\def\aap{Astron. \& Astrophys.}

\def\jcap{Journal of Cosmology and Astroparticle Physics}
\def\na{New Astronomy}

\defcitealias{Bernardeau2002}{BCGS02}
\defcitealias{Weinberg1992}{W92}

\begin{document}


\title{One-point remapping of Lagrangian perturbation theory in the mildly non-linear regime of cosmic structure formation}


\author{Florent Leclercq}
\email{florent.leclercq@polytechnique.org}
\affiliation{Institut d'Astrophysique de Paris (IAP), UMR 7095, CNRS -- UPMC Universit\'e Paris 6, Sorbonne Universit\'es, 98bis boulevard Arago, F-75014 Paris, France}
\affiliation{Institut Lagrange de Paris (ILP), Sorbonne Universit\'es,\\ 98bis boulevard Arago, F-75014 Paris, France}
\affiliation{\'Ecole polytechnique ParisTech,\\ Route de Saclay, F-91128 Palaiseau, France}

\author{Jens Jasche}
\affiliation{Institut d'Astrophysique de Paris (IAP), UMR 7095, CNRS -- UPMC Universit\'e Paris 6, Sorbonne Universit\'es, 98bis boulevard Arago, F-75014 Paris, France}
\affiliation{Institut Lagrange de Paris (ILP), Sorbonne Universit\'es,\\ 98bis boulevard Arago, F-75014 Paris, France}

\author{H{\'e}ctor Gil-Mar{\'i}n}
\affiliation{Institute of Cosmology \& Gravitation (ICG), University of Portsmouth,\\ Dennis Sciama Building, Portsmouth PO1 3FX, UK}
\affiliation{Institut de Ci\`encies del Cosmos (ICC), Universitat de Barcelona (IEEC-UB),\\ Mart\'i i Franqu\'es 1, E-08028, Spain}

\author{Benjamin Wandelt}
\affiliation{Institut d'Astrophysique de Paris (IAP), UMR 7095, CNRS -- UPMC Universit\'e Paris 6, Sorbonne Universit\'es, 98bis boulevard Arago, F-75014 Paris, France}
\affiliation{Institut Lagrange de Paris (ILP), Sorbonne Universit\'es,\\ 98bis boulevard Arago, F-75014 Paris, France}
\affiliation{Department of Physics, University of Illinois at Urbana-Champaign,\\ 1110 West Green Street, Urbana, IL~61801, USA}
\affiliation{Department of Astronomy, University of Illinois at Urbana-Champaign,\\ 1002 West Green Street, Urbana, IL~61801, USA}


\date{\today}

\begin{abstract}
\noindent On the smallest scales, three-dimensional large-scale structure surveys contain a wealth of cosmological information which cannot be trivially extracted due to the non-linear dynamical evolution of the density field. Lagrangian perturbation theory (LPT) is widely applied to the generation of mock halo catalogs and data analysis. In this work, we compare topological features of the cosmic web such as voids, sheets, filaments and clusters, in the density fields predicted by LPT and full numerical simulation of gravitational large-scale structure formation. We propose a method designed to improve the correspondence between these density fields, in the mildly non-linear regime. We develop a computationally fast and flexible tool for a variety of cosmological applications. Our method is based on a remapping of the approximately-evolved density field, using information extracted from $N$-body simulations. The remapping procedure consists of replacing the one-point distribution of the density contrast by one which better accounts for the full gravitational dynamics. As a result, we obtain a physically more pertinent density field on a point-by-point basis, while also improving higher-order statistics predicted by LPT. We quantify the approximation error in the power spectrum and in the bispectrum as a function of scale and redshift. Our remapping procedures improves one-, two- and three-point statistics at scales down to 8~Mpc/$h$.
\end{abstract}


\maketitle



\section{Introduction}

At present, observations of the three-dimensional large-scale structure (LSS) are major sources of information on the origin and evolution of the Universe. According to the current paradigm of cosmological structure formation, the presently observed structures formed via gravitational clustering of cold dark matter particles and condensation of baryonic matter in gravitational potential wells. Consequently, the large-scale matter distribution retains a memory of its formation history, enabling us to study the homogeneous as well as the inhomogeneous evolution of our Universe.

Due to non-linearities involved in the formation process, at present there exists just limited analytic understanding of structure formation in terms of perturbative expansions in Eulerian or Lagrangian representations. Both of these approaches rely on a truncated sequence of momentum moments of the Vlasov-Poisson system, completed by fluid dynamic assumptions \citep[see e.g.][\citetalias{Bernardeau2002} hereafter, and references therein]{Bernardeau2002}. For this reason, the validity of these approaches ceases, once the evolution of the large-scale structure enters the multi-stream regime \citep[see e.g.][]{Pueblas2009}.

Nevertheless, Eulerian and Lagrangian approximations have been successfully applied to the analysis of three-dimensional density fields in regimes where they are still applicable, either at large scales or in the early Universe. Particularly, Lagrangian perturbation theory (LPT) captures significant mode-coupling information that is encoded beyond linear theory, such as large-scale flows and free-streaming, yielding three-dimensional matter distributions approximating those of full scale numerical simulations with reasonable accuracy \citep{Moutarde1991,Buchert1994,Bouchet1995,Scoccimarro1998,Scoccimarro2000,Scoccimarro2002,Yoshisato2006}. Especially, second-order Lagrangian perturbation theory has been widely applied in data analysis and for fast generation of galaxy mock catalogs (e.g. \textsc{PTHalos}: \citealp{Scoccimarro2002,Manera2013}; \textsc{Pinocchio}: \citealp{Monaco2002a,Monaco2002b,Taffoni2002,Heisenberg2011,Monaco2013}) that can be useful to estimate error bounds when analyzing observations.

Modern cosmological data analysis has an increasing demand for analytic and computationally inexpensive models providing accurate representations of the mildly non-linear regime of structure formation. Over the years, various non-linear approximations and attempts to extend the validity of LPT have been proposed. These include the spherical collapse model \citep{Gunn1972}, the truncated Zel'dovich approximation \citep{Melott1994} and models with various forms for the velocity potential \citep{Coles1993,Munshi1994} or the addition of a viscosity term in the Euler equation \citep[the adhesion model,][]{Gurbatov1989}. Analytical techniques to improve the convergence and behavior of standard perturbation theory, successfully employed in quantum field theory and statistical physics, have also been applied in the context of gravitational clustering. These include renormalized perturbation theory \citep{Crocce2006a}, the path integral formalism \citep{Valageas2007}, and the renormalization group flow \citep{Matarrese1997}. More recently, \citet{Tassev2012a,Tassev2012b} constructed a physical picture of the matter distribution in the mildly non-linear regime, and developed a method yielding improvements over LPT \citep{Tassev2013}, in particular at the scales relevant for baryon acoustic peak reconstruction \citep{Tassev2012c}.

In this paper, we propose a numerically efficient method designed to improve the correspondence between approximate models and full numerical simulations of gravitational large-scale structure formation. Generally, it can be applied to any approximate model of gravitational instability, but it is especially targeted to improving Lagrangian methods. We will illustrate both these methods on fields evolved with LPT: at order one, the Zel'dovich approximation \citep[][ZA in the following]{Zeldovich1970,Shandarin1989} and second-order Lagrangian perturbation theory (2LPT in the following).

LPT and $N$-body density fields are visually similar, which suggests that the properties of LPT could be improved by one-to-one mapping in voxel space, following a similar line of thoughts as the ``Gaussianization'' idea originally proposed by \citet[][\citetalias{Weinberg1992} hereafter]{Weinberg1992} and inspired existing techniques, widely used in cosmology \citep{Melott1993,Croft1998,Narayanan1998,Croft1999,Feng2000,Neyrinck2011,Yu2011,Yu2012,Neyrinck2013b}. The method described in this paper is based on a \textit{remapping} of the approximately evolved particle distribution using information extracted from $N$-body simulations. It basically consists of replacing the one-point distribution of the approximately evolved distribution by one which better accounts for the full gravitational system. In this fashion, we adjust the one-point distribution to construct a physically more reasonable representation of the three-dimensional matter distribution, while retaining or improving higher order statistics, described already reasonably well by the ZA \citep{Zeldovich1970,Doroshkevich1970,Shandarin1989,Buchert1989,Moutarde1991,Yoshisato2006} and by 2LPT \citep{Moutarde1991,Buchert1994,Bouchet1995,Scoccimarro1998,Scoccimarro2000,Scoccimarro2002}.

Major problems with naive approaches to remapping LPT density fields arise from minor deviations in structure types represented by LPT models and full gravity. For this reason, we compare clusters, voids, sheets, and filaments predicted by LPT and $N$-body simulations. Besides being of general interest for LSS data analysis, the insights gained from this comparison allow us to improve the remapping procedure.

Implementing and testing the accuracy and the regime of validity of our method is essential, and is subject of the present paper. Our study will quantify the approximation error as a function of scale in terms of a set of statistical diagnostics. From cosmographic measurements, $\sigma_8$ is known to be of order unity, which means that gravity becomes highly non-linear at some scale around 8~Mpc/$h$. Our method is expected to break down due to shell-crossing in LPT, at some scale larger than 8~Mpc/$h$. Achieving a resolution of 16~Mpc/$h$ would already constitute substantial improvement with respect to existing methods, since non-linearities begin to affect even large-scale cosmographic measurements such as the determination of the baryon acoustic oscillations scale from galaxy surveys \citep[about 125 Mpc/$h$, e.g.][]{Eisenstein2005}. However, we will explore the validity of the improvement at 8~Mpc/$h$ down to 4~Mpc/$h$, to see to what extent we can push the limit for possible data analysis applications into the non-linear regime. Recall that in three-dimensional LSS surveys, the number of modes usable for cosmological exploitation scales as the cube of the largest wavenumber, $k^3$, meaning that even minor improvements in the mildly non-linear regime would give access to much more cosmological information from existing and upcoming observations.

As will be demonstrated, this method can be used to generate realizations of density fields much faster than $N$-body simulations. Even though approximate, these fast realizations of mock density fields may be sufficient to model the salient features of the non-linear density field for certain applications.

This paper is structured as follows. In section \ref{par:remapping}, we describe the remapping procedure for the density contrast of present-day density fields, analyze the obstacles to straightforward application and present an improved method. In section \ref{par:results}, we apply the procedure to cosmological models using data from numerical simulations, we study the statistics of remapped fields and quantify the approximation error. We discuss our results and give our conclusions in section \ref{par:conclusion}.

Corresponding LPT and $N$-body simulations presented in this paper were run from the same initial conditions, generated at redshift $z=63$ using second-order Lagrangian perturbation theory. The $N$-body simulations were run with the \textsc{Gadget-2} cosmological code \citep{Springel2001,Springel2005}. Evolutions of the Zel'dovich approximation were performed with \textsc{N-GenIC} \citep{Springel2005}, and of second-order Lagrangian perturbation theory with \textsc{2LPTic} \citep{Crocce2006b}. To ensure sufficient statistical significance, we used eight realizations of the same cosmology, changing the seed used to generate respective initial conditions. All computations are done after binning the density fields with a Cloud in Cell (CiC) method. The simulations contain $512^3$ dark matter halos particles in a 1024 Mpc/$h$ cubic box with periodic boundary conditions. We checked that with this setup, the power spectrum of dark matter particles agrees with the non-linear power spectrum of simulations run with higher mass resolution, provided by \textsc{Cosmic Emulator} tools \citep{Heitmann2009,Heitmann2010,Lawrence2010} (deviations are at most sub-percent level for $k~\lesssim~1~(\mathrm{Mpc}/h)^{-1}$). Therefore, at the scales of interest of this work, $k~\leq~0.4~(\mathrm{Mpc}/h)^{-1}$ (corresponding to the linear and mildly non-linear regime at redshift zero), the clustering of dark matter is exactly reproduced by our set of simulations.

The cosmological parameters used are WMAP-7 fiducial values \citep{Komatsu2011},
\begin{eqnarray}
\label{eq:comological-parameters}
& \Omega_\Lambda = 0.728, \Omega_\mathrm{m}~=~0.2715, \Omega_\mathrm{b}~=~0.0455, \nonumber \\
& \sigma_8 = 0.810, h = 0.704, n_{\mathrm{s}} = 0.967 .
\end{eqnarray}

\section{Remapping Lagrangian perturbation theory}
\label{par:remapping}

In this section, we discuss the remapping procedure and apply it to cosmological density fields evolved with LPT. Naively following the approach of \citetalias{Weinberg1992} for present-day density fields yields the procedure described in section \ref{par:remapping-procedure}. This approach is not entirely satisfactory and we analyze the reasons for its shortcomings in section \ref{par:comparison-structure-types}. In consequence, we propose a improvement of the remapping procedure in section \ref{par:improvement-remapping-procedure}. The properties of the remapping function are examined in section \ref{par:remapping-function}.

\subsection{Remapping procedure}
\label{par:remapping-procedure}

\begin{figure*}
\begin{center}
\includegraphics[width=0.87\textwidth]{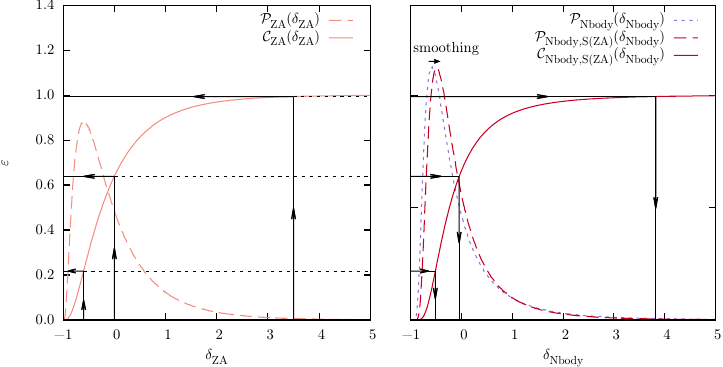} \\
\includegraphics[width=0.87\textwidth]{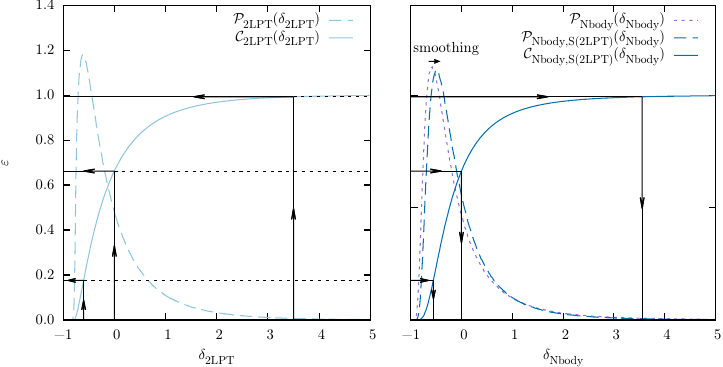}
\caption{A graphical illustration of the improved remapping procedure at redshift zero, for the ZA (upper panel) and 2LPT (lower panel). On the right panels, the dotted purple curves are the probability distribution function for the density contrast in full $N$-body dynamics. The first step of the procedure is to smooth the $N$-body density field using the transfer function given by eq. \eqref{eq:transfer-function}, which yields slightly different PDFs (dashed dark red and dark blue curves on the right panels). On the left panels, the dashed curves are the PDFs for the density contrast in the final density field, evolved from the same initial conditions with LPT (ZA: red line and 2LPT: blue line). The PDFs are computed on a 8~Mpc/$h$ mesh ($128^3$-voxel grid) using the full statistics from eight realizations of $512^3$-particles in a 1024~Mpc/$h$ box with periodic boundary conditions. The solid curves are their respective integrals, the cumulative distribution functions. The second step is remapping, which assigns a voxel with density contrast $\delta_\mathrm{LPT}$ and fractional rank $\varepsilon$ the value of $\delta_\mathrm{Nbody}$ that would have the same fractional rank in the smoothed $N$-body distribution (eq. \eqref{eq:remapping-eulerian-improved}). This remapping is illustrated for 2LPT with three sample points: $\delta_\mathrm{2LPT}=-0.60$ maps to $\delta_\mathrm{Nbody}=-0.56$, $\delta_\mathrm{2LPT}=0.00$ maps to $\delta_\mathrm{Nbody}=-0.01$, and $\delta_\mathrm{2LPT}=3.50$ maps to $\delta_\mathrm{Nbody}=3.56$. The remapping procedure imposes the one-point distribution of the smoothed $N$-body field while maintaining the rank order of the LPT-evolved density fields. The last step is an increase of small-scale power in the remapped distribution using the reciprocal transfer function, eq. \eqref{eq:transfer-function-reciprocal}.}
\label{fig:remapping_procedure_eulerian}
\end{center}
\end{figure*}

In this section, we describe the remapping algorithm used to go from a low-redshift realization of a density field evolved with LPT to one evolved with full $N$-body gravitational dynamics. Note that both fields obey the same initial conditions but are evolved by different physical models.

Density fields are defined on Cartesian grids of cubic voxels. Linear gravitational evolution exactly maintains the relative amplitude of fluctuations in different voxels. Due to mode coupling, positive and negative fluctuations grow at different rates in the non-linear regime, but even non-linear evolution tends to preserve the \textit{rank order} of the voxels, sorted by density.

The one-point probability distribution functions (PDF) and the cumulative distribution functions (CDF) of the final density fields, evolved with either LPT or full $N$-body gravitational dynamics, exhibit similar, but not identical shapes. This result suggests a way to improve the approximation with information extracted from the $N$-body simulation: maintain the rank order of the voxels, but reassign densities so that the two CDFs match. The method therefore resembles the ``Gaussianization'' procedure proposed by \citetalias{Weinberg1992}, an attempt to reconstruct the initial conditions of a density field from its final CDF.

Let $\mathcal{P}_{\mathrm{LPT}}$ and $\mathcal{P}_{\mathrm{Nbody}}$ denote the probability distribution functions for the density contrast in the LPT and in the full $N$-body density fields, respectively. Let $\mathcal{C}_{\mathrm{LPT}}$ and $\mathcal{C}_{\mathrm{Nbody}}$ be their integrals, the cumulative distribution functions. $\mathcal{C}_{\mathrm{LPT}}(\delta_\mathrm{LPT})$ is the \textit{fractional rank} for $\delta_\mathrm{LPT}$ i.e. the probability that the density contrast at a given voxel is smaller than $\delta_\mathrm{LPT}$, $\mathcal{P}_{\mathrm{LPT}}(\delta \leq \delta_\mathrm{LPT})$, and the analogous for the $N$-body field. The remapping procedure works as follows. A voxel with rank order $\delta_\mathrm{LPT}$ is assigned a new density $\delta_\mathrm{Nbody}$ such that
\begin{equation}
\label{eq:remapping-eulerian}
\mathcal{C}_{\mathrm{LPT}}(\delta_\mathrm{LPT}) = \mathcal{C}_{\mathrm{Nbody}}(\delta_\mathrm{Nbody})
\end{equation}
(also see figure \ref{fig:remapping_procedure_eulerian} for a schematic outline of this method). The left panels of figure \ref{fig:remapping_procedure_eulerian} show $\mathcal{P}_{\mathrm{LPT}}$ (dashed curves) and the corresponding cumulative distributions, $\mathcal{C}_{\mathrm{LPT}}$ (solid curves). On the right panels, the dotted curves represent the PDF of the corresponding $N$-body realization, $\mathcal{P}_{\mathrm{Nbody}}$. Remapping assigns to a voxel with density contrast $\delta_\mathrm{LPT}$ and fractional rank $\varepsilon~=~\mathcal{C}_{\mathrm{LPT}}(\delta_\mathrm{LPT})$ the value of $\delta_\mathrm{Nbody}$ that would have the same fractional rank in the $N$-body distribution.

\begin{figure*}
\begin{center}
\includegraphics[width=0.85\textwidth]{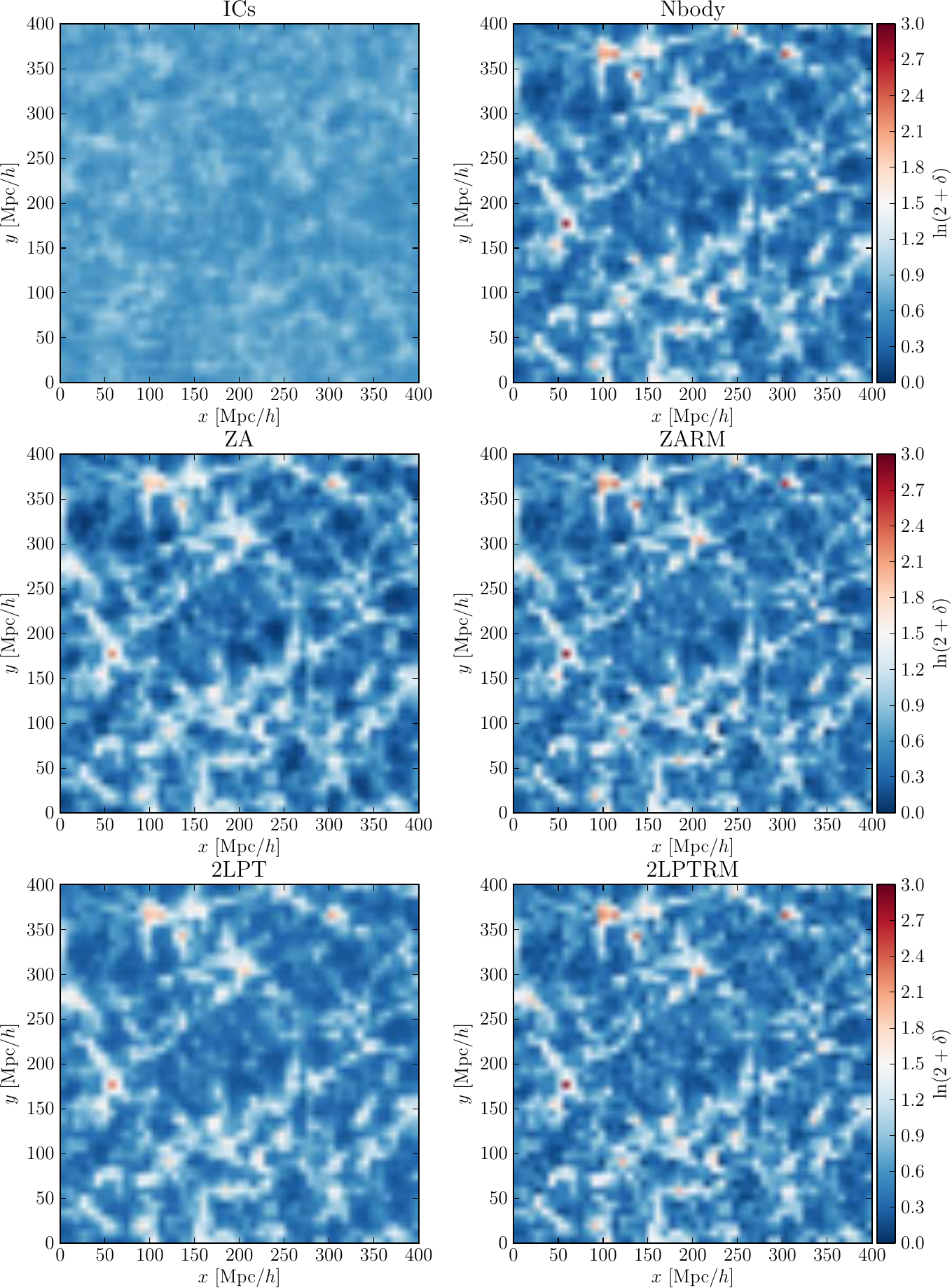}\\
\caption{Redshift-zero density contrast on a $128^2$-pixel slice of a $512^3$-particles realization in a 1024~Mpc/$h$ box with periodic boundary conditions. For clarity, the slice is limited to a square of 400~Mpc/$h$ side, and the quantity shown is the log-density field, $\ln(2+\delta)$. For comparison with the initial conditions, the density field at high redshift ($z=5$) is shown in the top left corner. The redshift-zero density fields are determined using, from top to bottom and from left to right: a full $N$-body simulation, the Zel'dovich approximation, alone (ZA) and after remapping (ZARM), second-order Lagrangian perturbation theory, alone (2LPT) and after remapping (2LPTRM). The remapping and the transfer function operations are performed on a $128^3$-voxel grid, corresponding to a mesh size of 8~Mpc/$h$.}
\label{fig:slices128}
\end{center}
\end{figure*}

Since $\mathcal{C}_{\mathrm{Nbody}}$ contains exactly the same information as $\mathcal{P}_{\mathrm{Nbody}}$, the remapping procedure imposes the one-point distribution taken from the $N$-body-evolved density field while maintaining the rank order of the LPT-evolved density field. In other words, only the weight of underdensities and overdensities is modified, while their locations remain unchanged (see also figure \ref{fig:slices128}). In this fashion, we seek to adjust the density field while maintaining higher-order statistics provided by LPT with reasonable accuracy. We checked numerically that mass is always conserved in this procedure.

\subsection{Comparison of structure types in LPT and in $N$-body dynamics}
\label{par:comparison-structure-types}

We implemented the remapping procedure described in the previous section and checked that we are able to modify LPT density fields so as to correctly reproduce the one-point distribution of a full $N$-body simulation. However, we experience a large-scale bias in the power spectrum, namely the amplitude of the spectrum of the remapped fields is slightly too high. Generally, a non-linear transformation in voxel space can change the variance of a field. This is consistent with the findings of \citeauthor{Weinberg1992} \citepalias{Weinberg1992}, who found a similar effect in his reconstructions of the initial conditions, and who addressed the issue by rescaling the density field in Fourier space. However, such an approach will generally destroy the remapped one-point distribution, and may even further lead to Gibbs ringing effects which will make the remapped field unphysical.

The bias indicates a stronger clustering in the remapped density field. Since remapping is a local operation in voxel space, this large-scale bias means that erroneous remapping of small-scale structures affects the clustering at larger scales. To identify the cause of this bias, we performed a study of differences in structure types in density fields predicted by LPT and $N$-body simulations. We employed the web-type classification algorithm proposed by \citet{Hahn2007}, which relies on estimating the eigenvalues of the Hessian of the gravitational potential. In particular, this algorithm dissects the voxels into four different web types (voids, sheets, filaments and clusters). With this analysis, we wanted to understand the effect of remapping in different density and dynamical regimes of the LSS. In particular, overdense clusters are objects in the strongly non-linear regime, far beyond shell-crossing, where predictions of LPT fail, while underdense voids are believed to be better apprehended \citepalias[e.g.][]{Bernardeau2002}.

As an indicator of the mismatch between the volume occupied by different structure types in LPT and $N$-body dynamics, we used the quantities $\gamma_t$ defined by
\begin{equation}
\label{eq:gamma}
\gamma_t \equiv \frac{N_{t}^{\mathrm{LPT}}-N_{t}^{\mathrm{Nbody}}}{N_{t}^{\mathrm{Nbody}}},
\end{equation}
where $t$ is one of the four structure types (voids, sheets, filaments, clusters), and $N_{t}^{\mathrm{LPT}}$ and $N_{t}^{\mathrm{Nbody}}$ are the numbers of voxels flagged as belonging to a structure of type $t$, in corresponding LPT and in $N$-body realizations, respectively.

\begin{figure*}
\begin{center}
\includegraphics[width=0.5\textwidth]{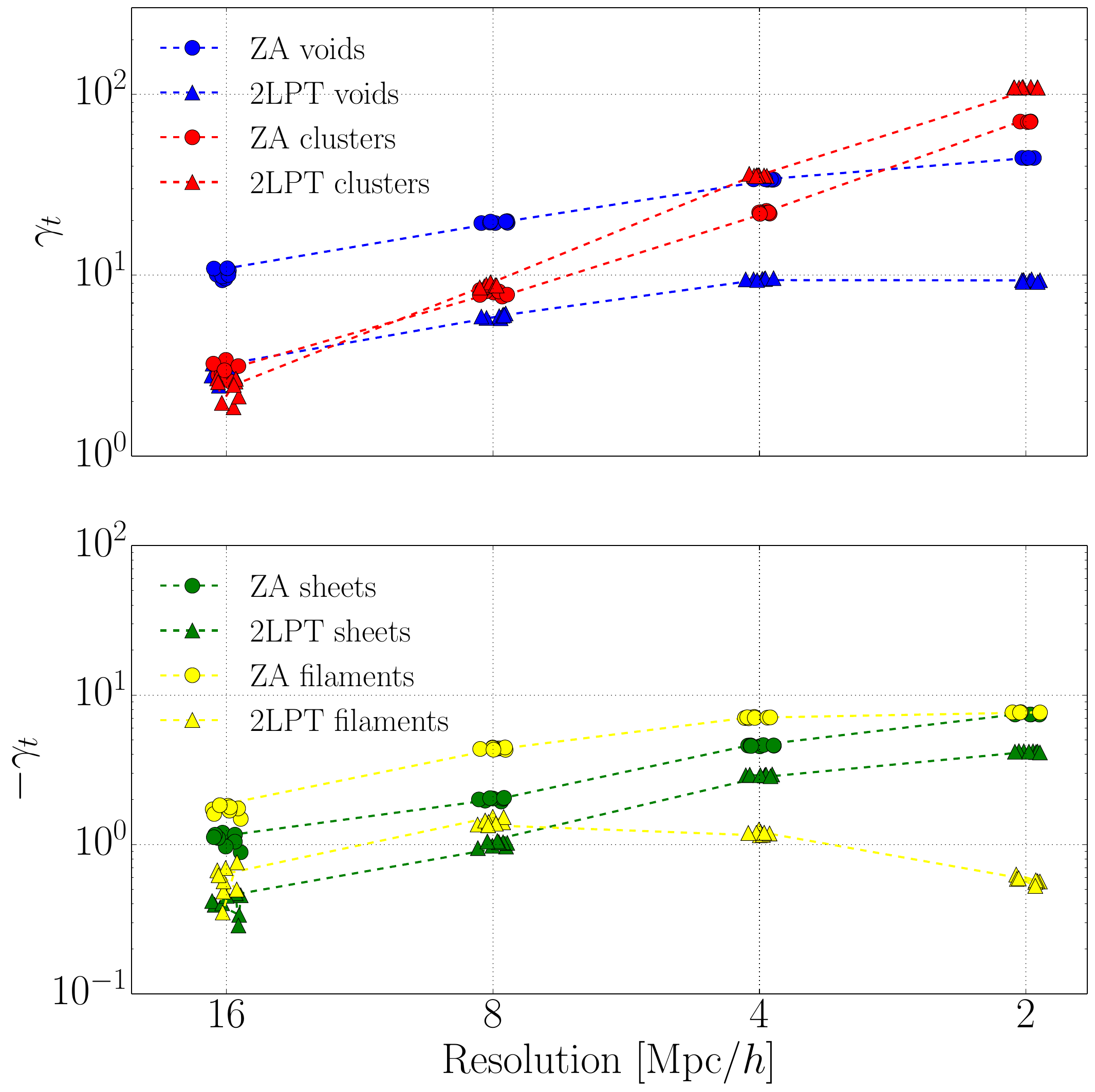}
\end{center}
\caption{Relative volume fraction of voids, sheets, filaments and clusters predicted by LPT, compared to $N$-body simulations, as a function of the resolution used for the definition of the density fields. The points are sightly randomized on the $x$-axis for clarity. The estimators $\gamma_t$ are defined by eq. \eqref{eq:gamma}. Eight realizations of the ZA (circles) and 2LPT (triangles) are compared to the corresponding $N$-body realization, for various resolutions. The volume fraction of incorrectly predicted structures in LPT generally increases with increasing resolution.}
\label{fig:structure_type_analysis}
\end{figure*}

In figure \ref{fig:structure_type_analysis}, we plot $\gamma_t$ as a function of the voxel size used to define the density fields. $\gamma_t$ is positive for clusters and voids, and negative for sheets and filaments, meaning that too large cluster and void regions are predicted in LPT, at the detriment of sheets and filaments. More specifically, LPT predicts fuzzier halos than $N$-body dynamics, and incorrectly predicts the surroundings of voids as part of them. This result indicates that even though LPT and $N$-body fields look visually similar, there are crucial differences in the representation of structure types. As demonstrated by figure \ref{fig:structure_type_analysis}, this mismatch increases with increasing resolution. This effect is of general interest when employing LPT in LSS data analysis.

\begin{figure*}
\begin{center}
\includegraphics[width=0.5\textwidth]{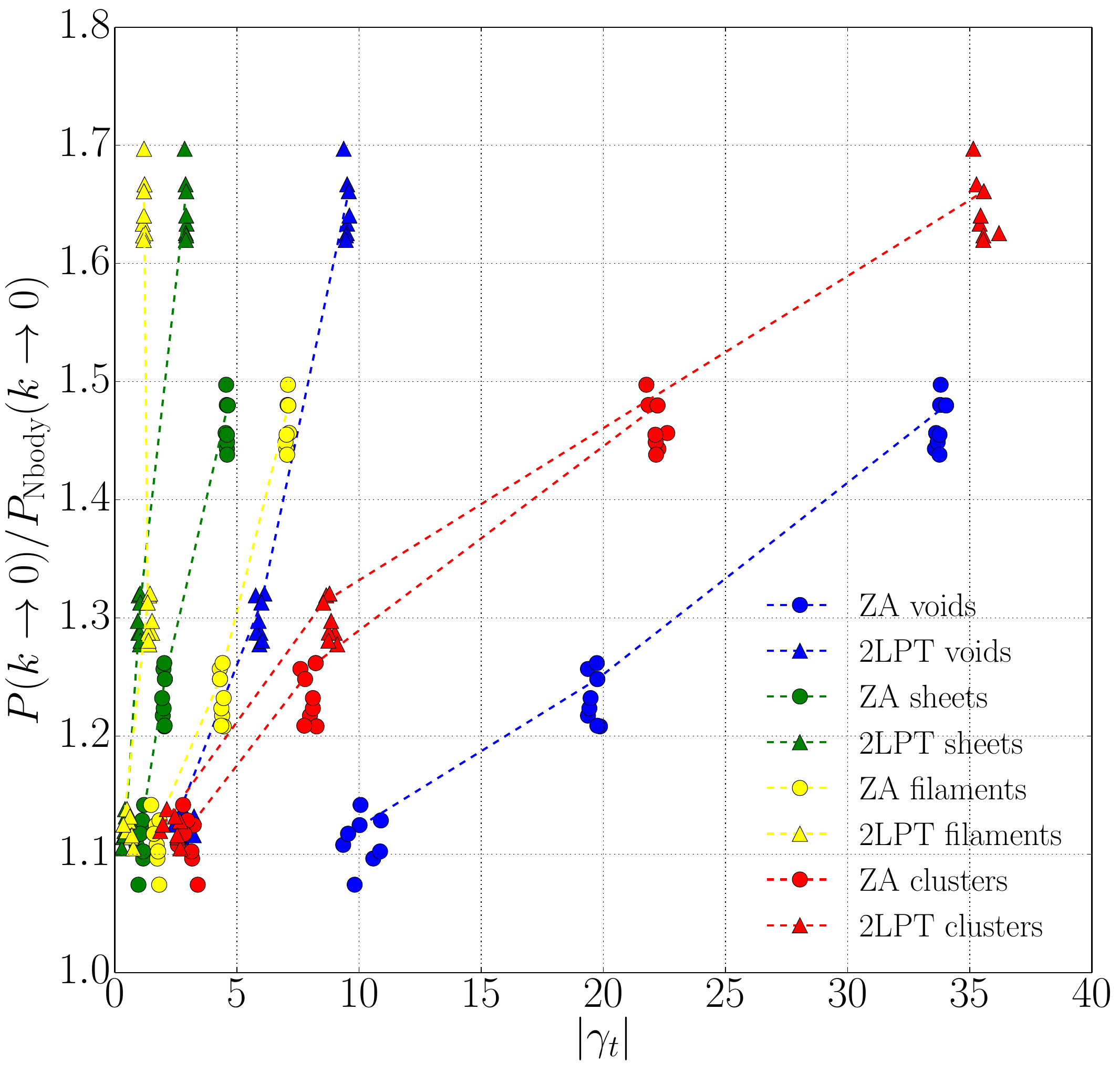}
\end{center}
\caption{The large-scale bias in the power spectrum of density fields, remapped with the procedure described in section \ref{par:remapping-procedure}, as a function of the mismatch between structure types. The estimators $\gamma_t$ are defined by eq. \eqref{eq:gamma}. Eight realizations of the ZA (circles) and 2LPT (triangles) are compared to the corresponding $N$-body realization, for various resolutions (from bottom to top: 16~Mpc/$h$, 8~Mpc/$h$, 4~Mpc/$h$). The large-scale bias in the power spectra of remapped fields is strongly correlated to the volume fraction of structures incorrectly predicted by LPT.}
\label{fig:bias_analysis}
\end{figure*}

In figure \ref{fig:bias_analysis}, we plot the large-scale bias observed in remapped fields obtained with the procedure of section \ref{par:remapping-procedure} as a function of $\gamma_t$, for various resolutions. A strong correlation is observed between the bias and the mismatch in the volume occupied by different structure types. The difference in the prediction of the volume of extended objects is the cause of the bias: in clusters and in voids, remapping enhances a larger volume than should be enhanced, which yields on average a stronger clustering in the box.

\subsection{Improvement of the remapping procedure}
\label{par:improvement-remapping-procedure}

In the previous section, we noted that because too much volume of LPT density fields is mapped to the tails of the $N$-body one-point distribution (dense clusters and deep voids), the average two-point correlations are larger after remapping. Given this result, minor modifications of the procedure described in section \ref{par:remapping-procedure} can solve the problem. We now propose to improve the remapping procedure of \citetalias{Weinberg1992} for present-day density fields, in a similar fashion as in \citet{Narayanan1998}, combining it with transfer function techniques to deal with the mildly non-linear modes as in \citet{Tassev2012a,Tassev2012b}. Our method works as follows.

\begin{enumerate}
\item We degrade the $N$-body density field to the same degree of smoothness as the LPT density field, by multiplying the Fourier modes of the density field by the transfer function
\begin{equation}
\label{eq:transfer-function}
T(k) \equiv \sqrt{\frac{P_{\mathrm{LPT}}(k)}{P_{\mathrm{Nbody}}(k)}} .
\end{equation}
This steps yields a new density field, noted Nbody,S(LPT), whose power spectrum matches that of the LPT density field.
\item We remap the LPT density field in the fashion described in section \ref{par:remapping-procedure}, but using as a reference the CDF of the smoothed density field, $\mathcal{C}_{\mathrm{Nbody,S(LPT)}}$, instead of the full $N$-body density field (see figure \ref{fig:remapping_procedure_eulerian}). The remapping condition, eq. \eqref{eq:remapping-eulerian}, now reads
\begin{equation}
\label{eq:remapping-eulerian-improved}
\mathcal{C}_{\mathrm{LPT}}(\delta_\mathrm{LPT}) = \mathcal{C}_{\mathrm{Nbody,S(LPT)}}(\delta_\mathrm{Nbody}) .
\end{equation}
\item We increase the power of small scales modes in the remapped distribution to the same value as in a full $N$-body simulation, using the reciprocal of the transfer function \eqref{eq:transfer-function}, namely
\begin{equation}
\label{eq:transfer-function-reciprocal}
T^{-1}(k) = \sqrt{\frac{P_{\mathrm{Nbody}}(k)}{P_{\mathrm{LPT}}(k)}} .
\end{equation}
\end{enumerate}

This procedure cures the large-scale bias issue experienced with the simple implementation of the remapping described in section \ref{par:remapping-procedure}, without requiring any prior knowledge on the corresponding $N$-body simulation. As we will demonstrate in section \ref{par:results}, it yields improvement of one-, two- and three-point statistics of LPT.

\subsection{Remapping function and transfer function}
\label{par:remapping-function}

Since $\mathcal{C}_{\mathrm{LPT}}$ and $\mathcal{C}_{\mathrm{Nbody,S(LPT)}}$ are monotonically increasing functions, there is no ambiguity in the choice of $\delta_\mathrm{Nbody}$, and this procedure defines a \textit{remapping function} $f$ such that
\begin{equation}
\label{eq:remapping-function-eulerian}
\delta_\mathrm{LPT} \mapsto \delta_\mathrm{Nbody} = \mathcal{C}_{\mathrm{Nbody,S(LPT)}}^{-1}(\mathcal{C}_{\mathrm{LPT}}(\delta_\mathrm{LPT})) \equiv f(\delta_\mathrm{LPT}) .
\end{equation}
Establishing a remapping function $f$ requires knowledge of both LPT and $N$-body density field statistics. Ideally, several realizations with different initial conditions should be combined in order to compute a precise remapping function. Indeed, a limited amount of available $N$-body simulations results in a lack of statistics and hence uncertainties for the remapping procedure in the high density regime. However, this effect is irrelevant from a practical point of view, since these high density events are very unlikely and affect only a negligible number of voxels. As a consequence this uncertainty will only affect to sub-percent level the usual statistical summaries of the density field. Note that in any case, if desired, the accuracy of the remapping function in the high density regime can be trivially enhanced by enlarging the size or number of $N$-body simulations used for its construction. For the analysis presented in this paper, the remapping functions have been computed using the full statistics from eight realizations of $512^3$ particles in a 1024 Mpc/$h$ box.

Note that once the relationship between the statistical behavior of the LPT fields and the full non-linear field is known, this procedure can be used on LPT realizations without the need of evolving corresponding $N$-body simulations. More specifically, the remapping function~$f$ (eq. \eqref{eq:remapping-function-eulerian}) and the transfer function~$T$ (eq. \eqref{eq:transfer-function}) can be tabulated and stored, then used for the fast construction of a large number of large-scale structure density fields. Since producing LPT realizations is computationally faster than running full gravitational simulations by a factor of several hundreds, our method can be used to produce a large set of $N$-body-like realizations in a short time.

\begin{figure*}
\begin{center}
\includegraphics[width=0.49\textwidth]{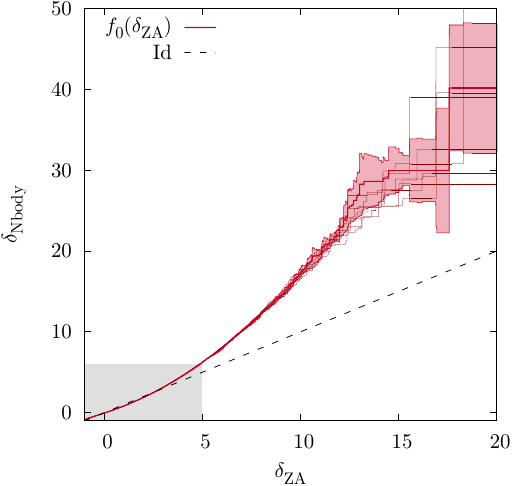}
\includegraphics[width=0.49\textwidth]{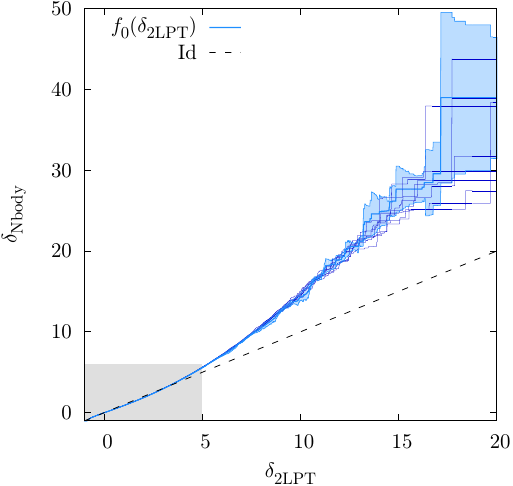} \\
\includegraphics[width=0.48\textwidth]{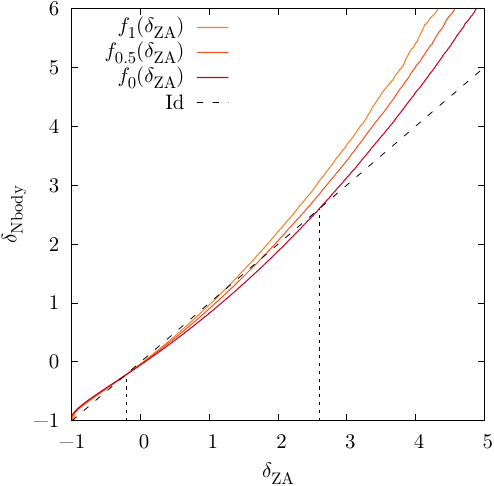}
\includegraphics[width=0.48\textwidth]{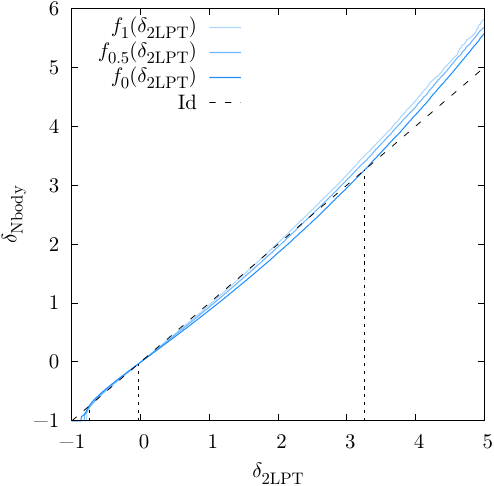} 
\end{center}
\caption{The remapping functions from LPT to smoothed $N$-body density fields, for the ZA (left panel) and 2LPT (right panel), all computed on a 8~Mpc/$h$ mesh. The precise redshift-zero remapping functions $f_0$ (red and blue solid curves) have been computed using the full statistics from eight realizations (darker red and blue solid curves). The error bars shown are the 1-$\sigma$ dispersion among the eight runs with reference to the full remapping function. The lower plots show the detail of the shaded area, in a density range containing most of the voxels. The redshift-dependence of the remapping function $f_z$ is shown for $z=1$, $z=0.5$ and $z=0$. The dashed line shows the identity function. Critical values of $\delta_{\mathrm{LPT}}$ for which remapping does not change the local density are identified.}
\label{fig:rE_function}
\end{figure*}

Some remapping functions are presented in figure \ref{fig:rE_function}. In each panel, the solid curves represent the remapping function $f_z$ at redshift $z$, computed with the LPT and $N$-body simulations. The dashed black line shows the identity function. We checked that the remapping function converges to the identity function with increasing redshift, as expected. Critical values where the remapping function crosses the identity function are identified. Between these critical values, remapping either increases or decreases the local density.

The PDFs for the density contrast are evaluated on a grid after a CiC assignment of particles. This means that the remapping function a priori depends on the size of voxels. The problem of choosing a voxel size for the computation of the remapping function is linked to the more general problem of choosing a mesh for the CiC evaluation of density. Choosing too coarse a binning will result in an underestimation of the clustering of particles, whereas choosing too fine a binning will also result in artifacts in overdensities (some voxels may be empty due to their too small size). The right choice of voxel size for the evaluation of the remapping function is the one giving the best evaluation of the density contrast. This choice has to be made depending on the desired application of the remapped data.

The remapping function describes how the PDF for the density contrast is affected by non-linear structure formation. For this reason, it depends on the nature of the gravitational interaction, as described by LPT and by full $N$-body dynamics, but weakly depends on the detail of the cosmological parameters. We checked the cosmology-dependence of the remapping function in simulations with the dark matter and baryon density in the Universe, $\Omega_\mathrm{m}$ and $\Omega_\mathrm{b}$, varying the fiducial values (eq. \eqref{eq:comological-parameters}) by $\pm$ $3 \sigma$ (still assuming a flat Universe):

\begin{eqnarray}
\label{eq:cosmological-parameters-1}
& \Omega_\Lambda = 0.750, \Omega_\mathrm{m} = 0.2494, \Omega_\mathrm{b} = 0.0428, \nonumber \\
& \sigma_8 = 0.810, h = 0.704, n_{\mathrm{s}} = 0.967;
\end{eqnarray}

\begin{eqnarray}
\label{eq:cosmological-parameters-2}
& \Omega_\Lambda = 0.700, \Omega_\mathrm{m} = 0.2992, \Omega_\mathrm{b} = 0.0488, \nonumber \\
& \sigma_8 = 0.810, h = 0.704, n_{\mathrm{s}} = 0.967.
\end{eqnarray}

\begin{figure*}
\begin{center}
\includegraphics[width=0.8\textwidth]{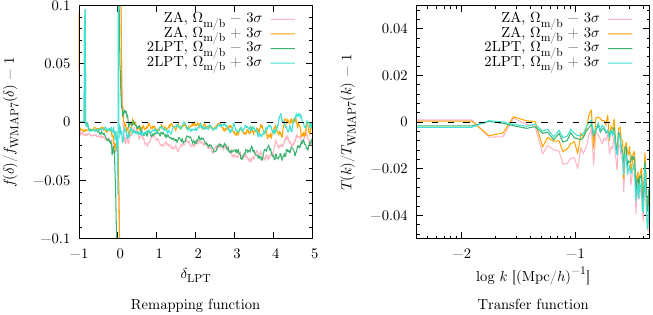}
\end{center}
\caption{Relative deviations of the remapping function (left panel) and of the transfer function (right panel) for varying cosmological parameters (eq. \eqref{eq:cosmological-parameters-1} and eq. \eqref{eq:cosmological-parameters-2}), with respect to their behaviour in a fiducial cosmology (eq. \eqref{eq:comological-parameters}).}
\label{fig:deviations_cosmology}
\end{figure*}

Even for these models notably far from the fiducial values, we found that the remapping function almost perfectly overlaps that of our main analysis, for the density range $\delta \in [-1;5]$, containing typically 98 to 99\% of the voxels. We found a difference of less than 5\% for $\delta = 5$ (see the left panel of figure \ref{fig:deviations_cosmology}).

The transfer function used in steps 1 and 3 of the improved procedure also exhibits very weak redshift-dependence, with deviations limited to a few percents at the smallest scales of interest of this work ($k~\approx~0.4$ (Mpc/$h$)$^{-1}$, see the right panel of figure \ref{fig:deviations_cosmology}).

\section{Statistics of remapped fields}
\label{par:results}

In this section, we discuss the validity of the improved remapping procedure described in section~\ref{par:improvement-remapping-procedure}, by studying the correlators of the remapped field in comparison to the input LPT and $N$-body fields. The remapping procedure based on the Eulerian density contrast essentially replaces the LPT one-point function by that of the smoothed $N$-body-evolved field. Since the position and shape of structures is left unchanged, we expect the higher-order correlators of the density field to be respected by the remapping procedure. Of particular interest is to check how remapping affects higher-order statistics and if possible improvements could be exploited in data analysis or artificial galaxy survey applications.

We implemented a numerical algorithm that computes and analyzes a remapped density field. The procedure can be divided in three steps:
\begin{enumerate}
\item We take as input two cosmological density fields, evolved from the same initial conditions with LPT (ZA or 2LPT) and with full $N$-body dynamics, and estimate the one-point statistics (PDF and CDF for $\delta$) and the transfer function for this particular realization. We repeat this step for the eight realizations used in our analysis.
\item We take as input all the one-point statistics computed with individual realizations, and we compute a precise remapping function using the full statistics of all available realizations, as described in section~\ref{par:remapping-procedure}. The transfer function used as a reference is the mean of all available realizations. At this point, the remapping function and the transfer function can be tabulated and stored for later use, and $N$-body simulations are no longer required.
\item For each realization, we remap the density field using the improved three-step procedure described in section~\ref{par:improvement-remapping-procedure} and we analyze its correlation functions.
\end{enumerate}

Our algorithm provides the one-point (section~\ref{par:eulerian-one-point-stats}) and two-point (section~\ref{par:eulerian-two-point-stats}) statistics. We used the code described in \citet{Gil-Marin2011,Gil-Marin2012} to study the three-point statistics (section~\ref{par:eulerian-three-point-stats}). The results are presented below.

\subsection{One-point statistics}
\label{par:eulerian-one-point-stats}

The remapping procedure, described in section~\ref{par:remapping}, is essentially a replacement of the cumulative distribution function of the density contrast $\delta$ of the input LPT-evolved field, $\mathcal{C}_{\mathrm{LPT}}$, by that of the reference $N$-body-evolved field after smoothing, $\mathcal{C}_{\mathrm{Nbody,S(LPT)}}$. After having applied the remapping procedure, we recomputed the PDF of the remapped field and verified that it matches that of the fiducial field as a sanity check.

Remapping and rescaling the density modes alters local density values but positions of structures remain unchanged. It is therefore important to check that remapping visually alters the LPT-evolved distribution in such a way that structures resemble more their $N$-body evolved counterparts. Figure \ref{fig:slices128} shows a slice of the density contrast $\delta$, measured at redshift zero, on a $128^2$-pixel sheet of a $512^3$-particles realization in a 1024~Mpc/$h$ box. The corresponding mesh size is 8~Mpc/$h$. Visually, remapped fields (ZARM and 2LPTRM) are closer to the full $N$-body result than their originals (ZA and 2LPT), with plausible particle distribution.

\begin{figure}
\begin{center}
\includegraphics[width=\columnwidth]{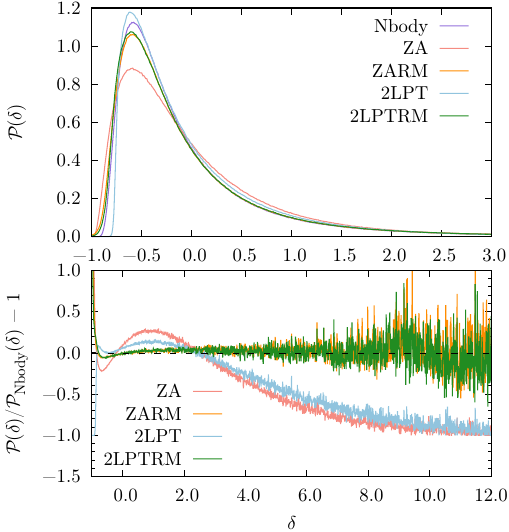}
\caption{\textit{Upper panel}. Redshift-zero probability distribution function for the density contrast $\delta$, computed from eight 1024~Mpc/$h$-box simulations of $512^3$ particles. The particle distribution is determined using: a full $N$-body simulation (purple curve), the Zel'dovich approximation, alone (ZA, light red curve) and after remapping (ZARM, orange curve), second-order Lagrangian perturbation theory, alone (2LPT, light blue curve) and after remapping (2LPTRM, green curve). \textit{Lower panel}. Relative deviations of the same PDFs with reference to $N$-body simulation results. Note that, contrary to standard LPT approaches, remapped fields follow the one-point distribution of full $N$-body dynamics in an unbiased way, especially in the high density regime.}
\label{fig:pdf}
\end{center}
\end{figure}

Since the improved remapping procedure involves a rescaling of the density modes in Fourier space (step 3), the PDF for the density contrast of the remapped fields is not guaranteed to be correct by construction, as would be the case with a naive remapping (section~\ref{par:remapping-procedure}). Therefore, the one-point distribution has to be carefully checked at this point. In figure \ref{fig:pdf}, we plot the probability distribution function for the density contrast at redshift zero for $N$-body simulations and the approximately evolved fields: with the ZA and 2LPT alone, and after remapping (ZARM and 2LPTRM). It can be observed that the peaks of the PDFs get closer to the reference set by $N$-body dynamics and that the PDF of remapped fields accurately follows that of full gravitational dynamics for $\delta > 0$. The procedure being successful on average for one-point statistics and accurate for the most common events in overdensities, we expect the number count of objects such as clusters predicted by LPT to be made more robust by our procedure.

\subsection{Two-point statistics}
\label{par:eulerian-two-point-stats}

\subsubsection{Power spectrum}
\label{par:eulerian-two-point-stats-power-spectrum}

\begin{figure}
\begin{center}
\includegraphics[width=\columnwidth]{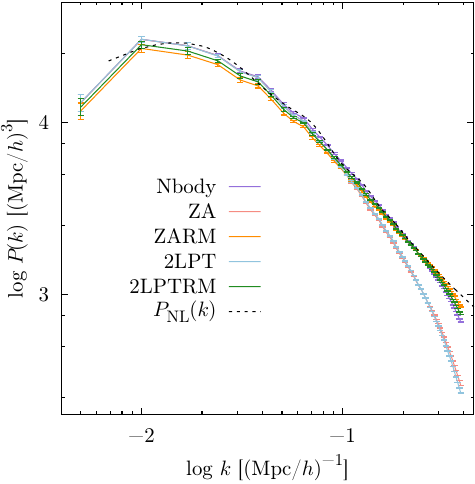}
\caption{Redshift-zero dark matter power spectra in a 1024~Mpc/$h$ simulation, with density fields computed with a mesh size of 8~Mpc/$h$. The particle distribution is determined using: a full $N$-body simulation (purple curve), the Zel'dovich approximation, alone (ZA, light red curve) and after remapping (ZARM, orange curve), second-order Lagrangian perturbation theory, alone (2LPT, light blue curve) and after remapping (2LPTRM, green curve). The dashed black curve represents $P_{\mathrm{NL}}(k)$, the theoretical power spectrum expected at $z=0$. Both ZARM and 2LPTRM show increased power in the mildly non-linear regime compared to ZA and 2LPT (at scales corresponding to $k~\gtrsim~0.1$~(Mpc/$h$)$^{-1}$ for this redshift), indicating an improvement of two-point statistics with the remapping procedure.}
\label{fig:PS}
\end{center}
\end{figure}

\begin{figure*}
\begin{center}
\includegraphics[width=\textwidth]{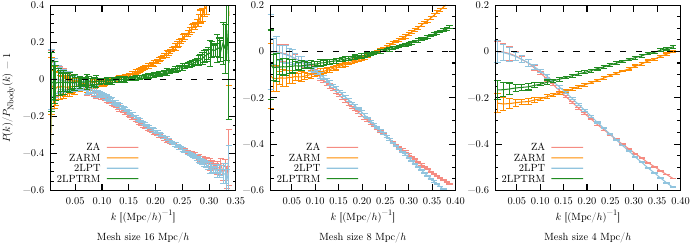}
\caption{\textit{Power spectrum: mesh size-dependence}. Relative deviations for the power spectra of various particle distributions, with reference to the density field computed with a full $N$-body simulation. The particle distribution is determined using: the Zel'dovich approximation, alone (ZA, light red curve) and after remapping (ZARM, orange curve), second-order Lagrangian perturbation theory, alone (2LPT, light blue curve) and after remapping (2LPTRM, green curve). The computation is done on different meshes: 16 Mpc/$h$ ($64^3$-voxel grid, left panel), 8 Mpc/$h$ ($128^3$-voxel grid, central panel) and 4 Mpc/$h$ ($256^3$-voxel grid, right panel). All results are shown at redshift $z=0$. LPT fields exhibit more small-scale correlations after remapping and their power spectra get closer to the shape of the full non-linear power spectrum.}
\label{fig:PS-deviations-mesh}
\end{center}
\end{figure*}

\begin{figure*}
\begin{center}
\includegraphics[width=\textwidth]{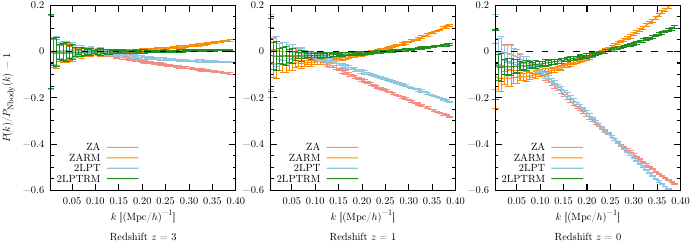}
\caption{\textit{Power spectrum: redshift-dependence}. Relative deviations for the power spectra of various particle distributions (see the caption of figure \ref{fig:PS-deviations-mesh}), with reference to the density field computed with a full $N$-body simulation. The computation is done on a 8 Mpc/$h$ mesh ($128^3$-voxel grid). Results at different redshifts are shown: $z=3$ (right panel), $z=1$ (central panel) and $z=0$ (left panel). The remapping procedure is increasingly successful with increasing redshift.}
\label{fig:PS-deviations-redshift}
\end{center}
\end{figure*}

We measured the power spectrum of dark matter particles, displaced according to each approximation and assigned to cells with a CiC scheme, for different mesh sizes. Power spectra were measured from theses meshes, with a correction for aliasing effects \citep{Jing2005}. Redshift-zero results computed on a 8~Mpc/$h$ mesh are presented in figure \ref{fig:PS}. There, the dashed line corresponds to the theoretical, non-linear power spectrum expected, computed with \textsc{Cosmic Emulator} tools \citep{Heitmann2009,Heitmann2010,Lawrence2010}. A deviation of full $N$-body simulations from this theoretical prediction can be observed at small scales. This discrepancy is a gridding artifact, completely due to the finite mesh size used for the analysis. The relative deviations of various power spectra with reference to the density field computed with a full $N$-body simulation are presented in figures \ref{fig:PS-deviations-mesh} and \ref{fig:PS-deviations-redshift}. In all the plots, the error bars represent the dispersion of the mean among eight independent realizations.

At high redshift ($z > 1$), we found no notable difference between the power spectrum of matter evolved with full $N$-body dynamics and that of LPT-remapped distributions. This indicates that our remapping procedure is increasingly successful as we go backwards in time towards the initial conditions, where LPT gives accurate results.

At low redshift, the power spectrum shape of remapped LPT fields is closer to the shape of the full non-linear power spectrum, turning down at smaller scales than the LPT power spectra. In particular, LPT fields exhibit more small-scale correlations after remapping, illustrating the success of our procedure in the mildly non-linear regime of large-scale structure formation. 

Contrary to the density fields obtained via a naive remapping approach, whose power spectra exhibit a positive artificial offset at large scales as discussed in section~\ref{par:comparison-structure-types}, the fields obtained with the improved procedure have correct two-point statistics  at all scales for coarse grids (down to 8~Mpc/$h$). For finer grids, a negative large-scale bias appears in the power spectrum, meaning that we have suppressed too much small-scale power in the $N$-body field in the first step of our procedure, which propagates to large scales with remapping. Comparing the panels of figures \ref{fig:PS-deviations-mesh} and \ref{fig:PS-deviations-redshift}, it can be observed that this effect is suppressed at higher redshift and for coarser binning. We found that a voxel size of 8~Mpc/$h$ is the best compromise, with a large-scale power spectrum compatible with the error bars and clear improvement at small scales, as can be seen in figure \ref{fig:PS}. This mesh size corresponds to the target resolution for analyses in the mildly non-linear regime, as discussed in the introduction.

\subsubsection{Fourier-space cross-correlation coefficient}
\label{par:eulerian-two-point-stats-cross-correlation}

\begin{figure}
\begin{center}
\includegraphics[width=\columnwidth]{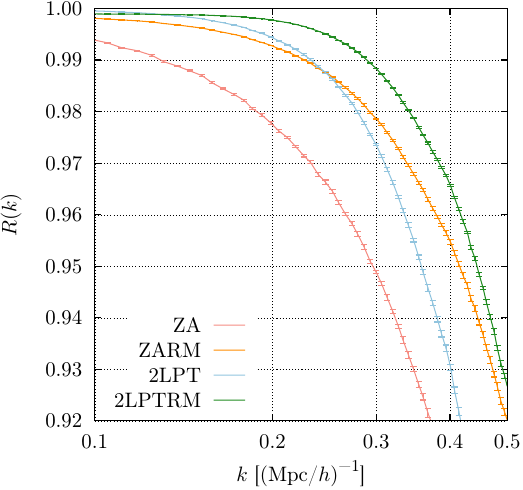}
\caption{Fourier-space cross-correlation coefficient between various approximately-evolved density fields and the particle distribution as evolved with full $N$-body dynamics, all at redshift zero. The binning of density fields is done on a 8~Mpc/$h$ mesh ($128^3$-voxel grid). At small scales, $k~\geq~0.2$~(Mpc/$h$)$^{-1}$, the cross-correlations with respect to the $N$-body-evolved field are notably better after remapping than with LPT alone.}
\label{fig:crosscorr}
\end{center}
\end{figure}

In figure \ref{fig:crosscorr}, we present the Fourier-space cross-correlation coefficient $R(k) \equiv P_{\delta \times \delta'}/\sqrt{P_\delta P_{\delta'}}$ between the redshift-zero density field in the $N$-body simulation and several other density fields. At this point, it is useful to recall that an approximation well-correlated with the non-linear density field can be used in a variety of cosmological applications, such as the reconstruction of the non-linear power spectrum \citep{Tassev2012b}. As pointed out by \citet{Neyrinck2013a}, the cross-correlation between 2LPT and full gravitational dynamics is higher at small $k$ than the cross-correlation between the ZA and the full dynamics, meaning that the position of structures is more correct when additional physics (non-local tidal effects) are taken into account.

At redshift zero and at small scales, the agreement is better with remapped LPT fields than with LPT alone, confirming the success of the remapping procedure to explore the mildly non-linear regime. In particular, the remapping of 2LPT predicts more than 96\% level accuracy at $k = 0.4$ (Mpc/$h$)$^{-1}$ (corresponding to scales of 16~Mpc/$h$), where 2LPT gives only 93\%. The cross-correlation coefficient indicates better agreements for the remapping of 2LPT than for the remapping of the ZA, which is consistent with the better performance of 2LPT in predicting the phases of the full $N$-body field.

\subsection{Three-point statistics}
\label{par:eulerian-three-point-stats}

\begin{figure}
\begin{center}
\includegraphics[width=\columnwidth]{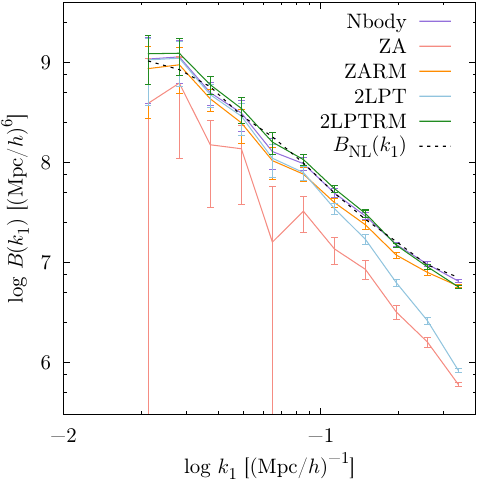}
\caption{Redshift-zero dark matter bispectra for equilateral triangle shape, in 1024~Mpc/$h$ simulations, with density fields computed on mesh of 8~Mpc/$h$ size. The particle distribution is determined using: a full $N$-body simulation (purple curve), the Zel'dovich approximation, alone (ZA, light red curve) and after remapping (ZARM, orange curve), second-order Lagrangian perturbation theory, alone (2LPT, light blue curve) and after remapping (2LPTRM, green curve). The dashed line, $B_{\mathrm{NL}}(k)$, corresponds to theoretical predictions for the bispectrum, found using the fitting formula of \citep{Gil-Marin2012}. Note that both ZARM and 2LPTRM show increased bispectrum in the mildly non-linear regime compared to ZA and 2LPT, indicating an improvement of three-point statistics with the remapping procedure.}
\label{fig:BS}
\end{center}
\end{figure}

We analyzed the accuracy of our method beyond second-order statistics, by studying the three-point correlation function in Fourier space, the bispectrum. The importance of three-point statistics relies in their ability to test the correctness of the shape of structures. Some of the natural applications are to test gravity \citep{Shiratra2007,Gil-Marin2011}, to break degeneracies in the galaxy bias \citep{Matarrese1997,Verde1998,Scoccimarro2001,Verde2002} or to test the existence of primordial non-Gaussianities in the initial matter density field \citep{Sefusatti2007,Jeong2009}.

As for the power spectrum, for the bispectrum we construct the dark matter density contrast field, namely $\delta(\textbf{k})$, putting particles in cells using a CiC scheme, and deconvolving them after performing the Fourier transform.

The algorithm used to compute the bispectrum $B(\textbf{k}_1,\textbf{k}_2,\textbf{k}_3)$ from this $\delta(\textbf{k})$ field consists of randomly drawing $k$-vectors from a specified bin, namely $\Delta k$, and randomly orientating the $(\textbf{k}_1,\textbf{k}_2,\textbf{k}_3)$ triangle in space. We chose the number of random triangles to depend on the number of fundamental triangle per bin, that scales as $k_1 k_2 k_3 \Delta k^3$ \citep{Scoccimarro1997}, where $\Delta k$ is the chosen $k$-binning: given $k_i$ we allow triangles whose $i$-side lies between $k_i-\Delta k/2$ and $k_i+\Delta k/2$. In this paper we always set $\Delta k = k_{\mathrm{min}} = 2\pi/L$, where $L$ is the size of the box. For the equilateral case, at scales of $k \approx 0.1$~(Mpc/$h$)$^{-1}$ we are generating $\sim 1.7 \times 10^6$ random triangles. We have verified that increasing the number of triangles beyond this value does not have any effect on the measurement.

As a rule of thumb, a maximum threshold in $k$ for trusting the simulation data is set by a quarter of the Nyquist frequency, defined by $k_{\mathrm{N}} = 2 \pi/L \times N_{\mathrm{v}}^{1/3}/2$, where $L$ is the size of the box and $N_{\mathrm{v}}$ is the number of voxels in the grid on which particles are placed in the initial conditions; which makes for our analysis ($L=1024$~Mpc/$h$, $N_{\mathrm{v}} = 512^3$), $k_\mathrm{N}/4~\approx~0.39$~(Mpc/$h$)$^{-1}$. At this scale, it has been observed that the power spectrum starts to deviate at the percent-level with respect to higher resolution simulations \citep{Heitmann2010}. For all plots and results shown in this paper, this limit in $k$ is never exceeded. Also, as a lower limit in $k$, we have observed that for scales larger than $\sim 3\,k_{\mathrm{min}}$ the effects of sample variance start to play an important role on the bispectrum and considerable deviations with respect to linear theory are observed. Therefore the largest scale for the bispectrum analysis is set to $3\,k_{\mathrm{min}} \approx 1.8 \times 10^{-2}~(\mathrm{Mpc}/h)^{-1}$.

The error bars presented in the bispectrum plots represent the dispersion of the mean among eight independent realizations, all of them with the same cosmological parameters. It has been tested \citep{Gil-Marin2012}, that this estimator for the error is in good agreement with theoretical predictions based on the Gaussianity of initial conditions \citep{Scoccimarro1998}.

Note that the subtracted shot noise is always assumed to be Poissonian:
\begin{equation}
B_{\mathrm{SN}}(\textbf{k}_1,\textbf{k}_2,\textbf{k}_3) = \frac{1}{\bar{n}} \left[ P(k_1)+P(k_2)+P(k_3) \right] + \frac{1}{\bar{n}^2} ,
\end{equation}
\citep[see e.g.][and references therein]{Peebles1980}, where $\bar{n}$ is the number density of particles in the box.

Throughout this paper, a triangle shape is defined by the relative length of vectors $\textbf{k}_1$ and $\textbf{k}_2$ and the inner angle $\theta_{12}$, in such a way that $\textbf{k}_1+\textbf{k}_2+\textbf{k}_3 = 0$ and $\textbf{k}_1~\cdot~\textbf{k}_2~=~k_1 k_2 \cos(\pi-\theta_{12})$. In figure \ref{fig:BS}, we plot the redshift-zero bispectrum, computed on a 8~Mpc/$h$ mesh, of the different density fields for equilateral triangles ($\theta_{12} = \pi/3$ and $k_2/k_1 = 1$). There, the dashed line corresponds to theoretical predictions for the non-linear bispectrum, found using the fitting formula of \citet{Gil-Marin2012}.

The overall result is a clear improvement of the bispectrum of LPT-evolved fields with the remapping procedure, especially on the small scales shown, probing the mildly non-linear regime, $k \gtrsim 0.1$ (Mpc/$h$)$^{-1}$ corresponding to scales $\lesssim 62$ Mpc/$h$, where LPT predicts less three-point correlation than full gravity. At large scales, the approximation error remains $\lesssim 1 \sigma$ of the estimated statistical uncertainty, even for a resolution of 8~Mpc/$h$ and at late times ($z=0$).

\begin{figure*}
\begin{center}
\includegraphics[width=\textwidth]{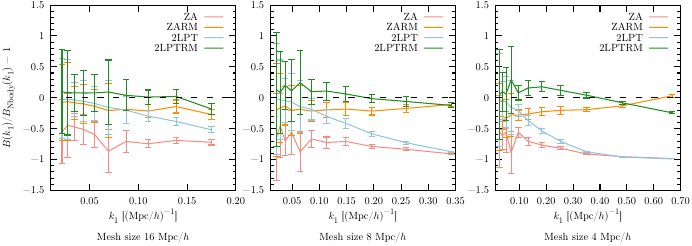}
\caption{\textit{Bispectrum: mesh size-dependence}. Relative deviations for the bispectra $B(k_1)$ of various particle distributions, with reference to the prediction from a full $N$-body simulation, $B_{\mathrm{Nbody}}(k_1)$. The particle distribution is determined using: the Zel'dovich approximation, alone (ZA, light red curve) and after remapping (ZARM, orange curve), second-order Lagrangian perturbation theory, alone (2LPT, light blue curve) and after remapping (2LPTRM, green curve). The computation of bispectra is done for equilateral triangles and on different meshes: 16 Mpc/$h$ ($64^3$-voxel grid, left panel), 8 Mpc/$h$ ($128^3$-voxel grid, central panel) and 4 Mpc/$h$ ($256^3$-voxel grid, right panel). All results are shown at redshift $z=0$. LPT fields exhibit more small-scale three-point correlations after remapping and their bispectra get closer to the shape of the full non-linear bispectrum.}
\label{fig:BS-deviations-mesh}
\end{center}
\end{figure*}

\begin{figure*}
\begin{center}
\includegraphics[width=\textwidth]{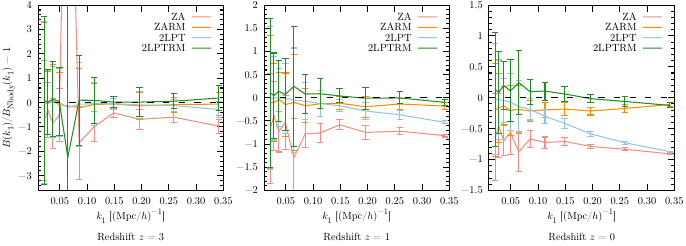}
\caption{\textit{Bispectrum: redshift-dependence}. Relative deviations for the bispectra $B(k_1)$ of various particle distributions (see the caption of figure \ref{fig:BS-deviations-mesh}), with reference to a full $N$-body simulation, $B_{\mathrm{Nbody}}(k_1)$. The computation of bispectra is done on a 8~Mpc/$h$ mesh ($128^3$-voxel grid) and for equilateral triangles. Results at different redshifts are shown: $z=3$ (right panel), $z=1$ (central panel) and $z=0$ (left panel).}
\label{fig:BS-deviations-redshift}
\end{center}
\end{figure*}

\begin{figure*}
\begin{center}
\includegraphics[width=\textwidth]{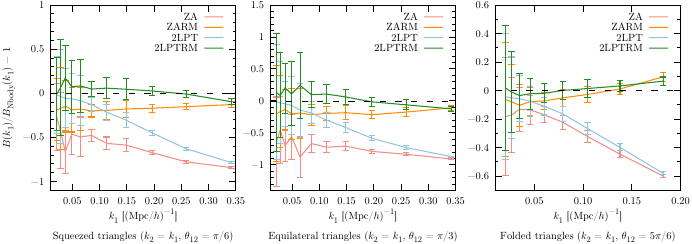}
\caption{\textit{Bispectrum: scale-dependence for different triangle shapes}. Relative deviations for the bispectra $B(k_1)$ of various particle distributions (see the caption of figure \ref{fig:BS-deviations-mesh}), with reference to a full $N$-body simulation, $B_{\mathrm{Nbody}}(k_1)$. The computation is done on a 8~Mpc/$h$ mesh ($128^3$-voxel grid) and results are shown at redshift $z=0$ for various triangle shapes as defined above.}
\label{fig:BS-deviations-triangle}
\end{center}
\end{figure*}

\begin{figure*}
\begin{center}
\includegraphics[width=\textwidth]{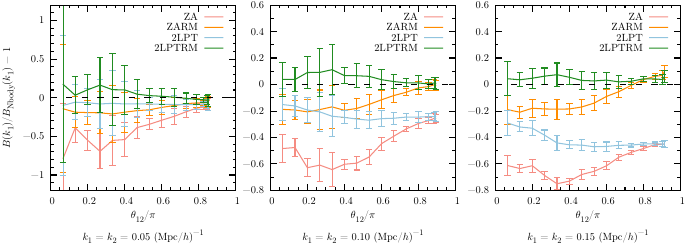}
\caption{\textit{Bispectrum: triangle shape-dependence}. Relative deviations for the bispectra $B(k_1)$ of various particle distributions (see the caption of figure \ref{fig:BS-deviations-mesh}), with reference to a full $N$-body simulation, $B_{\mathrm{Nbody}}(k_1)$. The computation is done on a 8~Mpc/$h$ mesh ($128^3$-voxel grid) and results are shown at redshift $z=0$. The dependence on the angle of the triangle $\theta_{12} = (\textbf{k}_1,\textbf{k}_2)$ is shown for different scales: $k_1=k_2=0.05~(\mathrm{Mpc}/h)^{-1}$ (corresponding to 125~Mpc/$h$), $k_1=k_2=0.10~(\mathrm{Mpc}/h)^{-1}$ (corresponding to 63~Mpc/$h$), $k_1=k_2=0.15~(\mathrm{Mpc}/h)^{-1}$ (corresponding to 42~Mpc/$h$).}
\label{fig:BS-deviations-theta}
\end{center}
\end{figure*}

The relative deviations of various bispectra with reference to full $N$-body simulations are shown in figures \ref{fig:BS-deviations-mesh}, \ref{fig:BS-deviations-redshift}, \ref{fig:BS-deviations-triangle} and \ref{fig:BS-deviations-theta}. As expected, the success of our remapping procedure in exploring small scales ($k \gtrsim 0.1$ (Mpc/$h$)$^{-1}$) is increased for more coarsely-binned density fields (see figure \ref{fig:BS-deviations-mesh}) and at higher redshift (see figure \ref{fig:BS-deviations-redshift}). In figure \ref{fig:BS-deviations-triangle} we examine the scale-dependence of the bispectrum for various triangle shapes. The precise dependence on the triangle shape at different scales is shown in figure \ref{fig:BS-deviations-theta}.

\section{Discussion and conclusion}
\label{par:conclusion}

The main subject of this paper is the development of a method designed to improve the correspondence between approximate models for gravitational dynamics and full numerical simulation of large-scale structure formation. Our methodology relies on a remapping of the one-point distribution of the density contrast of the approximately evolved particle distribution using information extracted from $N$-body simulations.

Due to the differences in the precise structure of the density fields in Lagrangian perturbation theory and in full gravity, the naive implementation of this procedure, inspired by \citetalias{Weinberg1992}, gives a large-scale bias in the power spectrum. This is not solved by a simple rescaling of Fourier modes, which leads to Gibbs ringing artifacts and an overall unphysical representation of the density fields. Smoothing LPT and $N$-body density fields with the same kernel is also unsuccessful, as smoothed $N$-body fields will always keep a sharper structure than smoothed LPT fields.

We figured out that the cause of this large-scale bias is not the different density predicted locally by LPT and $N$-body dynamics on a point-by-point basis, but a problem of mismatch between the volume of extended objects. Our findings question the reliability of LPT for LSS data analysis and generation of mock catalogs at low redshift and high mass resolution. They are also a likely explanation for the discrepancy between the power spectrum of initial conditions reconstructed via Gaussianization and linear theory expectations, encountered by \citetalias{Weinberg1992}.

Considering these results, we improved \citeauthor{Weinberg1992}'s remapping procedure for present-day density fields by the use of a transfer function. In this fashion, we obtain a physically more accurate representation of the three-dimensional matter distribution in the mildly non-linear regime, while improving higher-order statistics. Since LPT captures well the cosmological dependence and remapping operates on small-scale aspects of the density field, we found that our procedure is nearly independent of cosmological parameters.

The aim of this method is to develop a fast, flexible and efficient way to generate realizations of LSS density fields, accurately representing the mildly non-linear regime. Our procedure, therefore, responds to the increasing demand for numerically inexpensive models of three-dimensional LSS, for applications to modern cosmological data analysis. At the level of statistical error in our numerical experiments, the approach provides a good method for producing mock halo catalogs and low-redshift initial conditions for simulations, if desired. The resulting information can also be used in a variety of cosmological analyses of present and upcoming observations.

We showed that our approach allows fast generation of cosmological density fields that correlate with $N$-body simulations at better than 96\% down to scales of $k~\approx~0.4~(\mathrm{Mpc}/h)^{-1}$ at redshift zero and are substantially better than standard LPT results at higher redshifts on the same comoving scales. Remapping improves the fast LPT bispectrum predictions on small scales while the large scale bispectrum remains accurate to within about 1$\sigma$ of the measurement in our $N$-body simulations. Since real observations will have larger statistical errors for the foreseeable future, our method provides an adequate fast model of the non-linear density field on scales down to $\sim~8$~Mpc/$h$. These results constitute a substantial improvement with respect to existing techniques, since non-linearities begin to affect even large-scale measurements in galaxy surveys. Since the number of modes usable for cosmological exploitation scale as $k^3$, even minor improvements in the smallest scale $k$ allow access to much more knowledge from existing and upcoming observations. This work is a step further in the non-linear regime, which contains a wealth of yet unexploited cosmological information. For possible applications, we provided a cosmographic and statistical characterization of approximation errors.

Our remapping procedure predicts the two-point correlation function at around 95\% level accuracy and three-point correlation function at around 80\% level accuracy at redshift 3, for $k$ between 0.1 and 0.4 $(\mathrm{Mpc}/$h$)^{-1}$, illustrating the increasing success of our methods as we go backwards in time towards the initial conditions, when LPT is an accurate description of early structure formation. This is of particular interest in several areas of high-redshift cosmology, such as forecasting 21~cm surveys \citep{Lidz2007}, analyzing the properties of the intergalactic medium via the Lyman-$\alpha$ forest \citep{Kitaura2012} or probing the reionization epoch \citep{Mesinger2007}. This work might also add to methods of data analysis of several ongoing (\textsc{Hetdex}\footnote{\href{http://hetdex.org/}{http://hetdex.org/}}) and upcoming high-redshift galaxy surveys (\textsc{SDSS IV: eBOSS\footnote{\href{http://www.sdss3.org/future/eboss.php}{http://www.sdss3.org/future/eboss.php}}, DESI\footnote{\href{http://desi.lbl.gov/}{http://desi.lbl.gov/}}}).

However, the realization of density fields with these procedures stays approximate, since the full non-linear gravitational physics involves information contained in the shape of structures, which cannot be captured from a one-point modification of LPT, especially after shell-crossing. We studied the performance of one-point remapping of LPT and presented a statistical characterization of the errors, but additional refinements, such as a non-linear, density-dependent smoothing of the $N$-body field, could further improve on these approximations, for an increased computational cost. This is, however, beyond the scope and intent of this paper. Generally, the complications at large scales that we encounter when applying a local remapping seem difficult to solve in a Eulerian density field approach and would favor a Lagrangian, particle-based perspective, which will be a subject of further study.

Fast and accurate methods to model the non-linearly evolved mass distribution in the Universe have the potential of profound influence on modern cosmological data analysis. Recently proposed full Bayesian large scale structure inference methods, which extract information on the matter distribution in the Universe from galaxy redshift surveys, rely on Lagrangian perturbation theory \citep{Jasche2013}. The methods proposed in this paper will provide a numerically efficient and flexible extension of these methods, permitting us to push dynamic analyses of the large scale structure further into the non-linear regime. In particular, we envision our techniques to add to methods of physical inference and reconstruction of initial conditions from which the large-scale structure of the Universe originates. We may hope that this approach will yield insights into galaxy formation, a more detailed view on cosmic structure formation, on the nature of dark energy, as well as on physics of the early Universe by studying, for example, primordial non-Gaussianity via higher-order statistics.

\appendix*

\vspace{30pt}
\noindent \textsf{\textbf{Addendum: One-point statistics of the Lagrangian displacement field}}
\vspace{10pt}

\begin{figure}
\begin{center}
\includegraphics[width=\columnwidth]{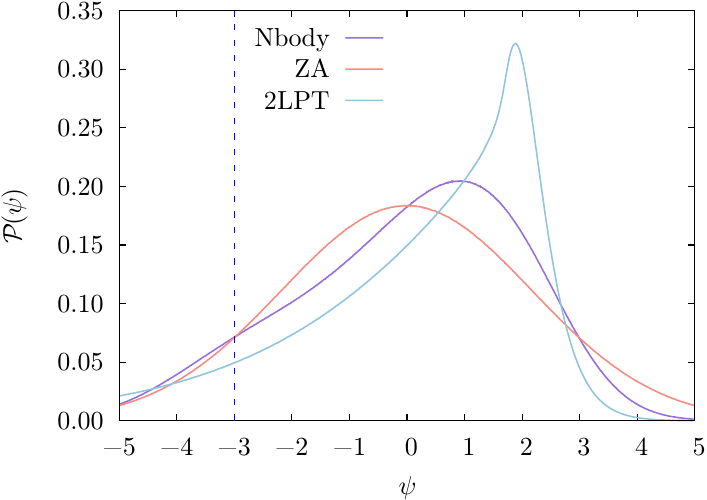}
\caption{Redshift-zero probability distribution function for the divergence of the displacement field $\psi$, computed from eight 1024~Mpc/$h$-box simulations of $512^3$ particles. A quantitative analysis of the deviation from Gaussianity of these PDFs is given in table \ref{tb:NGparam}. The particle distribution is determined using: a full $N$-body simulation (purple curve), the Zel'dovich approximation (ZA, light red curve) and second-order Lagrangian perturbation theory (2LPT, light blue curve). The vertical line at $\psi =-3$ represents the collapse barrier about which $\psi$ values bob around after gravitational collapse. A bump at this value is visible with full gravity, but LPT is unable to reproduce this feature. This regime corresponds to virialized, overdense clusters.}
\label{fig:divpsi_distrib}
\end{center}
\end{figure}

\begin{table}\centering
\begin{tabular}{lcc}
\hline\hline
Model & $\mathcal{P}_\delta$ & $\mathcal{P}_\psi$ \\
\hline
\multicolumn{1}{c}{} & \multicolumn{2}{c}{Skewness $\gamma_1$} \\
ZA & $2.36 \pm 0.01$ & $-0.0067 \pm 0.0001$ \\
2LPT & $2.83 \pm 0.01$ & $-1.5750 \pm 0.0002$ \\
$N$-body & $5.14 \pm 0.05$ & $-0.4274 \pm 0.0001$ \\
\hline
\multicolumn{1}{c}{} & \multicolumn{2}{c}{Excess kurtosis $\gamma_2$} \\
ZA & $9.95 \pm 0.09$ & $-2.2154 \times 10^{-6} \pm 0.0003$ \\
2LPT & $13.91 \pm 0.15$ & $3.544 \pm 0.0011$ \\
$N$-body & $62.60 \pm 2.75$ & $-0.2778 \pm 0.0004$ \\
\hline\hline
\end{tabular}
\caption{Non-Gaussianity parameters (the skewness $\gamma_1$ and the excess kurtosis $\gamma_2$) of the redshift-zero probability distribution functions $\mathcal{P}_\delta$ and $\mathcal{P}_\psi$ of the density contrast $\delta$ and the divergence of the displacement field $\psi$, respectively. The confidence intervals given correspond to the 1-$\sigma$ standard deviations among eight realizations. In all cases, $\gamma_1$ and $\gamma_2$ are reduced when measured from $\psi$ instead of $\delta$.}
\label{tb:NGparam}
\end{table}

\begin{figure*}
\begin{center}
\includegraphics[width=0.85\textwidth]{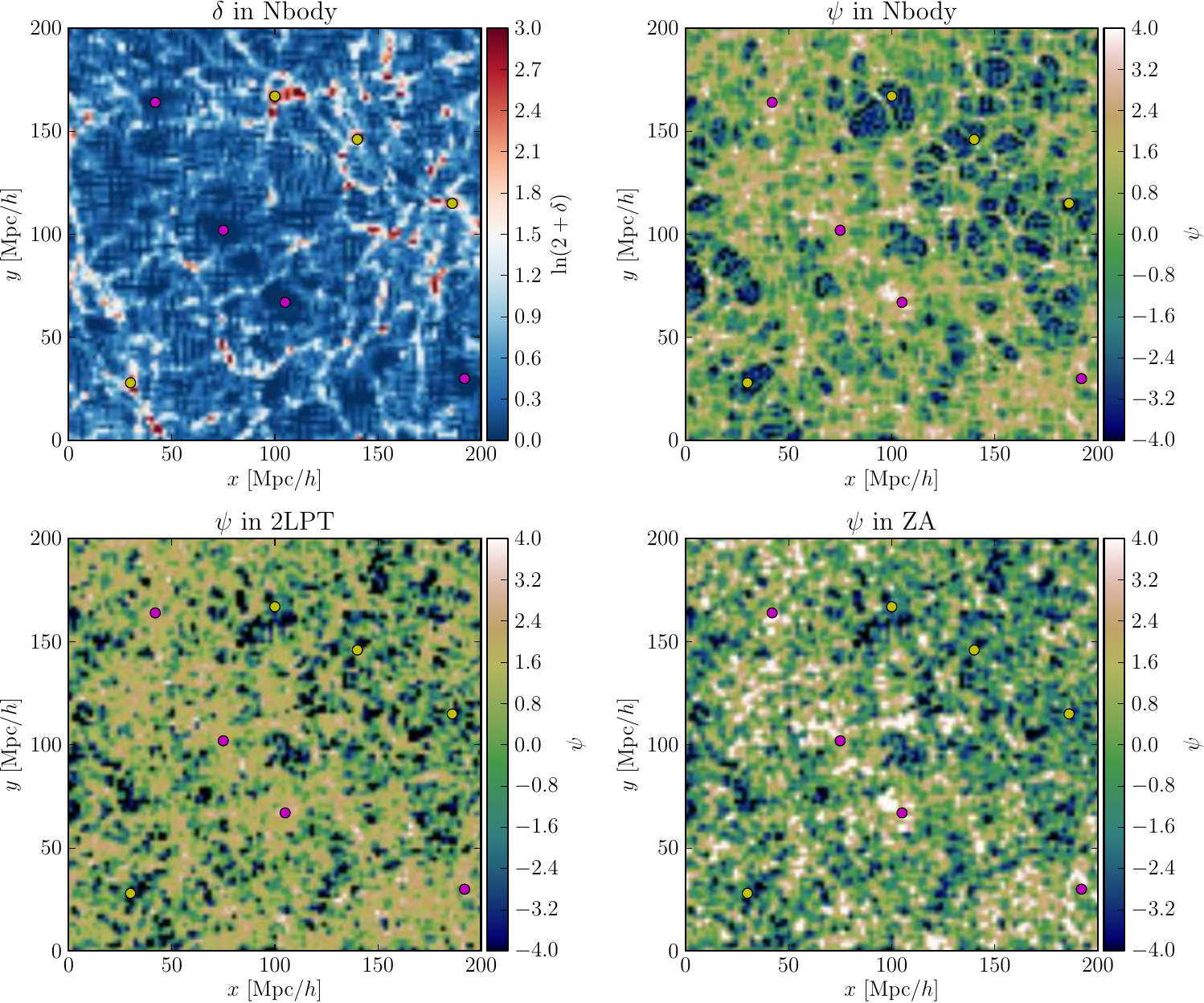}
\caption{Slices of the divergence of the displacement field, $\psi$, on a Lagrangian sheet of $512^2$ particles from a $512^3$-particle simulation of box size 1024 Mpc/$h$, run to redshift zero. For clarity we show only a 200~Mpc/$h$ region. Each pixel corresponds to a particle. The particle distribution is determined using respectively a full $N$-body simulation, the Zel'dovich approximation (ZA) and second-order Lagrangian perturbation theory (2LPT). In the upper left panel, the density contrast $\delta$ in the $N$-body simulation is shown, after binning on a $512^3$-voxel grid. To guide the eye, some clusters and voids are identified by yellow and purple dots, respectively. The ``lakes'', Lagrangian regions that have collapsed to form halos, are only visible in the $N$-body simulation, while the ``mountains'', Lagrangian regions corresponding to cosmic voids, are well reproduced by LPT.}
\label{fig:slices_divpsi}
\end{center}
\end{figure*}

\begin{figure*}
\begin{center}
\includegraphics[width=0.35\textwidth]{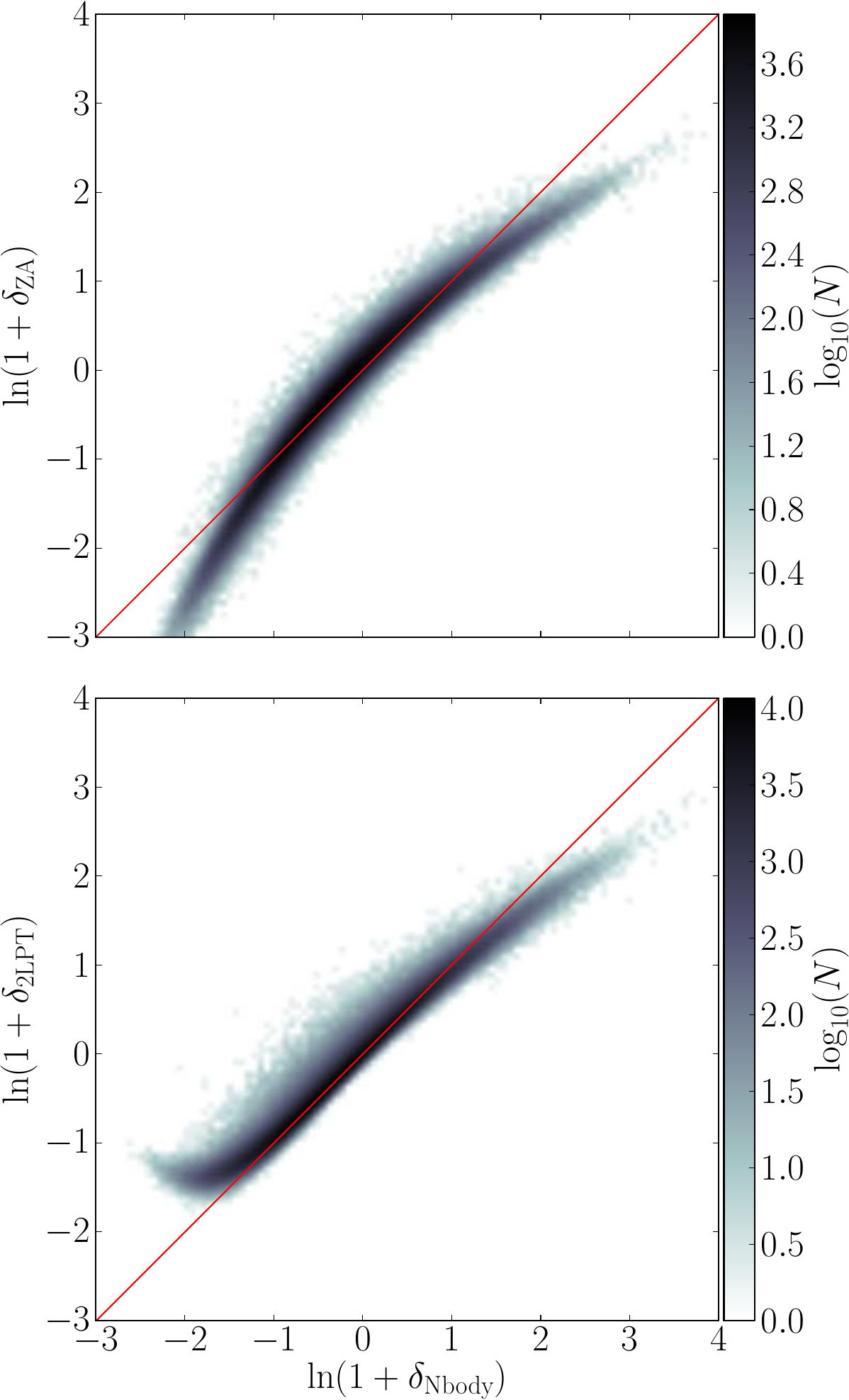} \quad \includegraphics[width=0.35\textwidth]{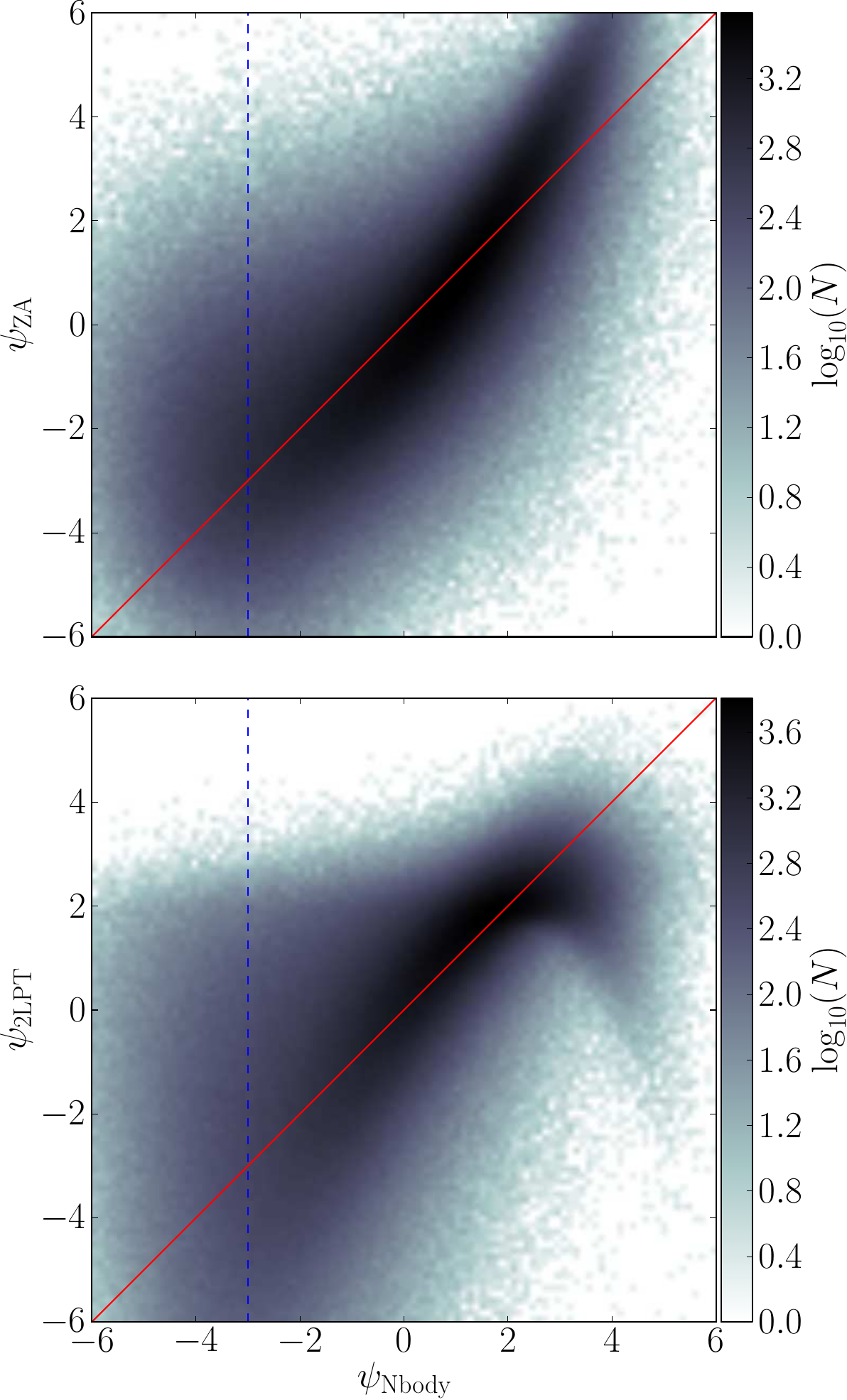}
\caption{\emph{Left panel}. Two-dimensional histograms comparing particle densities evolved with full $N$-body dynamics (the $x$-axis) to densities in the LPT-evolved particle distributions (the $y$-axis). The red lines show the ideal $y = x$ locus. A turn-up at low densities is visible with 2LPT, meaning that some overdense regions are predicted where there should be deep voids. \emph{Right panel}. Same plot for the divergence of the displacement field $\psi$. Negative $\psi$ corresponds to overdensities and positive $\psi$ correspond to underdensities. The dotted blue line shows the collapse barrier at $\psi = -3$ where particle get clustered in full gravity. The scatter is bigger with $\psi$ than with $\delta$, in particular in overdensities, since with LPT, particles do not cluster. The turn-up at low densities with 2LPT, observed with the density contrast, is also visible with the divergence of the displacement field.}
\label{fig:scatters}
\end{center}
\end{figure*}

The remapping procedure, described in section \ref{par:remapping}, relies on the Eulerian density contrast. As noted by previous authors \citep[see in particular][]{Neyrinck2013a}, in the Lagrangian representation of the LSS, it is natural to use the divergence of the displacement field $\psi$ instead of the Eulerian density contrast $\delta$. This addendum provides additional comments on the one-point statistics of $\psi$ and comparatively analyzes key features of $\psi$ and $\delta$.

In the Lagrangian frame, the quantity of interest is not the position, but the displacement field $\Psi(\textbf{q})$ which maps the initial comoving particle position $\textbf{q}$ to its final comoving Eulerian position \textbf{x} (see e.g. \citealp{Bouchet1995} or \citealp{Bernardeau2002} for overviews),
\begin{equation}
\label{eq:Lagrangian-Eulerian-mapping}
\textbf{x} \equiv \textbf{q} + \Psi(\textbf{q}) .
\end{equation}
It is important to note that, though $\Psi(\textbf{q})$ is \emph{a priori} a full three-dimensional vector field, it is curl-free up to second order in Lagrangian perturbation theory (appendix D in \citealp{Bernardeau1994} or \citealp{Bernardeau2002} for a review). We did not consider contributions beyond 2LPT. After publication of the main paper, \citet{Chan2014} analyzed the non-linear evolution of $\Psi$, splitting it into its scalar and vector parts (the so-called ``Helmholtz decomposition''). Looking at two-point statistics, he found that shell-crossing leads to a suppression of small-scale power in the scalar part, and, subdominantly, to the generation of a vector contribution.

Let $\psi(\textbf{q}) \equiv \nabla_{\textbf{q}} \cdot \Psi(\textbf{q})$ denote the divergence of the displacement field, where $\nabla_{\textbf{q}}$ is the divergence operator in Lagrangian coordinates. $\psi$ quantifies the angle-averaged spatial-stretching of the Lagrangian dark matter ``sheet'' in comoving coordinates \citep{Neyrinck2013a}. Let $\mathcal{P}_{\psi,\mathrm{LPT}}$ and $\mathcal{P}_{\psi,\mathrm{Nbody}}$ be the one-point probability distribution functions for the divergence of the displacement field in LPT and in $N$-body fields, respectively. We denote by $\mathcal{P}_\delta$ the corresponding PDFs for the Eulerian density contrast.

In figure \ref{fig:divpsi_distrib}, we show the PDFs of $\psi$ for the ZA, 2LPT and full $N$-body gravity. The most important feature of $\psi$ is that, whatever the model for structure formation, the PDF exhibits reduced non-Gaussianity compared to the PDF for the density contrast $\delta$ (see the upper panel of figure \ref{fig:pdf} for comparison). The main reason is that $\mathcal{P}_{\delta}$, unlike $\mathcal{P}_\psi$, is tied down to zero at $\delta = -1$. It is highly non-Gaussian in the final conditions, both in $N$-body simulations and in approximations to the true dynamics. For a quantitative analysis, we looked at the first and second-order non-Gaussianity statistics: the skewness $\gamma_1$ and the excess kurtosis $\gamma_2$,
\begin{equation}
\gamma_1 \equiv \frac{\mu_3}{\sigma^3} \quad \mathrm{and} \quad \gamma_2 \equiv \frac{\mu_4}{\sigma^4}-3,
\end{equation}
where $\mu_n$ is the $n$-th moment about the mean and $\sigma$ is the standard deviation. We estimated $\gamma_1$ and $\gamma_2$ at redshift zero in our simulations, in the one-point statistics of the density contrast $\delta$ and of the divergence of the displacement field $\psi$. The results are shown in table \ref{tb:NGparam}. In all cases, we found that both $\gamma_1$ and $\gamma_2$ are much smaller when measured from $\mathcal{P}_\psi$ instead of $\mathcal{P}_\delta$.

At linear order in Lagrangian perturbation theory (the Zel'dovich approximation), the divergence of the displacement field is proportional to the density contrast in the initial conditions, $\delta(\mathbf{q})$, scaling with the negative growth factor, $-D_1(\tau)$:
\begin{equation}
\label{eq:mapping-ZA}
\psi^{(1)}(\textbf{q},\tau) = \nabla_\textbf{q} \cdot \Psi^{(1)}(\textbf{q},\tau) = -D_1(\tau) \delta(\textbf{q}) .
\end{equation}
Since we take Gaussian initial conditions, the PDF for $\psi$ is Gaussian at any time with the ZA. In full gravity, non-linear evolution slightly breaks Gaussianity. $\mathcal{P}_{\psi,\mathrm{Nbody}}$ is slightly skewed towards negative values while its mode gets shifted around $\psi \approx 1$. Taking into account non-local effects, 2LPT tries to get closer to the shape observed in $N$-body simulations, but the correct skewness is overshot and the PDF is exceedingly peaked.

Figure \ref{fig:slices_divpsi} shows a slice of the divergence of the displacement field, measured at redshift zero for particles occupying a flat $512^2$-pixel Lagrangian sheet from one of our simulations. For comparison, see also the figures in \citet{Mohayaee2006,Pueblas2009,Neyrinck2013a}. We used the color scheme of the latter paper, suggesting a topographical analogy when working in Lagrangian coordinates. As structures take shape, $\psi$ departs from its initial value; it takes positive values in underdensities and negative values in overdensities. The shape of voids (the ``mountains'') is found to be reasonably similar in LPT and in the $N$-body simulation. For this reason, the influence of late-time non-linear effects in voids is milder as compared to overdense structures, which makes them easier to relate to the initial conditions. However, in overdense regions where $\psi$ decreases, it is not allowed to take arbitrary values: where gravitational collapse occurs, ``lakes'' form and $\psi$ gets stuck around a collapse barrier, $\psi \approx -3$. As expected, these ``lakes'', corresponding to virialized clusters, can only be found in $N$-body simulations, since LPT fails to accurately describe the highly non-linear physics involved. A small bump at $\psi=-3$ is visible in $\mathcal{P}_{\psi,\mathrm{Nbody}}$ (see figure \ref{fig:divpsi_distrib}). We checked that this bump gets more visible in higher mass-resolution simulations (200 Mpc/$h$ box for $256^3$ particles), where matter is more clustered. This means that part of the information about gravitational clustering can be found in the one-point statistics of $\psi$. Of course, the complete description of halos requires to precisely account for the shape of the ``lakes'', which can only be done via higher-order correlation functions. More generally, it is possible to use Lagrangian information in order to classify structures of the cosmic web. In particular, \textsc{diva} \citep{Lavaux2010} uses the shear of the displacement field and \textsc{origami} \citep{Falck2012} the number of phase-space folds. As pointed out by \citet{Falck2015}, while these techniques cannot be straightforwardly used for the analysis of galaxy surveys, where we lack Lagrangian information, recently proposed techniques for physical inference of the initial conditions \citep{Jasche2013,Jasche2015} should allow their use with observational data.

Figure \ref{fig:scatters} shows two-dimensional histograms comparing $N$-body simulations to the LPT realizations for the density contrast $\delta$ and the divergence of the displacement field $\psi$. At this point, it is useful to note that a good mapping exists in the case where the relation shown is monotonic and the scatter is narrow. As pointed out by \citet{Sahni1996} and \citet{Neyrinck2013a}, matter in the substructure of 2LPT-voids has incorrect statistical properties: there are overdense particles in the low density region of the 2LPT $\delta$-scatter plot. This degeneracy is also visible in the $\psi > 0$ region of the 2LPT $\psi$-scatter plot. On average, the scatter is bigger with $\psi$ than with $\delta$, in particular in overdensities ($\psi <0$), since with LPT, particles do not cluster: $\psi$ takes any value between 2 and $-3$ where it should remain around $-3$.

Summing up our discussions in this addendum, we analyzed the relative merits of the Lagrangian divergence of the displacement field $\psi$, and the Eulerian density contrast $\delta$ at the level of one-point statistics. The important differences are the following:

\begin{enumerate}
\item $\Psi$ being irrotational up to order two, its divergence $\psi$ contains nearly all information on the displacement field in one dimension, instead of three. The collapse barrier at $\psi=-3$ is visible in $\mathcal{P}_\psi$ for $N$-body simulations but not for LPT. A part of the information about non-linear gravitational clustering is therefore encoded in the one-point statistics of $\psi$.
\item $\psi$ exhibits much fewer gravitationally-induced non-Gaussian features than $\delta$ in the final conditions (figure \ref{fig:divpsi_distrib} and table \ref{tb:NGparam}).
\item However, the values of $\psi$ are more scattered than the values of $\delta$ with respect to the true dynamics (figure \ref{fig:scatters}), meaning that an unambiguous mapping is more difficult.
\end{enumerate}

\textit{Note added}. While this addendum was being refeered, the work of \citet{Neyrinck2015} appeared. It uses the spherical collapse prescription for $\psi$ while checking various scales for the initial density field. The result is a fast scheme for producing approximate particle realizations.

\acknowledgments

We are grateful to Francis Bernardeau, Nico Hamaus, David Langlois, Mark Neyrinck, Alice Pisani, Marcel Schmittfull, Svetlin Tassev and Mat\'ias Zaldarriaga for useful discussions. We express our gratitude to the anonymous referee for a number of helpful comments which helped improve the quality of this paper. We thank St\'ephane Rouberol for his support during the course of this work. FL, JJ and BW gratefully acknowledge the hospitality of the Institute for Advanced Study at Princeton during the late stages of this work. FL acknowledges funding from an AMX grant (\'Ecole polytechnique/Universit\'e Pierre et Marie Curie). JJ is partially supported by a Feodor Lynen Fellowship by the Alexander von Humboldt foundation. HGM is grateful for support from the UK Science and Technology Facilities Council through the grant ST/I001204/1. BW acknowledges support from NSF grants AST 07-08849 and AST 09-08693 ARRA and from the Agence Nationale de la Recherche via the Chaire d'Excellence ANR-10-CEXC-004-01, and computational resources provided through XSEDE grant AST100029. This material is based upon work supported in part by the National Science Foundation under Grant No. PHYS-1066293 and the hospitality of the Aspen Center for Physics.

\bibliography{Remapping_v4}

\begin{thebibliography}{75}%
\makeatletter
\providecommand \@ifxundefined [1]{%
 \@ifx{#1\undefined}
}%
\providecommand \@ifnum [1]{%
 \ifnum #1\expandafter \@firstoftwo
 \else \expandafter \@secondoftwo
 \fi
}%
\providecommand \@ifx [1]{%
 \ifx #1\expandafter \@firstoftwo
 \else \expandafter \@secondoftwo
 \fi
}%
\providecommand \natexlab [1]{#1}%
\providecommand \enquote  [1]{``#1''}%
\providecommand \bibnamefont  [1]{#1}%
\providecommand \bibfnamefont [1]{#1}%
\providecommand \citenamefont [1]{#1}%
\providecommand \href@noop [0]{\@secondoftwo}%
\providecommand \href [0]{\begingroup \@sanitize@url \@href}%
\providecommand \@href[1]{\@@startlink{#1}\@@href}%
\providecommand \@@href[1]{\endgroup#1\@@endlink}%
\providecommand \@sanitize@url [0]{\catcode `\\12\catcode `\$12\catcode
  `\&12\catcode `\#12\catcode `\^12\catcode `\_12\catcode `\%12\relax}%
\providecommand \@@startlink[1]{}%
\providecommand \@@endlink[0]{}%
\newcommand{\PineGreen}[1]{\textcolor{PineGreen}{#1}}%
\providecommand \url  [0]{\begingroup\@sanitize@url \@url }%
\providecommand \@url [1]{\endgroup\@href {#1}{\urlprefix }}%
\providecommand \urlprefix  [0]{URL }%
\providecommand \Eprint [0]{\href }%
\providecommand \doibase [0]{http://dx.doi.org/}%
\providecommand \selectlanguage [0]{\@gobble}%
\providecommand \bibinfo  [0]{\@secondoftwo}%
\providecommand \bibfield  [0]{\@secondoftwo}%
\providecommand \translation [1]{[#1]}%
\providecommand \BibitemOpen [0]{}%
\providecommand \bibitemStop [0]{}%
\providecommand \bibitemNoStop [0]{.\EOS\space}%
\providecommand \EOS [0]{\spacefactor3000\relax}%
\providecommand \BibitemShut  [1]{\csname bibitem#1\endcsname}%
\let\auto@bib@innerbib\@empty
\bibitem [{{Bernardeau}(1994)\citenamefont {{Bernardeau}}}]{Bernardeau1994}%
{(\PineGreen{{Bernardeau}}, \PineGreen{1994})}  \BibitemOpen
  \bibfield  {author} {\bibinfo {author} {\bibfnamefont {F.}~\bibnamefont
  {{Bernardeau}}},\ }\emph {{The nonlinear evolution of rare events}},\ \href
  {\doibase 10.1086/174121} {\bibfield  {journal} {\bibinfo  {journal} {\apj}\
  }\textbf {\bibinfo {volume} {427}},\ \bibinfo {pages} {51} (\bibinfo {year}
  {1994})},\ \Eprint {http://arxiv.org/abs/astro-ph/9311066}
  {astro-ph/9311066}\BibitemShut {NoStop}%
\bibitem [{{Bernardeau} {\textit{et~al}}\mbox{.}(2002)\citenamefont
  {{Bernardeau}}, \citenamefont {{Colombi}}, \citenamefont {{Gazta{\~n}aga}},\
  \&\ \citenamefont {{Scoccimarro}}}]{Bernardeau2002}%
{(\PineGreen{{Bernardeau} {\textit{et~al}}\mbox{.}}, \PineGreen{2002})}
  \BibitemOpen
  \bibfield  {author} {\bibinfo {author} {\bibfnamefont {F.}~\bibnamefont
  {{Bernardeau}}}, \bibinfo {author} {\bibfnamefont {S.}~\bibnamefont
  {{Colombi}}}, \bibinfo {author} {\bibfnamefont {E.}~\bibnamefont
  {{Gazta{\~n}aga}}}, \bibinfo {author} {\bibfnamefont {R.}~\bibnamefont
  {{Scoccimarro}}},\ }\emph {{Large-scale structure of the Universe and
  cosmological perturbation theory}},\ \href {\doibase
  10.1016/S0370-1573(02)00135-7} {\bibfield  {journal} {\bibinfo  {journal}
  {\physrep}\ }\textbf {\bibinfo {volume} {367}},\ \bibinfo {pages} {1}
  (\bibinfo {year} {2002})},\ \Eprint {http://arxiv.org/abs/astro-ph/0112551}
  {astro-ph/0112551}\BibitemShut {NoStop}%
\bibitem [{{Bouchet} {\textit{et~al}}\mbox{.}(1995)\citenamefont {{Bouchet}},
  \citenamefont {{Colombi}}, \citenamefont {{Hivon}},\ \&\ \citenamefont
  {{Juszkiewicz}}}]{Bouchet1995}%
{(\PineGreen{{Bouchet} {\textit{et~al}}\mbox{.}}, \PineGreen{1995})}
  \BibitemOpen
  \bibfield  {author} {\bibinfo {author} {\bibfnamefont {F.~R.}\ \bibnamefont
  {{Bouchet}}}, \bibinfo {author} {\bibfnamefont {S.}~\bibnamefont
  {{Colombi}}}, \bibinfo {author} {\bibfnamefont {E.}~\bibnamefont {{Hivon}}},
  \bibinfo {author} {\bibfnamefont {R.}~\bibnamefont {{Juszkiewicz}}},\ }\emph
  {{Perturbative Lagrangian approach to gravitational instability}},\
  \href@noop {} {\bibfield  {journal} {\bibinfo  {journal} {\aap}\ }\textbf
  {\bibinfo {volume} {296}},\ \bibinfo {pages} {575} (\bibinfo {year}
  {1995})},\ \Eprint {http://arxiv.org/abs/astro-ph/9406013}
  {astro-ph/9406013}\BibitemShut {NoStop}%
\bibitem [{{Buchert}(1989)\citenamefont {{Buchert}}}]{Buchert1989}%
{(\PineGreen{{Buchert}}, \PineGreen{1989})}  \BibitemOpen
  \bibfield  {author} {\bibinfo {author} {\bibfnamefont {T.}~\bibnamefont
  {{Buchert}}},\ }\emph {{A class of solutions in Newtonian cosmology and the
  pancake theory}},\ \href@noop {} {\bibfield  {journal} {\bibinfo  {journal}
  {\aap}\ }\textbf {\bibinfo {volume} {223}},\ \bibinfo {pages} {9} (\bibinfo
  {year} {1989})}\BibitemShut {NoStop}%
\bibitem [{{Buchert}, {Melott} \& {Weiss}(1994)\citenamefont {{Buchert}},
  \citenamefont {{Melott}},\ \&\ \citenamefont {{Weiss}}}]{Buchert1994}%
{(\PineGreen{{Buchert}, {Melott} \& {Weiss}}, \PineGreen{1994})}  \BibitemOpen
  \bibfield  {author} {\bibinfo {author} {\bibfnamefont {T.}~\bibnamefont
  {{Buchert}}}, \bibinfo {author} {\bibfnamefont {A.~L.}\ \bibnamefont
  {{Melott}}}, \bibinfo {author} {\bibfnamefont {A.~G.}\ \bibnamefont
  {{Weiss}}},\ }\emph {{Testing higher-order Lagrangian perturbation theory
  against numerical simulations I. Pancake models}},\ \href@noop {} {\bibfield
  {journal} {\bibinfo  {journal} {\aap}\ }\textbf {\bibinfo {volume} {288}},\
  \bibinfo {pages} {349} (\bibinfo {year} {1994})},\ \Eprint
  {http://arxiv.org/abs/astro-ph/9309056} {astro-ph/9309056}\BibitemShut
  {NoStop}%
\bibitem [{{Chan}(2014)\citenamefont {{Chan}}}]{Chan2014}%
{(\PineGreen{{Chan}}, \PineGreen{2014})}  \BibitemOpen
  \bibfield  {author} {\bibinfo {author} {\bibfnamefont {K.~C.}\ \bibnamefont
  {{Chan}}},\ }\emph {{Helmholtz decomposition of the Lagrangian
  displacement}},\ \href {\doibase 10.1103/PhysRevD.89.083515} {\bibfield
  {journal} {\bibinfo  {journal} {\prd}\ }\textbf {\bibinfo {volume} {89}},\
  \bibinfo {eid} {083515} (\bibinfo {year} {2014})},\ \Eprint
  {http://arxiv.org/abs/1309.2243} {arXiv:1309.2243}\BibitemShut {NoStop}%
\bibitem [{{Coles}, {Melott} \& {Shandarin}(1993)\citenamefont {{Coles}},
  \citenamefont {{Melott}},\ \&\ \citenamefont {{Shandarin}}}]{Coles1993}%
{(\PineGreen{{Coles}, {Melott} \& {Shandarin}}, \PineGreen{1993})}
  \BibitemOpen
  \bibfield  {author} {\bibinfo {author} {\bibfnamefont {P.}~\bibnamefont
  {{Coles}}}, \bibinfo {author} {\bibfnamefont {A.~L.}\ \bibnamefont
  {{Melott}}}, \bibinfo {author} {\bibfnamefont {S.~F.}\ \bibnamefont
  {{Shandarin}}},\ }\emph {{Testing approximations for non-linear gravitational
  clustering}},\ \href@noop {} {\bibfield  {journal} {\bibinfo  {journal}
  {\mnras}\ }\textbf {\bibinfo {volume} {260}},\ \bibinfo {pages} {765}
  (\bibinfo {year} {1993})}\BibitemShut {NoStop}%
\bibitem [{{Crocce}, {Pueblas} \& {Scoccimarro}(2006)\citenamefont {{Crocce}},
  \citenamefont {{Pueblas}},\ \&\ \citenamefont {{Scoccimarro}}}]{Crocce2006b}%
{(\PineGreen{{Crocce}, {Pueblas} \& {Scoccimarro}}, \PineGreen{2006})}
  \BibitemOpen
  \bibfield  {author} {\bibinfo {author} {\bibfnamefont {M.}~\bibnamefont
  {{Crocce}}}, \bibinfo {author} {\bibfnamefont {S.}~\bibnamefont {{Pueblas}}},
  \bibinfo {author} {\bibfnamefont {R.}~\bibnamefont {{Scoccimarro}}},\ }\emph
  {{Transients from initial conditions in cosmological simulations}},\ \href
  {\doibase 10.1111/j.1365-2966.2006.11040.x} {\bibfield  {journal} {\bibinfo
  {journal} {\mnras}\ }\textbf {\bibinfo {volume} {373}},\ \bibinfo {pages}
  {369} (\bibinfo {year} {2006})},\ \Eprint
  {http://arxiv.org/abs/astro-ph/0606505} {astro-ph/0606505}\BibitemShut
  {NoStop}%
\bibitem [{{Crocce} \& {Scoccimarro}(2006)\citenamefont {{Crocce}}\ \&\
  \citenamefont {{Scoccimarro}}}]{Crocce2006a}%
{(\PineGreen{{Crocce} \& {Scoccimarro}}, \PineGreen{2006})}  \BibitemOpen
  \bibfield  {author} {\bibinfo {author} {\bibfnamefont {M.}~\bibnamefont
  {{Crocce}}}, \bibinfo {author} {\bibfnamefont {R.}~\bibnamefont
  {{Scoccimarro}}},\ }\emph {{Renormalized cosmological perturbation theory}},\
  \href {\doibase 10.1103/PhysRevD.73.063519} {\bibfield  {journal} {\bibinfo
  {journal} {\prd}\ }\textbf {\bibinfo {volume} {73}},\ \bibinfo {eid} {063519}
  (\bibinfo {year} {2006})},\ \Eprint {http://arxiv.org/abs/astro-ph/0509418}
  {astro-ph/0509418}\BibitemShut {NoStop}%
\bibitem [{{Croft} {\textit{et~al}}\mbox{.}(1998)\citenamefont {{Croft}},
  \citenamefont {{Weinberg}}, \citenamefont {{Katz}},\ \&\ \citenamefont
  {{Hernquist}}}]{Croft1998}%
{(\PineGreen{{Croft} {\textit{et~al}}\mbox{.}}, \PineGreen{1998})}
  \BibitemOpen
  \bibfield  {author} {\bibinfo {author} {\bibfnamefont {R.~A.~C.}\
  \bibnamefont {{Croft}}}, \bibinfo {author} {\bibfnamefont {D.~H.}\
  \bibnamefont {{Weinberg}}}, \bibinfo {author} {\bibfnamefont
  {N.}~\bibnamefont {{Katz}}}, \bibinfo {author} {\bibfnamefont
  {L.}~\bibnamefont {{Hernquist}}},\ }\emph {{Recovery of the Power Spectrum of
  Mass Fluctuations from Observations of the Ly alpha Forest}},\ \href
  {\doibase 10.1086/305289} {\bibfield  {journal} {\bibinfo  {journal} {\apj}\
  }\textbf {\bibinfo {volume} {495}},\ \bibinfo {pages} {44} (\bibinfo {year}
  {1998})},\ \Eprint {http://arxiv.org/abs/arXiv:astro-ph/9708018}
  {arXiv:astro-ph/9708018}\BibitemShut {NoStop}%
\bibitem [{{Croft} {\textit{et~al}}\mbox{.}(1999)\citenamefont {{Croft}},
  \citenamefont {{Weinberg}}, \citenamefont {{Pettini}}, \citenamefont
  {{Hernquist}},\ \&\ \citenamefont {{Katz}}}]{Croft1999}%
{(\PineGreen{{Croft} {\textit{et~al}}\mbox{.}}, \PineGreen{1999})}
  \BibitemOpen
  \bibfield  {author} {\bibinfo {author} {\bibfnamefont {R.~A.~C.}\
  \bibnamefont {{Croft}}}, \bibinfo {author} {\bibfnamefont {D.~H.}\
  \bibnamefont {{Weinberg}}}, \bibinfo {author} {\bibfnamefont
  {M.}~\bibnamefont {{Pettini}}}, \bibinfo {author} {\bibfnamefont
  {L.}~\bibnamefont {{Hernquist}}}, \bibinfo {author} {\bibfnamefont
  {N.}~\bibnamefont {{Katz}}},\ }\emph {{The Power Spectrum of Mass
  Fluctuations Measured from the LYalpha Forest at Redshift Z=2.5}},\ \href
  {\doibase 10.1086/307438} {\bibfield  {journal} {\bibinfo  {journal} {\apj}\
  }\textbf {\bibinfo {volume} {520}},\ \bibinfo {pages} {1} (\bibinfo {year}
  {1999})},\ \Eprint {http://arxiv.org/abs/arXiv:astro-ph/9809401}
  {arXiv:astro-ph/9809401}\BibitemShut {NoStop}%
\bibitem [{{Doroshkevich}(1970)\citenamefont
  {{Doroshkevich}}}]{Doroshkevich1970}%
{(\PineGreen{{Doroshkevich}}, \PineGreen{1970})}  \BibitemOpen
  \bibfield  {author} {\bibinfo {author} {\bibfnamefont {A.~G.}\ \bibnamefont
  {{Doroshkevich}}},\ }\emph {{The space structure of perturbations and the
  origin of rotation of galaxies in the theory of fluctuation.}},\ \href@noop
  {} {\bibfield  {journal} {\bibinfo  {journal} {Astrofizika}\ }\textbf
  {\bibinfo {volume} {6}},\ \bibinfo {pages} {581} (\bibinfo {year}
  {1970})}\BibitemShut {NoStop}%
\bibitem [{{Eisenstein} {\textit{et~al}}\mbox{.}(2005)\citenamefont
  {{Eisenstein}}, \citenamefont {{Zehavi}}, \citenamefont {{Hogg}},
  \citenamefont {{Scoccimarro}}, \citenamefont {{Blanton}}, \citenamefont
  {{Nichol}}, \citenamefont {{Scranton}}, \citenamefont {{Seo}}, \citenamefont
  {{Tegmark}}, \citenamefont {{Zheng}}, \citenamefont {{Anderson}},
  \citenamefont {{Annis}}, \citenamefont {{Bahcall}}, \citenamefont
  {{Brinkmann}}, \citenamefont {{Burles}}, \citenamefont {{Castander}},
  \citenamefont {{Connolly}}, \citenamefont {{Csabai}}, \citenamefont {{Doi}},
  \citenamefont {{Fukugita}}, \citenamefont {{Frieman}}, \citenamefont
  {{Glazebrook}}, \citenamefont {{Gunn}}, \citenamefont {{Hendry}},
  \citenamefont {{Hennessy}}, \citenamefont {{Ivezi{\'c}}}, \citenamefont
  {{Kent}}, \citenamefont {{Knapp}}, \citenamefont {{Lin}}, \citenamefont
  {{Loh}}, \citenamefont {{Lupton}}, \citenamefont {{Margon}}, \citenamefont
  {{McKay}}, \citenamefont {{Meiksin}}, \citenamefont {{Munn}}, \citenamefont
  {{Pope}}, \citenamefont {{Richmond}}, \citenamefont {{Schlegel}},
  \citenamefont {{Schneider}}, \citenamefont {{Shimasaku}}, \citenamefont
  {{Stoughton}}, \citenamefont {{Strauss}}, \citenamefont {{SubbaRao}},
  \citenamefont {{Szalay}}, \citenamefont {{Szapudi}}, \citenamefont
  {{Tucker}}, \citenamefont {{Yanny}},\ \&\ \citenamefont
  {{York}}}]{Eisenstein2005}%
{(\PineGreen{{Eisenstein} {\textit{et~al}}\mbox{.}}, \PineGreen{2005})}
  \BibitemOpen
  \bibfield  {author} {\bibinfo {author} {\bibfnamefont {D.~J.}\ \bibnamefont
  {{Eisenstein}}}, \bibinfo {author} {\bibfnamefont {I.}~\bibnamefont
  {{Zehavi}}}, \bibinfo {author} {\bibfnamefont {D.~W.}\ \bibnamefont
  {{Hogg}}}, \bibinfo {author} {\bibfnamefont {R.}~\bibnamefont
  {{Scoccimarro}}}, \bibinfo {author} {\bibfnamefont {M.~R.}\ \bibnamefont
  {{Blanton}}}, \bibinfo {author} {\bibfnamefont {R.~C.}\ \bibnamefont
  {{Nichol}}}, \bibinfo {author} {\bibfnamefont {R.}~\bibnamefont
  {{Scranton}}}, \bibinfo {author} {\bibfnamefont {H.-J.}\ \bibnamefont
  {{Seo}}}, \bibinfo {author} {\bibfnamefont {M.}~\bibnamefont {{Tegmark}}},
  \bibinfo {author} {\bibfnamefont {Z.}~\bibnamefont {{Zheng}}}, \bibinfo
  {author} {\bibfnamefont {S.~F.}\ \bibnamefont {{Anderson}}}, \bibinfo
  {author} {\bibfnamefont {J.}~\bibnamefont {{Annis}}}, \bibinfo {author}
  {\bibfnamefont {N.}~\bibnamefont {{Bahcall}}}, \bibinfo {author}
  {\bibfnamefont {J.}~\bibnamefont {{Brinkmann}}}, \bibinfo {author}
  {\bibfnamefont {S.}~\bibnamefont {{Burles}}}, \bibinfo {author}
  {\bibfnamefont {F.~J.}\ \bibnamefont {{Castander}}}, \bibinfo {author}
  {\bibfnamefont {A.}~\bibnamefont {{Connolly}}}, \bibinfo {author}
  {\bibfnamefont {I.}~\bibnamefont {{Csabai}}}, \bibinfo {author}
  {\bibfnamefont {M.}~\bibnamefont {{Doi}}}, \bibinfo {author} {\bibfnamefont
  {M.}~\bibnamefont {{Fukugita}}}, \bibinfo {author} {\bibfnamefont {J.~A.}\
  \bibnamefont {{Frieman}}}, \bibinfo {author} {\bibfnamefont {K.}~\bibnamefont
  {{Glazebrook}}}, \bibinfo {author} {\bibfnamefont {J.~E.}\ \bibnamefont
  {{Gunn}}}, \bibinfo {author} {\bibfnamefont {J.~S.}\ \bibnamefont
  {{Hendry}}}, \bibinfo {author} {\bibfnamefont {G.}~\bibnamefont
  {{Hennessy}}}, \bibinfo {author} {\bibfnamefont {Z.}~\bibnamefont
  {{Ivezi{\'c}}}}, \bibinfo {author} {\bibfnamefont {S.}~\bibnamefont
  {{Kent}}}, \bibinfo {author} {\bibfnamefont {G.~R.}\ \bibnamefont {{Knapp}}},
  \bibinfo {author} {\bibfnamefont {H.}~\bibnamefont {{Lin}}}, \bibinfo
  {author} {\bibfnamefont {Y.-S.}\ \bibnamefont {{Loh}}}, \bibinfo {author}
  {\bibfnamefont {R.~H.}\ \bibnamefont {{Lupton}}}, \bibinfo {author}
  {\bibfnamefont {B.}~\bibnamefont {{Margon}}}, \bibinfo {author}
  {\bibfnamefont {T.~A.}\ \bibnamefont {{McKay}}}, \bibinfo {author}
  {\bibfnamefont {A.}~\bibnamefont {{Meiksin}}}, \bibinfo {author}
  {\bibfnamefont {J.~A.}\ \bibnamefont {{Munn}}}, \bibinfo {author}
  {\bibfnamefont {A.}~\bibnamefont {{Pope}}}, \bibinfo {author} {\bibfnamefont
  {M.~W.}\ \bibnamefont {{Richmond}}}, \bibinfo {author} {\bibfnamefont
  {D.}~\bibnamefont {{Schlegel}}}, \bibinfo {author} {\bibfnamefont {D.~P.}\
  \bibnamefont {{Schneider}}}, \bibinfo {author} {\bibfnamefont
  {K.}~\bibnamefont {{Shimasaku}}}, \bibinfo {author} {\bibfnamefont
  {C.}~\bibnamefont {{Stoughton}}}, \bibinfo {author} {\bibfnamefont {M.~A.}\
  \bibnamefont {{Strauss}}}, \bibinfo {author} {\bibfnamefont {M.}~\bibnamefont
  {{SubbaRao}}}, \bibinfo {author} {\bibfnamefont {A.~S.}\ \bibnamefont
  {{Szalay}}}, \bibinfo {author} {\bibfnamefont {I.}~\bibnamefont {{Szapudi}}},
  \bibinfo {author} {\bibfnamefont {D.~L.}\ \bibnamefont {{Tucker}}}, \bibinfo
  {author} {\bibfnamefont {B.}~\bibnamefont {{Yanny}}}, \bibinfo {author}
  {\bibfnamefont {D.~G.}\ \bibnamefont {{York}}},\ }\emph {{Detection of the
  Baryon Acoustic Peak in the Large-Scale Correlation Function of SDSS Luminous
  Red Galaxies}},\ \href {\doibase 10.1086/466512} {\bibfield  {journal}
  {\bibinfo  {journal} {\apj}\ }\textbf {\bibinfo {volume} {633}},\ \bibinfo
  {pages} {560} (\bibinfo {year} {2005})},\ \Eprint
  {http://arxiv.org/abs/astro-ph/0501171} {astro-ph/0501171}\BibitemShut
  {NoStop}%
\bibitem [{{Falck} \& {Neyrinck}(2015)\citenamefont {{Falck}}\ \&\
  \citenamefont {{Neyrinck}}}]{Falck2015}%
{(\PineGreen{{Falck} \& {Neyrinck}}, \PineGreen{2015})}  \BibitemOpen
  \bibfield  {author} {\bibinfo {author} {\bibfnamefont {B.}~\bibnamefont
  {{Falck}}}, \bibinfo {author} {\bibfnamefont {M.~C.}\ \bibnamefont
  {{Neyrinck}}},\ }\emph {{The persistent percolation of single-stream
  voids}},\ \href {\doibase 10.1093/mnras/stv879} {\bibfield  {journal}
  {\bibinfo  {journal} {\mnras}\ }\textbf {\bibinfo {volume} {450}},\ \bibinfo
  {pages} {3239} (\bibinfo {year} {2015})},\ \Eprint
  {http://arxiv.org/abs/1410.4751} {arXiv:1410.4751}\BibitemShut {NoStop}%
\bibitem [{{Falck}, {Neyrinck} \& {Szalay}(2012)\citenamefont {{Falck}},
  \citenamefont {{Neyrinck}},\ \&\ \citenamefont {{Szalay}}}]{Falck2012}%
{(\PineGreen{{Falck}, {Neyrinck} \& {Szalay}}, \PineGreen{2012})}  \BibitemOpen
  \bibfield  {author} {\bibinfo {author} {\bibfnamefont {B.~L.}\ \bibnamefont
  {{Falck}}}, \bibinfo {author} {\bibfnamefont {M.~C.}\ \bibnamefont
  {{Neyrinck}}}, \bibinfo {author} {\bibfnamefont {A.~S.}\ \bibnamefont
  {{Szalay}}},\ }\emph {{ORIGAMI: Delineating Halos Using Phase-space Folds}},\
  \href {\doibase 10.1088/0004-637X/754/2/126} {\bibfield  {journal} {\bibinfo
  {journal} {\apj}\ }\textbf {\bibinfo {volume} {754}},\ \bibinfo {eid} {126}
  (\bibinfo {year} {2012})},\ \Eprint {http://arxiv.org/abs/1201.2353}
  {arXiv:1201.2353 [astro-ph.CO]}\BibitemShut {NoStop}%
\bibitem [{{Feng} \& {Fang}(2000)\citenamefont {{Feng}}\ \&\ \citenamefont
  {{Fang}}}]{Feng2000}%
{(\PineGreen{{Feng} \& {Fang}}, \PineGreen{2000})}  \BibitemOpen
  \bibfield  {author} {\bibinfo {author} {\bibfnamefont {L.-L.}\ \bibnamefont
  {{Feng}}}, \bibinfo {author} {\bibfnamefont {L.-Z.}\ \bibnamefont {{Fang}}},\
  }\emph {{Non-Gaussianity and the Recovery of the Mass Power Spectrum from the
  Ly{$\alpha$} Forest}},\ \href {\doibase 10.1086/308874} {\bibfield  {journal}
  {\bibinfo  {journal} {\apj}\ }\textbf {\bibinfo {volume} {535}},\ \bibinfo
  {pages} {519} (\bibinfo {year} {2000})},\ \Eprint
  {http://arxiv.org/abs/arXiv:astro-ph/0001348}
  {arXiv:astro-ph/0001348}\BibitemShut {NoStop}%
\bibitem [{{Gil-Mar{\'i}n} {\textit{et~al}}\mbox{.}(2011)\citenamefont
  {{Gil-Mar{\'i}n}}, \citenamefont {{Schmidt}}, \citenamefont {{Hu}},
  \citenamefont {{Jimenez}},\ \&\ \citenamefont {{Verde}}}]{Gil-Marin2011}%
{(\PineGreen{{Gil-Mar{\'i}n} {\textit{et~al}}\mbox{.}}, \PineGreen{2011})}
  \BibitemOpen
  \bibfield  {author} {\bibinfo {author} {\bibfnamefont {H.}~\bibnamefont
  {{Gil-Mar{\'i}n}}}, \bibinfo {author} {\bibfnamefont {F.}~\bibnamefont
  {{Schmidt}}}, \bibinfo {author} {\bibfnamefont {W.}~\bibnamefont {{Hu}}},
  \bibinfo {author} {\bibfnamefont {R.}~\bibnamefont {{Jimenez}}}, \bibinfo
  {author} {\bibfnamefont {L.}~\bibnamefont {{Verde}}},\ }\emph {{The
  bispectrum of f(R) cosmologies}},\ \href {\doibase
  10.1088/1475-7516/2011/11/019} {\bibfield  {journal} {\bibinfo  {journal}
  {\jcap}\ }\textbf {\bibinfo {volume} {11}},\ \bibinfo {eid} {019} (\bibinfo
  {year} {2011})},\ \Eprint {http://arxiv.org/abs/1109.2115} {arXiv:1109.2115
  [astro-ph.CO]}\BibitemShut {NoStop}%
\bibitem [{{Gil-Mar{\'i}n} {\textit{et~al}}\mbox{.}(2012)\citenamefont
  {{Gil-Mar{\'i}n}}, \citenamefont {{Wagner}}, \citenamefont {{Fragkoudi}},
  \citenamefont {{Jimenez}},\ \&\ \citenamefont {{Verde}}}]{Gil-Marin2012}%
{(\PineGreen{{Gil-Mar{\'i}n} {\textit{et~al}}\mbox{.}}, \PineGreen{2012})}
  \BibitemOpen
  \bibfield  {author} {\bibinfo {author} {\bibfnamefont {H.}~\bibnamefont
  {{Gil-Mar{\'i}n}}}, \bibinfo {author} {\bibfnamefont {C.}~\bibnamefont
  {{Wagner}}}, \bibinfo {author} {\bibfnamefont {F.}~\bibnamefont
  {{Fragkoudi}}}, \bibinfo {author} {\bibfnamefont {R.}~\bibnamefont
  {{Jimenez}}}, \bibinfo {author} {\bibfnamefont {L.}~\bibnamefont {{Verde}}},\
  }\emph {{An improved fitting formula for the dark matter bispectrum}},\ \href
  {\doibase 10.1088/1475-7516/2012/02/047} {\bibfield  {journal} {\bibinfo
  {journal} {\jcap}\ }\textbf {\bibinfo {volume} {2}},\ \bibinfo {eid} {047}
  (\bibinfo {year} {2012})},\ \Eprint {http://arxiv.org/abs/1111.4477}
  {arXiv:1111.4477 [astro-ph.CO]}\BibitemShut {NoStop}%
\bibitem [{{Gunn} \& {Gott}(1972)\citenamefont {{Gunn}}\ \&\ \citenamefont
  {{Gott}}}]{Gunn1972}%
{(\PineGreen{{Gunn} \& {Gott}}, \PineGreen{1972})}  \BibitemOpen
  \bibfield  {author} {\bibinfo {author} {\bibfnamefont {J.~E.}\ \bibnamefont
  {{Gunn}}}, \bibinfo {author} {\bibfnamefont {J.~R.}\ \bibnamefont {{Gott}},
  \bibfnamefont {III}},\ }\emph {{On the Infall of Matter Into Clusters of
  Galaxies and Some Effects on Their Evolution}},\ \href {\doibase
  10.1086/151605} {\bibfield  {journal} {\bibinfo  {journal} {\apj}\ }\textbf
  {\bibinfo {volume} {176}},\ \bibinfo {pages} {1} (\bibinfo {year}
  {1972})}\BibitemShut {NoStop}%
\bibitem [{{Gurbatov}, {Saichev} \& {Shandarin}(1989)\citenamefont
  {{Gurbatov}}, \citenamefont {{Saichev}},\ \&\ \citenamefont
  {{Shandarin}}}]{Gurbatov1989}%
{(\PineGreen{{Gurbatov}, {Saichev} \& {Shandarin}}, \PineGreen{1989})}
  \BibitemOpen
  \bibfield  {author} {\bibinfo {author} {\bibfnamefont {S.~N.}\ \bibnamefont
  {{Gurbatov}}}, \bibinfo {author} {\bibfnamefont {A.~I.}\ \bibnamefont
  {{Saichev}}}, \bibinfo {author} {\bibfnamefont {S.~F.}\ \bibnamefont
  {{Shandarin}}},\ }\emph {{The large-scale structure of the universe in the
  frame of the model equation of non-linear diffusion}},\ \href@noop {}
  {\bibfield  {journal} {\bibinfo  {journal} {\mnras}\ }\textbf {\bibinfo
  {volume} {236}},\ \bibinfo {pages} {385} (\bibinfo {year}
  {1989})}\BibitemShut {NoStop}%
\bibitem [{{Hahn} {\textit{et~al}}\mbox{.}(2007)\citenamefont {{Hahn}},
  \citenamefont {{Porciani}}, \citenamefont {{Carollo}},\ \&\ \citenamefont
  {{Dekel}}}]{Hahn2007}%
{(\PineGreen{{Hahn} {\textit{et~al}}\mbox{.}}, \PineGreen{2007})}  \BibitemOpen
  \bibfield  {author} {\bibinfo {author} {\bibfnamefont {O.}~\bibnamefont
  {{Hahn}}}, \bibinfo {author} {\bibfnamefont {C.}~\bibnamefont {{Porciani}}},
  \bibinfo {author} {\bibfnamefont {C.~M.}\ \bibnamefont {{Carollo}}}, \bibinfo
  {author} {\bibfnamefont {A.}~\bibnamefont {{Dekel}}},\ }\emph {{Properties of
  dark matter haloes in clusters, filaments, sheets and voids}},\ \href
  {\doibase 10.1111/j.1365-2966.2006.11318.x} {\bibfield  {journal} {\bibinfo
  {journal} {\mnras}\ }\textbf {\bibinfo {volume} {375}},\ \bibinfo {pages}
  {489} (\bibinfo {year} {2007})},\ \Eprint
  {http://arxiv.org/abs/astro-ph/0610280} {astro-ph/0610280}\BibitemShut
  {NoStop}%
\bibitem [{{Heisenberg}, {Sch{\"a}fer} \& {Bartelmann}(2011)\citenamefont
  {{Heisenberg}}, \citenamefont {{Sch{\"a}fer}},\ \&\ \citenamefont
  {{Bartelmann}}}]{Heisenberg2011}%
{(\PineGreen{{Heisenberg}, {Sch{\"a}fer} \& {Bartelmann}}, \PineGreen{2011})}
  \BibitemOpen
  \bibfield  {author} {\bibinfo {author} {\bibfnamefont {L.}~\bibnamefont
  {{Heisenberg}}}, \bibinfo {author} {\bibfnamefont {B.~M.}\ \bibnamefont
  {{Sch{\"a}fer}}}, \bibinfo {author} {\bibfnamefont {M.}~\bibnamefont
  {{Bartelmann}}},\ }\emph {{A study of relative velocity statistics in
  Lagrangian perturbation theory with PINOCCHIO}},\ \href {\doibase
  10.1111/j.1365-2966.2011.19252.x} {\bibfield  {journal} {\bibinfo  {journal}
  {\mnras}\ }\textbf {\bibinfo {volume} {416}},\ \bibinfo {pages} {3057}
  (\bibinfo {year} {2011})},\ \Eprint {http://arxiv.org/abs/1011.1559}
  {arXiv:1011.1559 [astro-ph.CO]}\BibitemShut {NoStop}%
\bibitem [{{Heitmann} {\textit{et~al}}\mbox{.}(2009)\citenamefont {{Heitmann}},
  \citenamefont {{Higdon}}, \citenamefont {{White}}, \citenamefont {{Habib}},
  \citenamefont {{Williams}}, \citenamefont {{Lawrence}},\ \&\ \citenamefont
  {{Wagner}}}]{Heitmann2009}%
{(\PineGreen{{Heitmann} {\textit{et~al}}\mbox{.}}, \PineGreen{2009})}
  \BibitemOpen
  \bibfield  {author} {\bibinfo {author} {\bibfnamefont {K.}~\bibnamefont
  {{Heitmann}}}, \bibinfo {author} {\bibfnamefont {D.}~\bibnamefont
  {{Higdon}}}, \bibinfo {author} {\bibfnamefont {M.}~\bibnamefont {{White}}},
  \bibinfo {author} {\bibfnamefont {S.}~\bibnamefont {{Habib}}}, \bibinfo
  {author} {\bibfnamefont {B.~J.}\ \bibnamefont {{Williams}}}, \bibinfo
  {author} {\bibfnamefont {E.}~\bibnamefont {{Lawrence}}}, \bibinfo {author}
  {\bibfnamefont {C.}~\bibnamefont {{Wagner}}},\ }\emph {{The Coyote Universe.
  II. Cosmological Models and Precision Emulation of the Nonlinear Matter Power
  Spectrum}},\ \href {\doibase 10.1088/0004-637X/705/1/156} {\bibfield
  {journal} {\bibinfo  {journal} {\apj}\ }\textbf {\bibinfo {volume} {705}},\
  \bibinfo {pages} {156} (\bibinfo {year} {2009})},\ \Eprint
  {http://arxiv.org/abs/0902.0429} {arXiv:0902.0429 [astro-ph.CO]}\BibitemShut
  {NoStop}%
\bibitem [{{Heitmann} {\textit{et~al}}\mbox{.}(2010)\citenamefont {{Heitmann}},
  \citenamefont {{White}}, \citenamefont {{Wagner}}, \citenamefont {{Habib}},\
  \&\ \citenamefont {{Higdon}}}]{Heitmann2010}%
{(\PineGreen{{Heitmann} {\textit{et~al}}\mbox{.}}, \PineGreen{2010})}
  \BibitemOpen
  \bibfield  {author} {\bibinfo {author} {\bibfnamefont {K.}~\bibnamefont
  {{Heitmann}}}, \bibinfo {author} {\bibfnamefont {M.}~\bibnamefont {{White}}},
  \bibinfo {author} {\bibfnamefont {C.}~\bibnamefont {{Wagner}}}, \bibinfo
  {author} {\bibfnamefont {S.}~\bibnamefont {{Habib}}}, \bibinfo {author}
  {\bibfnamefont {D.}~\bibnamefont {{Higdon}}},\ }\emph {{The Coyote Universe.
  I. Precision Determination of the Nonlinear Matter Power Spectrum}},\ \href
  {\doibase 10.1088/0004-637X/715/1/104} {\bibfield  {journal} {\bibinfo
  {journal} {\apj}\ }\textbf {\bibinfo {volume} {715}},\ \bibinfo {pages} {104}
  (\bibinfo {year} {2010})},\ \Eprint {http://arxiv.org/abs/0812.1052}
  {arXiv:0812.1052}\BibitemShut {NoStop}%
\bibitem [{{Jasche}, {Leclercq} \& {Wandelt}(2015)\citenamefont {{Jasche}},
  \citenamefont {{Leclercq}},\ \&\ \citenamefont {{Wandelt}}}]{Jasche2015}%
{(\PineGreen{{Jasche}, {Leclercq} \& {Wandelt}}, \PineGreen{2015})}
  \BibitemOpen
  \bibfield  {author} {\bibinfo {author} {\bibfnamefont {J.}~\bibnamefont
  {{Jasche}}}, \bibinfo {author} {\bibfnamefont {F.}~\bibnamefont
  {{Leclercq}}}, \bibinfo {author} {\bibfnamefont {B.~D.}\ \bibnamefont
  {{Wandelt}}},\ }\emph {{Past and present cosmic structure in the SDSS DR7
  main sample}},\ \href {\doibase 10.1088/1475-7516/2015/01/036} {\bibfield
  {journal} {\bibinfo  {journal} {\jcap}\ }\textbf {\bibinfo {volume} {1}},\
  \bibinfo {eid} {036} (\bibinfo {year} {2015})},\ \Eprint
  {http://arxiv.org/abs/1409.6308} {arXiv:1409.6308}\BibitemShut {NoStop}%
\bibitem [{{Jasche} \& {Wandelt}(2013)\citenamefont {{Jasche}}\ \&\
  \citenamefont {{Wandelt}}}]{Jasche2013}%
{(\PineGreen{{Jasche} \& {Wandelt}}, \PineGreen{2013})}  \BibitemOpen
  \bibfield  {author} {\bibinfo {author} {\bibfnamefont {J.}~\bibnamefont
  {{Jasche}}}, \bibinfo {author} {\bibfnamefont {B.~D.}\ \bibnamefont
  {{Wandelt}}},\ }\emph {{Bayesian physical reconstruction of initial
  conditions from large-scale structure surveys}},\ \href {\doibase
  10.1093/mnras/stt449} {\bibfield  {journal} {\bibinfo  {journal} {\mnras}\
  }\textbf {\bibinfo {volume} {432}},\ \bibinfo {pages} {894} (\bibinfo {year}
  {2013})},\ \Eprint {http://arxiv.org/abs/1203.3639} {arXiv:1203.3639
  [astro-ph.CO]}\BibitemShut {NoStop}%
\bibitem [{{Jeong} \& {Komatsu}(2009)\citenamefont {{Jeong}}\ \&\ \citenamefont
  {{Komatsu}}}]{Jeong2009}%
{(\PineGreen{{Jeong} \& {Komatsu}}, \PineGreen{2009})}  \BibitemOpen
  \bibfield  {author} {\bibinfo {author} {\bibfnamefont {D.}~\bibnamefont
  {{Jeong}}}, \bibinfo {author} {\bibfnamefont {E.}~\bibnamefont {{Komatsu}}},\
  }\emph {{Primordial Non-Gaussianity, Scale-dependent Bias, and the Bispectrum
  of Galaxies}},\ \href {\doibase 10.1088/0004-637X/703/2/1230} {\bibfield
  {journal} {\bibinfo  {journal} {\apj}\ }\textbf {\bibinfo {volume} {703}},\
  \bibinfo {pages} {1230} (\bibinfo {year} {2009})},\ \Eprint
  {http://arxiv.org/abs/0904.0497} {arXiv:0904.0497 [astro-ph.CO]}\BibitemShut
  {NoStop}%
\bibitem [{{Jing}(2005)\citenamefont {{Jing}}}]{Jing2005}%
{(\PineGreen{{Jing}}, \PineGreen{2005})}  \BibitemOpen
  \bibfield  {author} {\bibinfo {author} {\bibfnamefont {Y.~P.}\ \bibnamefont
  {{Jing}}},\ }\emph {{Correcting for the Alias Effect When Measuring the Power
  Spectrum Using a Fast Fourier Transform}},\ \href {\doibase 10.1086/427087}
  {\bibfield  {journal} {\bibinfo  {journal} {\apj}\ }\textbf {\bibinfo
  {volume} {620}},\ \bibinfo {pages} {559} (\bibinfo {year} {2005})},\ \Eprint
  {http://arxiv.org/abs/astro-ph/0409240} {astro-ph/0409240}\BibitemShut
  {NoStop}%
\bibitem [{{Kitaura}, {Gallerani} \& {Ferrara}(2012)\citenamefont {{Kitaura}},
  \citenamefont {{Gallerani}},\ \&\ \citenamefont {{Ferrara}}}]{Kitaura2012}%
{(\PineGreen{{Kitaura}, {Gallerani} \& {Ferrara}}, \PineGreen{2012})}
  \BibitemOpen
  \bibfield  {author} {\bibinfo {author} {\bibfnamefont {F.-S.}\ \bibnamefont
  {{Kitaura}}}, \bibinfo {author} {\bibfnamefont {S.}~\bibnamefont
  {{Gallerani}}}, \bibinfo {author} {\bibfnamefont {A.}~\bibnamefont
  {{Ferrara}}},\ }\emph {{Multiscale inference of matter fields and baryon
  acoustic oscillations from the Ly{$\alpha$} forest}},\ \href {\doibase
  10.1111/j.1365-2966.2011.19997.x} {\bibfield  {journal} {\bibinfo  {journal}
  {\mnras}\ }\textbf {\bibinfo {volume} {420}},\ \bibinfo {pages} {61}
  (\bibinfo {year} {2012})},\ \Eprint {http://arxiv.org/abs/1011.6233}
  {arXiv:1011.6233 [astro-ph.CO]}\BibitemShut {NoStop}%
\bibitem [{{Komatsu} {\textit{et~al}}\mbox{.}(2011)\citenamefont {{Komatsu}},
  \citenamefont {{Smith}}, \citenamefont {{Dunkley}}, \citenamefont
  {{Bennett}}, \citenamefont {{Gold}}, \citenamefont {{Hinshaw}}, \citenamefont
  {{Jarosik}}, \citenamefont {{Larson}}, \citenamefont {{Nolta}}, \citenamefont
  {{Page}}, \citenamefont {{Spergel}}, \citenamefont {{Halpern}}, \citenamefont
  {{Hill}}, \citenamefont {{Kogut}}, \citenamefont {{Limon}}, \citenamefont
  {{Meyer}}, \citenamefont {{Odegard}}, \citenamefont {{Tucker}}, \citenamefont
  {{Weiland}}, \citenamefont {{Wollack}},\ \&\ \citenamefont
  {{Wright}}}]{Komatsu2011}%
{(\PineGreen{{Komatsu} {\textit{et~al}}\mbox{.}}, \PineGreen{2011})}
  \BibitemOpen
  \bibfield  {author} {\bibinfo {author} {\bibfnamefont {E.}~\bibnamefont
  {{Komatsu}}}, \bibinfo {author} {\bibfnamefont {K.~M.}\ \bibnamefont
  {{Smith}}}, \bibinfo {author} {\bibfnamefont {J.}~\bibnamefont {{Dunkley}}},
  \bibinfo {author} {\bibfnamefont {C.~L.}\ \bibnamefont {{Bennett}}}, \bibinfo
  {author} {\bibfnamefont {B.}~\bibnamefont {{Gold}}}, \bibinfo {author}
  {\bibfnamefont {G.}~\bibnamefont {{Hinshaw}}}, \bibinfo {author}
  {\bibfnamefont {N.}~\bibnamefont {{Jarosik}}}, \bibinfo {author}
  {\bibfnamefont {D.}~\bibnamefont {{Larson}}}, \bibinfo {author}
  {\bibfnamefont {M.~R.}\ \bibnamefont {{Nolta}}}, \bibinfo {author}
  {\bibfnamefont {L.}~\bibnamefont {{Page}}}, \bibinfo {author} {\bibfnamefont
  {D.~N.}\ \bibnamefont {{Spergel}}}, \bibinfo {author} {\bibfnamefont
  {M.}~\bibnamefont {{Halpern}}}, \bibinfo {author} {\bibfnamefont {R.~S.}\
  \bibnamefont {{Hill}}}, \bibinfo {author} {\bibfnamefont {A.}~\bibnamefont
  {{Kogut}}}, \bibinfo {author} {\bibfnamefont {M.}~\bibnamefont {{Limon}}},
  \bibinfo {author} {\bibfnamefont {S.~S.}\ \bibnamefont {{Meyer}}}, \bibinfo
  {author} {\bibfnamefont {N.}~\bibnamefont {{Odegard}}}, \bibinfo {author}
  {\bibfnamefont {G.~S.}\ \bibnamefont {{Tucker}}}, \bibinfo {author}
  {\bibfnamefont {J.~L.}\ \bibnamefont {{Weiland}}}, \bibinfo {author}
  {\bibfnamefont {E.}~\bibnamefont {{Wollack}}}, \bibinfo {author}
  {\bibfnamefont {E.~L.}\ \bibnamefont {{Wright}}},\ }\emph {{Seven-year
  Wilkinson Microwave Anisotropy Probe (WMAP) Observations: Cosmological
  Interpretation}},\ \href {\doibase 10.1088/0067-0049/192/2/18} {\bibfield
  {journal} {\bibinfo  {journal} {\apjs}\ }\textbf {\bibinfo {volume} {192}},\
  \bibinfo {eid} {18} (\bibinfo {year} {2011})},\ \Eprint
  {http://arxiv.org/abs/1001.4538} {arXiv:1001.4538 [astro-ph.CO]}\BibitemShut
  {NoStop}%
\bibitem [{{Lavaux} \& {Wandelt}(2010)\citenamefont {{Lavaux}}\ \&\
  \citenamefont {{Wandelt}}}]{Lavaux2010}%
{(\PineGreen{{Lavaux} \& {Wandelt}}, \PineGreen{2010})}  \BibitemOpen
  \bibfield  {author} {\bibinfo {author} {\bibfnamefont {G.}~\bibnamefont
  {{Lavaux}}}, \bibinfo {author} {\bibfnamefont {B.~D.}\ \bibnamefont
  {{Wandelt}}},\ }\emph {{Precision cosmology with voids: definition, methods,
  dynamics}},\ \href {\doibase 10.1111/j.1365-2966.2010.16197.x} {\bibfield
  {journal} {\bibinfo  {journal} {\mnras}\ }\textbf {\bibinfo {volume} {403}},\
  \bibinfo {pages} {1392} (\bibinfo {year} {2010})},\ \Eprint
  {http://arxiv.org/abs/0906.4101} {arXiv:0906.4101 [astro-ph.CO]}\BibitemShut
  {NoStop}%
\bibitem [{{Lawrence} {\textit{et~al}}\mbox{.}(2010)\citenamefont {{Lawrence}},
  \citenamefont {{Heitmann}}, \citenamefont {{White}}, \citenamefont
  {{Higdon}}, \citenamefont {{Wagner}}, \citenamefont {{Habib}},\ \&\
  \citenamefont {{Williams}}}]{Lawrence2010}%
{(\PineGreen{{Lawrence} {\textit{et~al}}\mbox{.}}, \PineGreen{2010})}
  \BibitemOpen
  \bibfield  {author} {\bibinfo {author} {\bibfnamefont {E.}~\bibnamefont
  {{Lawrence}}}, \bibinfo {author} {\bibfnamefont {K.}~\bibnamefont
  {{Heitmann}}}, \bibinfo {author} {\bibfnamefont {M.}~\bibnamefont {{White}}},
  \bibinfo {author} {\bibfnamefont {D.}~\bibnamefont {{Higdon}}}, \bibinfo
  {author} {\bibfnamefont {C.}~\bibnamefont {{Wagner}}}, \bibinfo {author}
  {\bibfnamefont {S.}~\bibnamefont {{Habib}}}, \bibinfo {author} {\bibfnamefont
  {B.}~\bibnamefont {{Williams}}},\ }\emph {{The Coyote Universe. III.
  Simulation Suite and Precision Emulator for the Nonlinear Matter Power
  Spectrum}},\ \href {\doibase 10.1088/0004-637X/713/2/1322} {\bibfield
  {journal} {\bibinfo  {journal} {\apj}\ }\textbf {\bibinfo {volume} {713}},\
  \bibinfo {pages} {1322} (\bibinfo {year} {2010})},\ \Eprint
  {http://arxiv.org/abs/0912.4490} {arXiv:0912.4490 [astro-ph.CO]}\BibitemShut
  {NoStop}%
\bibitem [{{Lidz} {\textit{et~al}}\mbox{.}(2007)\citenamefont {{Lidz}},
  \citenamefont {{Zahn}}, \citenamefont {{McQuinn}}, \citenamefont
  {{Zaldarriaga}}, \citenamefont {{Dutta}},\ \&\ \citenamefont
  {{Hernquist}}}]{Lidz2007}%
{(\PineGreen{{Lidz} {\textit{et~al}}\mbox{.}}, \PineGreen{2007})}  \BibitemOpen
  \bibfield  {author} {\bibinfo {author} {\bibfnamefont {A.}~\bibnamefont
  {{Lidz}}}, \bibinfo {author} {\bibfnamefont {O.}~\bibnamefont {{Zahn}}},
  \bibinfo {author} {\bibfnamefont {M.}~\bibnamefont {{McQuinn}}}, \bibinfo
  {author} {\bibfnamefont {M.}~\bibnamefont {{Zaldarriaga}}}, \bibinfo {author}
  {\bibfnamefont {S.}~\bibnamefont {{Dutta}}}, \bibinfo {author} {\bibfnamefont
  {L.}~\bibnamefont {{Hernquist}}},\ }\emph {{Higher Order Contributions to the
  21 cm Power Spectrum}},\ \href {\doibase 10.1086/511670} {\bibfield
  {journal} {\bibinfo  {journal} {\apj}\ }\textbf {\bibinfo {volume} {659}},\
  \bibinfo {pages} {865} (\bibinfo {year} {2007})},\ \Eprint
  {http://arxiv.org/abs/astro-ph/0610054} {astro-ph/0610054}\BibitemShut
  {NoStop}%
\bibitem [{{Manera} {\textit{et~al}}\mbox{.}(2013)\citenamefont {{Manera}},
  \citenamefont {{Scoccimarro}}, \citenamefont {{Percival}}, \citenamefont
  {{Samushia}}, \citenamefont {{McBride}}, \citenamefont {{Ross}},
  \citenamefont {{Sheth}}, \citenamefont {{White}}, \citenamefont {{Reid}},
  \citenamefont {{S{\'a}nchez}}, \citenamefont {{de Putter}}, \citenamefont
  {{Xu}}, \citenamefont {{Berlind}}, \citenamefont {{Brinkmann}}, \citenamefont
  {{Maraston}}, \citenamefont {{Nichol}}, \citenamefont {{Montesano}},
  \citenamefont {{Padmanabhan}}, \citenamefont {{Skibba}}, \citenamefont
  {{Tojeiro}},\ \&\ \citenamefont {{Weaver}}}]{Manera2013}%
{(\PineGreen{{Manera} {\textit{et~al}}\mbox{.}}, \PineGreen{2013})}
  \BibitemOpen
  \bibfield  {author} {\bibinfo {author} {\bibfnamefont {M.}~\bibnamefont
  {{Manera}}}, \bibinfo {author} {\bibfnamefont {R.}~\bibnamefont
  {{Scoccimarro}}}, \bibinfo {author} {\bibfnamefont {W.~J.}\ \bibnamefont
  {{Percival}}}, \bibinfo {author} {\bibfnamefont {L.}~\bibnamefont
  {{Samushia}}}, \bibinfo {author} {\bibfnamefont {C.~K.}\ \bibnamefont
  {{McBride}}}, \bibinfo {author} {\bibfnamefont {A.~J.}\ \bibnamefont
  {{Ross}}}, \bibinfo {author} {\bibfnamefont {R.~K.}\ \bibnamefont {{Sheth}}},
  \bibinfo {author} {\bibfnamefont {M.}~\bibnamefont {{White}}}, \bibinfo
  {author} {\bibfnamefont {B.~A.}\ \bibnamefont {{Reid}}}, \bibinfo {author}
  {\bibfnamefont {A.~G.}\ \bibnamefont {{S{\'a}nchez}}}, \bibinfo {author}
  {\bibfnamefont {R.}~\bibnamefont {{de Putter}}}, \bibinfo {author}
  {\bibfnamefont {X.}~\bibnamefont {{Xu}}}, \bibinfo {author} {\bibfnamefont
  {A.~A.}\ \bibnamefont {{Berlind}}}, \bibinfo {author} {\bibfnamefont
  {J.}~\bibnamefont {{Brinkmann}}}, \bibinfo {author} {\bibfnamefont
  {C.}~\bibnamefont {{Maraston}}}, \bibinfo {author} {\bibfnamefont
  {B.}~\bibnamefont {{Nichol}}}, \bibinfo {author} {\bibfnamefont
  {F.}~\bibnamefont {{Montesano}}}, \bibinfo {author} {\bibfnamefont
  {N.}~\bibnamefont {{Padmanabhan}}}, \bibinfo {author} {\bibfnamefont {R.~A.}\
  \bibnamefont {{Skibba}}}, \bibinfo {author} {\bibfnamefont {R.}~\bibnamefont
  {{Tojeiro}}}, \bibinfo {author} {\bibfnamefont {B.~A.}\ \bibnamefont
  {{Weaver}}},\ }\emph {{The clustering of galaxies in the SDSS-III Baryon
  Oscillation Spectroscopic Survey: a large sample of mock galaxy
  catalogues}},\ \href {\doibase 10.1093/mnras/sts084} {\bibfield  {journal}
  {\bibinfo  {journal} {\mnras}\ }\textbf {\bibinfo {volume} {428}},\ \bibinfo
  {pages} {1036} (\bibinfo {year} {2013})},\ \Eprint
  {http://arxiv.org/abs/1203.6609} {arXiv:1203.6609 [astro-ph.CO]}\BibitemShut
  {NoStop}%
\bibitem [{{Matarrese}, {Verde} \& {Heavens}(1997)\citenamefont {{Matarrese}},
  \citenamefont {{Verde}},\ \&\ \citenamefont {{Heavens}}}]{Matarrese1997}%
{(\PineGreen{{Matarrese}, {Verde} \& {Heavens}}, \PineGreen{1997})}
  \BibitemOpen
  \bibfield  {author} {\bibinfo {author} {\bibfnamefont {S.}~\bibnamefont
  {{Matarrese}}}, \bibinfo {author} {\bibfnamefont {L.}~\bibnamefont
  {{Verde}}}, \bibinfo {author} {\bibfnamefont {A.~F.}\ \bibnamefont
  {{Heavens}}},\ }\emph {{Large-scale bias in the Universe: bispectrum
  method}},\ \href@noop {} {\bibfield  {journal} {\bibinfo  {journal} {\mnras}\
  }\textbf {\bibinfo {volume} {290}},\ \bibinfo {pages} {651} (\bibinfo {year}
  {1997})},\ \Eprint {http://arxiv.org/abs/astro-ph/9706059}
  {astro-ph/9706059}\BibitemShut {NoStop}%
\bibitem [{Melott(1993)\citenamefont {Melott}}]{Melott1993}%
{(\PineGreen{Melott}, \PineGreen{1993})}  \BibitemOpen
  \bibfield  {author} {\bibinfo {author} {\bibfnamefont {A.~L.}\ \bibnamefont
  {Melott}},\ }\emph {{Improving the reconstruction of the velocity potential
  and primordial density fluctuations by choice of smoothing windows}},\ \href
  {\doibase 10.1086/186999} {\bibfield  {journal} {\bibinfo  {journal} {\apj}\
  }\textbf {\bibinfo {volume} {414}},\ \bibinfo {pages} {L73} (\bibinfo {year}
  {1993})}\BibitemShut {NoStop}%
\bibitem [{{Melott}, {Pellman} \& {Shandarin}(1994)\citenamefont {{Melott}},
  \citenamefont {{Pellman}},\ \&\ \citenamefont {{Shandarin}}}]{Melott1994}%
{(\PineGreen{{Melott}, {Pellman} \& {Shandarin}}, \PineGreen{1994})}
  \BibitemOpen
  \bibfield  {author} {\bibinfo {author} {\bibfnamefont {A.~L.}\ \bibnamefont
  {{Melott}}}, \bibinfo {author} {\bibfnamefont {T.~F.}\ \bibnamefont
  {{Pellman}}}, \bibinfo {author} {\bibfnamefont {S.~F.}\ \bibnamefont
  {{Shandarin}}},\ }\emph {{Optimizing the Zeldovich Approximation}},\
  \href@noop {} {\bibfield  {journal} {\bibinfo  {journal} {\mnras}\ }\textbf
  {\bibinfo {volume} {269}},\ \bibinfo {pages} {626} (\bibinfo {year}
  {1994})},\ \Eprint {http://arxiv.org/abs/astro-ph/9312044}
  {astro-ph/9312044}\BibitemShut {NoStop}%
\bibitem [{{Mesinger} \& {Furlanetto}(2007)\citenamefont {{Mesinger}}\ \&\
  \citenamefont {{Furlanetto}}}]{Mesinger2007}%
{(\PineGreen{{Mesinger} \& {Furlanetto}}, \PineGreen{2007})}  \BibitemOpen
  \bibfield  {author} {\bibinfo {author} {\bibfnamefont {A.}~\bibnamefont
  {{Mesinger}}}, \bibinfo {author} {\bibfnamefont {S.}~\bibnamefont
  {{Furlanetto}}},\ }\emph {{Efficient Simulations of Early Structure Formation
  and Reionization}},\ \href {\doibase 10.1086/521806} {\bibfield  {journal}
  {\bibinfo  {journal} {\apj}\ }\textbf {\bibinfo {volume} {669}},\ \bibinfo
  {pages} {663} (\bibinfo {year} {2007})},\ \Eprint
  {http://arxiv.org/abs/0704.0946} {arXiv:0704.0946}\BibitemShut {NoStop}%
\bibitem [{{Mohayaee} {\textit{et~al}}\mbox{.}(2006)\citenamefont {{Mohayaee}},
  \citenamefont {{Mathis}}, \citenamefont {{Colombi}},\ \&\ \citenamefont
  {{Silk}}}]{Mohayaee2006}%
{(\PineGreen{{Mohayaee} {\textit{et~al}}\mbox{.}}, \PineGreen{2006})}
  \BibitemOpen
  \bibfield  {author} {\bibinfo {author} {\bibfnamefont {R.}~\bibnamefont
  {{Mohayaee}}}, \bibinfo {author} {\bibfnamefont {H.}~\bibnamefont
  {{Mathis}}}, \bibinfo {author} {\bibfnamefont {S.}~\bibnamefont {{Colombi}}},
  \bibinfo {author} {\bibfnamefont {J.}~\bibnamefont {{Silk}}},\ }\emph
  {{Reconstruction of primordial density fields}},\ \href {\doibase
  10.1111/j.1365-2966.2005.09774.x} {\bibfield  {journal} {\bibinfo  {journal}
  {\mnras}\ }\textbf {\bibinfo {volume} {365}},\ \bibinfo {pages} {939}
  (\bibinfo {year} {2006})},\ \Eprint {http://arxiv.org/abs/astro-ph/0501217}
  {astro-ph/0501217}\BibitemShut {NoStop}%
\bibitem [{{Monaco} {\textit{et~al}}\mbox{.}(2013)\citenamefont {{Monaco}},
  \citenamefont {{Sefusatti}}, \citenamefont {{Borgani}}, \citenamefont
  {{Crocce}}, \citenamefont {{Fosalba}}, \citenamefont {{Sheth}},\ \&\
  \citenamefont {{Theuns}}}]{Monaco2013}%
{(\PineGreen{{Monaco} {\textit{et~al}}\mbox{.}}, \PineGreen{2013})}
  \BibitemOpen
  \bibfield  {author} {\bibinfo {author} {\bibfnamefont {P.}~\bibnamefont
  {{Monaco}}}, \bibinfo {author} {\bibfnamefont {E.}~\bibnamefont
  {{Sefusatti}}}, \bibinfo {author} {\bibfnamefont {S.}~\bibnamefont
  {{Borgani}}}, \bibinfo {author} {\bibfnamefont {M.}~\bibnamefont {{Crocce}}},
  \bibinfo {author} {\bibfnamefont {P.}~\bibnamefont {{Fosalba}}}, \bibinfo
  {author} {\bibfnamefont {R.~K.}\ \bibnamefont {{Sheth}}}, \bibinfo {author}
  {\bibfnamefont {T.}~\bibnamefont {{Theuns}}},\ }\emph {{An accurate tool for
  the fast generation of dark matter halo catalogues}},\ \href {\doibase
  10.1093/mnras/stt907} {\bibfield  {journal} {\bibinfo  {journal} {\mnras}\
  }\textbf {\bibinfo {volume} {433}},\ \bibinfo {pages} {2389} (\bibinfo {year}
  {2013})},\ \Eprint {http://arxiv.org/abs/1305.1505} {arXiv:1305.1505
  [astro-ph.CO]}\BibitemShut {NoStop}%
\bibitem [{{Monaco}, {Theuns} \& {Taffoni}(2002)\citenamefont {{Monaco}},
  \citenamefont {{Theuns}},\ \&\ \citenamefont {{Taffoni}}}]{Monaco2002b}%
{(\PineGreen{{Monaco}, {Theuns} \& {Taffoni}}, \PineGreen{2002})}  \BibitemOpen
  \bibfield  {author} {\bibinfo {author} {\bibfnamefont {P.}~\bibnamefont
  {{Monaco}}}, \bibinfo {author} {\bibfnamefont {T.}~\bibnamefont {{Theuns}}},
  \bibinfo {author} {\bibfnamefont {G.}~\bibnamefont {{Taffoni}}},\ }\emph
  {{The pinocchio algorithm: pinpointing orbit-crossing collapsed hierarchical
  objects in a linear density field}},\ \href {\doibase
  10.1046/j.1365-8711.2002.05162.x} {\bibfield  {journal} {\bibinfo  {journal}
  {\mnras}\ }\textbf {\bibinfo {volume} {331}},\ \bibinfo {pages} {587}
  (\bibinfo {year} {2002})},\ \Eprint {http://arxiv.org/abs/astro-ph/0109323}
  {astro-ph/0109323}\BibitemShut {NoStop}%
\bibitem [{{Monaco} {\textit{et~al}}\mbox{.}(2002)\citenamefont {{Monaco}},
  \citenamefont {{Theuns}}, \citenamefont {{Taffoni}}, \citenamefont
  {{Governato}}, \citenamefont {{Quinn}},\ \&\ \citenamefont
  {{Stadel}}}]{Monaco2002a}%
{(\PineGreen{{Monaco} {\textit{et~al}}\mbox{.}}, \PineGreen{2002})}
  \BibitemOpen
  \bibfield  {author} {\bibinfo {author} {\bibfnamefont {P.}~\bibnamefont
  {{Monaco}}}, \bibinfo {author} {\bibfnamefont {T.}~\bibnamefont {{Theuns}}},
  \bibinfo {author} {\bibfnamefont {G.}~\bibnamefont {{Taffoni}}}, \bibinfo
  {author} {\bibfnamefont {F.}~\bibnamefont {{Governato}}}, \bibinfo {author}
  {\bibfnamefont {T.}~\bibnamefont {{Quinn}}}, \bibinfo {author} {\bibfnamefont
  {J.}~\bibnamefont {{Stadel}}},\ }\emph {{Predicting the Number, Spatial
  Distribution, and Merging History of Dark Matter Halos}},\ \href {\doibase
  10.1086/324182} {\bibfield  {journal} {\bibinfo  {journal} {\apj}\ }\textbf
  {\bibinfo {volume} {564}},\ \bibinfo {pages} {8} (\bibinfo {year} {2002})},\
  \Eprint {http://arxiv.org/abs/astro-ph/0109322}
  {astro-ph/0109322}\BibitemShut {NoStop}%
\bibitem [{{Moutarde} {\textit{et~al}}\mbox{.}(1991)\citenamefont {{Moutarde}},
  \citenamefont {{Alimi}}, \citenamefont {{Bouchet}}, \citenamefont
  {{Pellat}},\ \&\ \citenamefont {{Ramani}}}]{Moutarde1991}%
{(\PineGreen{{Moutarde} {\textit{et~al}}\mbox{.}}, \PineGreen{1991})}
  \BibitemOpen
  \bibfield  {author} {\bibinfo {author} {\bibfnamefont {F.}~\bibnamefont
  {{Moutarde}}}, \bibinfo {author} {\bibfnamefont {J.-M.}\ \bibnamefont
  {{Alimi}}}, \bibinfo {author} {\bibfnamefont {F.~R.}\ \bibnamefont
  {{Bouchet}}}, \bibinfo {author} {\bibfnamefont {R.}~\bibnamefont {{Pellat}}},
  \bibinfo {author} {\bibfnamefont {A.}~\bibnamefont {{Ramani}}},\ }\emph
  {{Precollapse scale invariance in gravitational instability}},\ \href
  {\doibase 10.1086/170728} {\bibfield  {journal} {\bibinfo  {journal} {\apj}\
  }\textbf {\bibinfo {volume} {382}},\ \bibinfo {pages} {377} (\bibinfo {year}
  {1991})}\BibitemShut {NoStop}%
\bibitem [{{Munshi} \& {Starobinsky}(1994)\citenamefont {{Munshi}}\ \&\
  \citenamefont {{Starobinsky}}}]{Munshi1994}%
{(\PineGreen{{Munshi} \& {Starobinsky}}, \PineGreen{1994})}  \BibitemOpen
  \bibfield  {author} {\bibinfo {author} {\bibfnamefont {D.}~\bibnamefont
  {{Munshi}}}, \bibinfo {author} {\bibfnamefont {A.~A.}\ \bibnamefont
  {{Starobinsky}}},\ }\emph {{Nonlinear approximations to gravitational
  instability: A comparison in second-order perturbation theory}},\ \href
  {\doibase 10.1086/174255} {\bibfield  {journal} {\bibinfo  {journal} {\apj}\
  }\textbf {\bibinfo {volume} {428}},\ \bibinfo {pages} {433} (\bibinfo {year}
  {1994})},\ \Eprint {http://arxiv.org/abs/astro-ph/9311056}
  {astro-ph/9311056}\BibitemShut {NoStop}%
\bibitem [{{Narayanan} \& {Weinberg}(1998)\citenamefont {{Narayanan}}\ \&\
  \citenamefont {{Weinberg}}}]{Narayanan1998}%
{(\PineGreen{{Narayanan} \& {Weinberg}}, \PineGreen{1998})}  \BibitemOpen
  \bibfield  {author} {\bibinfo {author} {\bibfnamefont {V.~K.}\ \bibnamefont
  {{Narayanan}}}, \bibinfo {author} {\bibfnamefont {D.~H.}\ \bibnamefont
  {{Weinberg}}},\ }\emph {{Reconstruction Analysis of Galaxy Redshift Surveys:
  A Hybrid Reconstruction Method}},\ \href {\doibase 10.1086/306429} {\bibfield
   {journal} {\bibinfo  {journal} {\apj}\ }\textbf {\bibinfo {volume} {508}},\
  \bibinfo {pages} {440} (\bibinfo {year} {1998})},\ \Eprint
  {http://arxiv.org/abs/arXiv:astro-ph/9806238}
  {arXiv:astro-ph/9806238}\BibitemShut {NoStop}%
\bibitem [{{Neyrinck}(2013)\citenamefont {{Neyrinck}}}]{Neyrinck2013a}%
{(\PineGreen{{Neyrinck}}, \PineGreen{2013})}  \BibitemOpen
  \bibfield  {author} {\bibinfo {author} {\bibfnamefont {M.~C.}\ \bibnamefont
  {{Neyrinck}}},\ }\emph {{Quantifying distortions of the Lagrangian
  dark-matter mesh in cosmology}},\ \href {\doibase 10.1093/mnras/sts027}
  {\bibfield  {journal} {\bibinfo  {journal} {\mnras}\ }\textbf {\bibinfo
  {volume} {428}},\ \bibinfo {pages} {141} (\bibinfo {year} {2013})},\ \Eprint
  {http://arxiv.org/abs/1204.1326} {arXiv:1204.1326 [astro-ph.CO]}\BibitemShut
  {NoStop}%
\bibitem [{{Neyrinck}(2015)\citenamefont {{Neyrinck}}}]{Neyrinck2015}%
{(\PineGreen{{Neyrinck}}, \PineGreen{2015})}  \BibitemOpen
  \bibfield  {author} {\bibinfo {author} {\bibfnamefont {M.~C.}\ \bibnamefont
  {{Neyrinck}}},\ }\emph {{Truthing the stretch: Non-perturbative cosmological
  realizations with multiscale spherical collapse}},\ \href@noop {} {\bibfield
  {journal} {\bibinfo  {journal} {ArXiv e-prints}\ } (\bibinfo {year}
  {2015})},\ \Eprint {http://arxiv.org/abs/1503.07534}
  {arXiv:1503.07534}\BibitemShut {NoStop}%
\bibitem [{{Neyrinck}, {Szapudi} \& {Szalay}(2011)\citenamefont {{Neyrinck}},
  \citenamefont {{Szapudi}},\ \&\ \citenamefont {{Szalay}}}]{Neyrinck2011}%
{(\PineGreen{{Neyrinck}, {Szapudi} \& {Szalay}}, \PineGreen{2011})}
  \BibitemOpen
  \bibfield  {author} {\bibinfo {author} {\bibfnamefont {M.~C.}\ \bibnamefont
  {{Neyrinck}}}, \bibinfo {author} {\bibfnamefont {I.}~\bibnamefont
  {{Szapudi}}}, \bibinfo {author} {\bibfnamefont {A.~S.}\ \bibnamefont
  {{Szalay}}},\ }\emph {{Rejuvenating Power Spectra. II. The Gaussianized
  Galaxy Density Field}},\ \href {\doibase 10.1088/0004-637X/731/2/116}
  {\bibfield  {journal} {\bibinfo  {journal} {\apj}\ }\textbf {\bibinfo
  {volume} {731}},\ \bibinfo {eid} {116} (\bibinfo {year} {2011})},\ \Eprint
  {http://arxiv.org/abs/1009.5680} {arXiv:1009.5680 [astro-ph.CO]}\BibitemShut
  {NoStop}%
\bibitem [{{Neyrinck} \& {Yang}(2013)\citenamefont {{Neyrinck}}\ \&\
  \citenamefont {{Yang}}}]{Neyrinck2013b}%
{(\PineGreen{{Neyrinck} \& {Yang}}, \PineGreen{2013})}  \BibitemOpen
  \bibfield  {author} {\bibinfo {author} {\bibfnamefont {M.~C.}\ \bibnamefont
  {{Neyrinck}}}, \bibinfo {author} {\bibfnamefont {L.~F.}\ \bibnamefont
  {{Yang}}},\ }\emph {{Ringing the initial Universe: the response of
  overdensity and transformed-density power spectra to initial spikes}},\ \href
  {\doibase 10.1093/mnras/stt949} {\bibfield  {journal} {\bibinfo  {journal}
  {\mnras}\ }\textbf {\bibinfo {volume} {433}},\ \bibinfo {pages} {1628}
  (\bibinfo {year} {2013})},\ \Eprint {http://arxiv.org/abs/1305.1629}
  {arXiv:1305.1629 [astro-ph.CO]}\BibitemShut {NoStop}%
\bibitem [{{Peebles}(1980)\citenamefont {{Peebles}}}]{Peebles1980}%
{(\PineGreen{{Peebles}}, \PineGreen{1980})}  \BibitemOpen
  \bibfield  {author} {\bibinfo {author} {\bibfnamefont {P.~J.~E.}\
  \bibnamefont {{Peebles}}},\ }\href@noop {} {\emph {\bibinfo {title} {Research
  supported by the National Science Foundation.~Princeton, N.J., Princeton
  University Press, 1980.~435 p.}}}\ (\bibinfo {year} {1980})\BibitemShut
  {NoStop}%
\bibitem [{{Pueblas} \& {Scoccimarro}(2009)\citenamefont {{Pueblas}}\ \&\
  \citenamefont {{Scoccimarro}}}]{Pueblas2009}%
{(\PineGreen{{Pueblas} \& {Scoccimarro}}, \PineGreen{2009})}  \BibitemOpen
  \bibfield  {author} {\bibinfo {author} {\bibfnamefont {S.}~\bibnamefont
  {{Pueblas}}}, \bibinfo {author} {\bibfnamefont {R.}~\bibnamefont
  {{Scoccimarro}}},\ }\emph {{Generation of vorticity and velocity dispersion
  by orbit crossing}},\ \href {\doibase 10.1103/PhysRevD.80.043504} {\bibfield
  {journal} {\bibinfo  {journal} {\prd}\ }\textbf {\bibinfo {volume} {80}},\
  \bibinfo {eid} {043504} (\bibinfo {year} {2009})},\ \Eprint
  {http://arxiv.org/abs/0809.4606} {arXiv:0809.4606}\BibitemShut {NoStop}%
\bibitem [{{Sahni} \& {Shandarin}(1996)\citenamefont {{Sahni}}\ \&\
  \citenamefont {{Shandarin}}}]{Sahni1996}%
{(\PineGreen{{Sahni} \& {Shandarin}}, \PineGreen{1996})}  \BibitemOpen
  \bibfield  {author} {\bibinfo {author} {\bibfnamefont {V.}~\bibnamefont
  {{Sahni}}}, \bibinfo {author} {\bibfnamefont {S.}~\bibnamefont
  {{Shandarin}}},\ }\emph {{Accuracy of Lagrangian approximations in voids}},\
  \href@noop {} {\bibfield  {journal} {\bibinfo  {journal} {\mnras}\ }\textbf
  {\bibinfo {volume} {282}},\ \bibinfo {pages} {641} (\bibinfo {year}
  {1996})},\ \Eprint {http://arxiv.org/abs/astro-ph/9510142}
  {astro-ph/9510142}\BibitemShut {NoStop}%
\bibitem [{{Scoccimarro}(1997)\citenamefont {{Scoccimarro}}}]{Scoccimarro1997}%
{(\PineGreen{{Scoccimarro}}, \PineGreen{1997})}  \BibitemOpen
  \bibfield  {author} {\bibinfo {author} {\bibfnamefont {R.}~\bibnamefont
  {{Scoccimarro}}},\ }\emph {{Cosmological Perturbations: Entering the
  Nonlinear Regime}},\ \href {\doibase 10.1086/304578} {\bibfield  {journal}
  {\bibinfo  {journal} {\apj}\ }\textbf {\bibinfo {volume} {487}},\ \bibinfo
  {pages} {1} (\bibinfo {year} {1997})},\ \Eprint
  {http://arxiv.org/abs/astro-ph/9612207} {astro-ph/9612207}\BibitemShut
  {NoStop}%
\bibitem [{{Scoccimarro}(1998)\citenamefont {{Scoccimarro}}}]{Scoccimarro1998}%
{(\PineGreen{{Scoccimarro}}, \PineGreen{1998})}  \BibitemOpen
  \bibfield  {author} {\bibinfo {author} {\bibfnamefont {R.}~\bibnamefont
  {{Scoccimarro}}},\ }\emph {{Transients from initial conditions: a
  perturbative analysis}},\ \href {\doibase 10.1046/j.1365-8711.1998.01845.x}
  {\bibfield  {journal} {\bibinfo  {journal} {\mnras}\ }\textbf {\bibinfo
  {volume} {299}},\ \bibinfo {pages} {1097} (\bibinfo {year} {1998})},\ \Eprint
  {http://arxiv.org/abs/astro-ph/9711187} {astro-ph/9711187}\BibitemShut
  {NoStop}%
\bibitem [{{Scoccimarro}(2000)\citenamefont {{Scoccimarro}}}]{Scoccimarro2000}%
{(\PineGreen{{Scoccimarro}}, \PineGreen{2000})}  \BibitemOpen
  \bibfield  {author} {\bibinfo {author} {\bibfnamefont {R.}~\bibnamefont
  {{Scoccimarro}}},\ }\emph {{The Bispectrum: From Theory to Observations}},\
  \href {\doibase 10.1086/317248} {\bibfield  {journal} {\bibinfo  {journal}
  {\apj}\ }\textbf {\bibinfo {volume} {544}},\ \bibinfo {pages} {597} (\bibinfo
  {year} {2000})},\ \Eprint {http://arxiv.org/abs/astro-ph/0004086}
  {astro-ph/0004086}\BibitemShut {NoStop}%
\bibitem [{{Scoccimarro} {\textit{et~al}}\mbox{.}(2001)\citenamefont
  {{Scoccimarro}}, \citenamefont {{Feldman}}, \citenamefont {{Fry}},\ \&\
  \citenamefont {{Frieman}}}]{Scoccimarro2001}%
{(\PineGreen{{Scoccimarro} {\textit{et~al}}\mbox{.}}, \PineGreen{2001})}
  \BibitemOpen
  \bibfield  {author} {\bibinfo {author} {\bibfnamefont {R.}~\bibnamefont
  {{Scoccimarro}}}, \bibinfo {author} {\bibfnamefont {H.~A.}\ \bibnamefont
  {{Feldman}}}, \bibinfo {author} {\bibfnamefont {J.~N.}\ \bibnamefont
  {{Fry}}}, \bibinfo {author} {\bibfnamefont {J.~A.}\ \bibnamefont
  {{Frieman}}},\ }\emph {{The Bispectrum of IRAS Redshift Catalogs}},\ \href
  {\doibase 10.1086/318284} {\bibfield  {journal} {\bibinfo  {journal} {\apj}\
  }\textbf {\bibinfo {volume} {546}},\ \bibinfo {pages} {652} (\bibinfo {year}
  {2001})},\ \Eprint {http://arxiv.org/abs/astro-ph/0004087}
  {astro-ph/0004087}\BibitemShut {NoStop}%
\bibitem [{{Scoccimarro} \& {Sheth}(2002)\citenamefont {{Scoccimarro}}\ \&\
  \citenamefont {{Sheth}}}]{Scoccimarro2002}%
{(\PineGreen{{Scoccimarro} \& {Sheth}}, \PineGreen{2002})}  \BibitemOpen
  \bibfield  {author} {\bibinfo {author} {\bibfnamefont {R.}~\bibnamefont
  {{Scoccimarro}}}, \bibinfo {author} {\bibfnamefont {R.~K.}\ \bibnamefont
  {{Sheth}}},\ }\emph {{PTHALOS: a fast method for generating mock galaxy
  distributions}},\ \href {\doibase 10.1046/j.1365-8711.2002.04999.x}
  {\bibfield  {journal} {\bibinfo  {journal} {\mnras}\ }\textbf {\bibinfo
  {volume} {329}},\ \bibinfo {pages} {629} (\bibinfo {year} {2002})},\ \Eprint
  {http://arxiv.org/abs/astro-ph/0106120} {astro-ph/0106120}\BibitemShut
  {NoStop}%
\bibitem [{{Sefusatti} \& {Komatsu}(2007)\citenamefont {{Sefusatti}}\ \&\
  \citenamefont {{Komatsu}}}]{Sefusatti2007}%
{(\PineGreen{{Sefusatti} \& {Komatsu}}, \PineGreen{2007})}  \BibitemOpen
  \bibfield  {author} {\bibinfo {author} {\bibfnamefont {E.}~\bibnamefont
  {{Sefusatti}}}, \bibinfo {author} {\bibfnamefont {E.}~\bibnamefont
  {{Komatsu}}},\ }\emph {{Bispectrum of galaxies from high-redshift galaxy
  surveys: Primordial non-Gaussianity and nonlinear galaxy bias}},\ \href
  {\doibase 10.1103/PhysRevD.76.083004} {\bibfield  {journal} {\bibinfo
  {journal} {\prd}\ }\textbf {\bibinfo {volume} {76}},\ \bibinfo {eid} {083004}
  (\bibinfo {year} {2007})},\ \Eprint {http://arxiv.org/abs/0705.0343}
  {arXiv:0705.0343}\BibitemShut {NoStop}%
\bibitem [{{Shandarin} \& {Zel'dovich}(1989)\citenamefont {{Shandarin}}\ \&\
  \citenamefont {{Zel'dovich}}}]{Shandarin1989}%
{(\PineGreen{{Shandarin} \& {Zel'dovich}}, \PineGreen{1989})}  \BibitemOpen
  \bibfield  {author} {\bibinfo {author} {\bibfnamefont {S.~F.}\ \bibnamefont
  {{Shandarin}}}, \bibinfo {author} {\bibfnamefont {Y.~B.}\ \bibnamefont
  {{Zel'dovich}}},\ }\emph {{The large-scale structure of the universe:
  Turbulence, intermittency, structures in a self-gravitating medium}},\ \href
  {\doibase 10.1103/RevModPhys.61.185} {\bibfield  {journal} {\bibinfo
  {journal} {Reviews of Modern Physics}\ }\textbf {\bibinfo {volume} {61}},\
  \bibinfo {pages} {185} (\bibinfo {year} {1989})}\BibitemShut {NoStop}%
\bibitem [{{Shirata} {\textit{et~al}}\mbox{.}(2007)\citenamefont {{Shirata}},
  \citenamefont {{Suto}}, \citenamefont {{Hikage}}, \citenamefont
  {{Shiromizu}},\ \&\ \citenamefont {{Yoshida}}}]{Shiratra2007}%
{(\PineGreen{{Shirata} {\textit{et~al}}\mbox{.}}, \PineGreen{2007})}
  \BibitemOpen
  \bibfield  {author} {\bibinfo {author} {\bibfnamefont {A.}~\bibnamefont
  {{Shirata}}}, \bibinfo {author} {\bibfnamefont {Y.}~\bibnamefont {{Suto}}},
  \bibinfo {author} {\bibfnamefont {C.}~\bibnamefont {{Hikage}}}, \bibinfo
  {author} {\bibfnamefont {T.}~\bibnamefont {{Shiromizu}}}, \bibinfo {author}
  {\bibfnamefont {N.}~\bibnamefont {{Yoshida}}},\ }\emph {{Galaxy clustering
  constraints on deviations from Newtonian gravity at cosmological scales. II.
  Perturbative and numerical analyses of power spectrum and bispectrum}},\
  \href {\doibase 10.1103/PhysRevD.76.044026} {\bibfield  {journal} {\bibinfo
  {journal} {\prd}\ }\textbf {\bibinfo {volume} {76}},\ \bibinfo {eid} {044026}
  (\bibinfo {year} {2007})},\ \Eprint {http://arxiv.org/abs/0705.1311}
  {arXiv:0705.1311}\BibitemShut {NoStop}%
\bibitem [{{Springel}(2005)\citenamefont {{Springel}}}]{Springel2005}%
{(\PineGreen{{Springel}}, \PineGreen{2005})}  \BibitemOpen
  \bibfield  {author} {\bibinfo {author} {\bibfnamefont {V.}~\bibnamefont
  {{Springel}}},\ }\emph {{The cosmological simulation code GADGET-2}},\ \href
  {\doibase 10.1111/j.1365-2966.2005.09655.x} {\bibfield  {journal} {\bibinfo
  {journal} {\mnras}\ }\textbf {\bibinfo {volume} {364}},\ \bibinfo {pages}
  {1105} (\bibinfo {year} {2005})},\ \Eprint
  {http://arxiv.org/abs/astro-ph/0505010} {astro-ph/0505010}\BibitemShut
  {NoStop}%
\bibitem [{{Springel}, {Yoshida} \& {White}(2001)\citenamefont {{Springel}},
  \citenamefont {{Yoshida}},\ \&\ \citenamefont {{White}}}]{Springel2001}%
{(\PineGreen{{Springel}, {Yoshida} \& {White}}, \PineGreen{2001})}
  \BibitemOpen
  \bibfield  {author} {\bibinfo {author} {\bibfnamefont {V.}~\bibnamefont
  {{Springel}}}, \bibinfo {author} {\bibfnamefont {N.}~\bibnamefont
  {{Yoshida}}}, \bibinfo {author} {\bibfnamefont {S.~D.~M.}\ \bibnamefont
  {{White}}},\ }\emph {{GADGET: a code for collisionless and gasdynamical
  cosmological simulations}},\ \href {\doibase 10.1016/S1384-1076(01)00042-2}
  {\bibfield  {journal} {\bibinfo  {journal} {\na}\ }\textbf {\bibinfo {volume}
  {6}},\ \bibinfo {pages} {79} (\bibinfo {year} {2001})},\ \Eprint
  {http://arxiv.org/abs/astro-ph/0003162} {astro-ph/0003162}\BibitemShut
  {NoStop}%
\bibitem [{{Taffoni}, {Monaco} \& {Theuns}(2002)\citenamefont {{Taffoni}},
  \citenamefont {{Monaco}},\ \&\ \citenamefont {{Theuns}}}]{Taffoni2002}%
{(\PineGreen{{Taffoni}, {Monaco} \& {Theuns}}, \PineGreen{2002})}  \BibitemOpen
  \bibfield  {author} {\bibinfo {author} {\bibfnamefont {G.}~\bibnamefont
  {{Taffoni}}}, \bibinfo {author} {\bibfnamefont {P.}~\bibnamefont {{Monaco}}},
  \bibinfo {author} {\bibfnamefont {T.}~\bibnamefont {{Theuns}}},\ }\emph
  {{PINOCCHIO and the hierarchical build-up of dark matter haloes}},\ \href
  {\doibase 10.1046/j.1365-8711.2002.05441.x} {\bibfield  {journal} {\bibinfo
  {journal} {\mnras}\ }\textbf {\bibinfo {volume} {333}},\ \bibinfo {pages}
  {623} (\bibinfo {year} {2002})},\ \Eprint
  {http://arxiv.org/abs/astro-ph/0109324} {astro-ph/0109324}\BibitemShut
  {NoStop}%
\bibitem [{{Tassev} \& {Zaldarriaga}(2012{\natexlab{a}})\citenamefont
  {{Tassev}}\ \&\ \citenamefont {{Zaldarriaga}}}]{Tassev2012a}%
{(\PineGreen{{Tassev} \& {Zaldarriaga}}, \PineGreen{2012{\natexlab{a}}})}
  \BibitemOpen
  \bibfield  {author} {\bibinfo {author} {\bibfnamefont {S.}~\bibnamefont
  {{Tassev}}}, \bibinfo {author} {\bibfnamefont {M.}~\bibnamefont
  {{Zaldarriaga}}},\ }\emph {{Estimating CDM particle trajectories in the
  mildly non-linear regime of structure formation. Implications for the density
  field in real and redshift space}},\ \href {\doibase
  10.1088/1475-7516/2012/12/011} {\bibfield  {journal} {\bibinfo  {journal}
  {\jcap}\ }\textbf {\bibinfo {volume} {12}},\ \bibinfo {eid} {011} (\bibinfo
  {year} {2012}{\natexlab{a}})},\ \Eprint {http://arxiv.org/abs/1203.5785}
  {arXiv:1203.5785 [astro-ph.CO]}\BibitemShut {NoStop}%
\bibitem [{{Tassev} \& {Zaldarriaga}(2012{\natexlab{b}})\citenamefont
  {{Tassev}}\ \&\ \citenamefont {{Zaldarriaga}}}]{Tassev2012b}%
{(\PineGreen{{Tassev} \& {Zaldarriaga}}, \PineGreen{2012{\natexlab{b}}})}
  \BibitemOpen
  \bibfield  {author} {\bibinfo {author} {\bibfnamefont {S.}~\bibnamefont
  {{Tassev}}}, \bibinfo {author} {\bibfnamefont {M.}~\bibnamefont
  {{Zaldarriaga}}},\ }\emph {{The mildly non-linear regime of structure
  formation}},\ \href {\doibase 10.1088/1475-7516/2012/04/013} {\bibfield
  {journal} {\bibinfo  {journal} {\jcap}\ }\textbf {\bibinfo {volume} {4}},\
  \bibinfo {eid} {013} (\bibinfo {year} {2012}{\natexlab{b}})},\ \Eprint
  {http://arxiv.org/abs/1109.4939} {arXiv:1109.4939 [astro-ph.CO]}\BibitemShut
  {NoStop}%
\bibitem [{{Tassev} \& {Zaldarriaga}(2012{\natexlab{c}})\citenamefont
  {{Tassev}}\ \&\ \citenamefont {{Zaldarriaga}}}]{Tassev2012c}%
{(\PineGreen{{Tassev} \& {Zaldarriaga}}, \PineGreen{2012{\natexlab{c}}})}
  \BibitemOpen
  \bibfield  {author} {\bibinfo {author} {\bibfnamefont {S.}~\bibnamefont
  {{Tassev}}}, \bibinfo {author} {\bibfnamefont {M.}~\bibnamefont
  {{Zaldarriaga}}},\ }\emph {{Towards an optimal reconstruction of baryon
  oscillations}},\ \href {\doibase 10.1088/1475-7516/2012/10/006} {\bibfield
  {journal} {\bibinfo  {journal} {\jcap}\ }\textbf {\bibinfo {volume} {10}},\
  \bibinfo {eid} {006} (\bibinfo {year} {2012}{\natexlab{c}})},\ \Eprint
  {http://arxiv.org/abs/1203.6066} {arXiv:1203.6066 [astro-ph.CO]}\BibitemShut
  {NoStop}%
\bibitem [{{Tassev}, {Zaldarriaga} \& {Eisenstein}(2013)\citenamefont
  {{Tassev}}, \citenamefont {{Zaldarriaga}},\ \&\ \citenamefont
  {{Eisenstein}}}]{Tassev2013}%
{(\PineGreen{{Tassev}, {Zaldarriaga} \& {Eisenstein}}, \PineGreen{2013})}
  \BibitemOpen
  \bibfield  {author} {\bibinfo {author} {\bibfnamefont {S.}~\bibnamefont
  {{Tassev}}}, \bibinfo {author} {\bibfnamefont {M.}~\bibnamefont
  {{Zaldarriaga}}}, \bibinfo {author} {\bibfnamefont {D.~J.}\ \bibnamefont
  {{Eisenstein}}},\ }\emph {{Solving large scale structure in ten easy steps
  with COLA}},\ \href {\doibase 10.1088/1475-7516/2013/06/036} {\bibfield
  {journal} {\bibinfo  {journal} {\jcap}\ }\textbf {\bibinfo {volume} {6}},\
  \bibinfo {eid} {036} (\bibinfo {year} {2013})},\ \Eprint
  {http://arxiv.org/abs/1301.0322} {arXiv:1301.0322 [astro-ph.CO]}\BibitemShut
  {NoStop}%
\bibitem [{{Valageas}(2007)\citenamefont {{Valageas}}}]{Valageas2007}%
{(\PineGreen{{Valageas}}, \PineGreen{2007})}  \BibitemOpen
  \bibfield  {author} {\bibinfo {author} {\bibfnamefont {P.}~\bibnamefont
  {{Valageas}}},\ }\emph {{Large-N expansions applied to gravitational
  clustering}},\ \href {\doibase 10.1051/0004-6361:20066832} {\bibfield
  {journal} {\bibinfo  {journal} {\aap}\ }\textbf {\bibinfo {volume} {465}},\
  \bibinfo {pages} {725} (\bibinfo {year} {2007})},\ \Eprint
  {http://arxiv.org/abs/astro-ph/0611849} {astro-ph/0611849}\BibitemShut
  {NoStop}%
\bibitem [{{Verde} {\textit{et~al}}\mbox{.}(1998)\citenamefont {{Verde}},
  \citenamefont {{Heavens}}, \citenamefont {{Matarrese}},\ \&\ \citenamefont
  {{Moscardini}}}]{Verde1998}%
{(\PineGreen{{Verde} {\textit{et~al}}\mbox{.}}, \PineGreen{1998})}
  \BibitemOpen
  \bibfield  {author} {\bibinfo {author} {\bibfnamefont {L.}~\bibnamefont
  {{Verde}}}, \bibinfo {author} {\bibfnamefont {A.~F.}\ \bibnamefont
  {{Heavens}}}, \bibinfo {author} {\bibfnamefont {S.}~\bibnamefont
  {{Matarrese}}}, \bibinfo {author} {\bibfnamefont {L.}~\bibnamefont
  {{Moscardini}}},\ }\emph {{Large-scale bias in the Universe - II.
  Redshift-space bispectrum}},\ \href {\doibase
  10.1046/j.1365-8711.1998.01937.x} {\bibfield  {journal} {\bibinfo  {journal}
  {\mnras}\ }\textbf {\bibinfo {volume} {300}},\ \bibinfo {pages} {747}
  (\bibinfo {year} {1998})},\ \Eprint {http://arxiv.org/abs/astro-ph/9806028}
  {astro-ph/9806028}\BibitemShut {NoStop}%
\bibitem [{{Verde} {\textit{et~al}}\mbox{.}(2002)\citenamefont {{Verde}},
  \citenamefont {{Heavens}}, \citenamefont {{Percival}}, \citenamefont
  {{Matarrese}}, \citenamefont {{Baugh}}, \citenamefont {{Bland-Hawthorn}},
  \citenamefont {{Bridges}}, \citenamefont {{Cannon}}, \citenamefont {{Cole}},
  \citenamefont {{Colless}}, \citenamefont {{Collins}}, \citenamefont
  {{Couch}}, \citenamefont {{Dalton}}, \citenamefont {{De Propris}},
  \citenamefont {{Driver}}, \citenamefont {{Efstathiou}}, \citenamefont
  {{Ellis}}, \citenamefont {{Frenk}}, \citenamefont {{Glazebrook}},
  \citenamefont {{Jackson}}, \citenamefont {{Lahav}}, \citenamefont {{Lewis}},
  \citenamefont {{Lumsden}}, \citenamefont {{Maddox}}, \citenamefont
  {{Madgwick}}, \citenamefont {{Norberg}}, \citenamefont {{Peacock}},
  \citenamefont {{Peterson}}, \citenamefont {{Sutherland}},\ \&\ \citenamefont
  {{Taylor}}}]{Verde2002}%
{(\PineGreen{{Verde} {\textit{et~al}}\mbox{.}}, \PineGreen{2002})}
  \BibitemOpen
  \bibfield  {author} {\bibinfo {author} {\bibfnamefont {L.}~\bibnamefont
  {{Verde}}}, \bibinfo {author} {\bibfnamefont {A.~F.}\ \bibnamefont
  {{Heavens}}}, \bibinfo {author} {\bibfnamefont {W.~J.}\ \bibnamefont
  {{Percival}}}, \bibinfo {author} {\bibfnamefont {S.}~\bibnamefont
  {{Matarrese}}}, \bibinfo {author} {\bibfnamefont {C.~M.}\ \bibnamefont
  {{Baugh}}}, \bibinfo {author} {\bibfnamefont {J.}~\bibnamefont
  {{Bland-Hawthorn}}}, \bibinfo {author} {\bibfnamefont {T.}~\bibnamefont
  {{Bridges}}}, \bibinfo {author} {\bibfnamefont {R.}~\bibnamefont {{Cannon}}},
  \bibinfo {author} {\bibfnamefont {S.}~\bibnamefont {{Cole}}}, \bibinfo
  {author} {\bibfnamefont {M.}~\bibnamefont {{Colless}}}, \bibinfo {author}
  {\bibfnamefont {C.}~\bibnamefont {{Collins}}}, \bibinfo {author}
  {\bibfnamefont {W.}~\bibnamefont {{Couch}}}, \bibinfo {author} {\bibfnamefont
  {G.}~\bibnamefont {{Dalton}}}, \bibinfo {author} {\bibfnamefont
  {R.}~\bibnamefont {{De Propris}}}, \bibinfo {author} {\bibfnamefont {S.~P.}\
  \bibnamefont {{Driver}}}, \bibinfo {author} {\bibfnamefont {G.}~\bibnamefont
  {{Efstathiou}}}, \bibinfo {author} {\bibfnamefont {R.~S.}\ \bibnamefont
  {{Ellis}}}, \bibinfo {author} {\bibfnamefont {C.~S.}\ \bibnamefont
  {{Frenk}}}, \bibinfo {author} {\bibfnamefont {K.}~\bibnamefont
  {{Glazebrook}}}, \bibinfo {author} {\bibfnamefont {C.}~\bibnamefont
  {{Jackson}}}, \bibinfo {author} {\bibfnamefont {O.}~\bibnamefont {{Lahav}}},
  \bibinfo {author} {\bibfnamefont {I.}~\bibnamefont {{Lewis}}}, \bibinfo
  {author} {\bibfnamefont {S.}~\bibnamefont {{Lumsden}}}, \bibinfo {author}
  {\bibfnamefont {S.}~\bibnamefont {{Maddox}}}, \bibinfo {author}
  {\bibfnamefont {D.}~\bibnamefont {{Madgwick}}}, \bibinfo {author}
  {\bibfnamefont {P.}~\bibnamefont {{Norberg}}}, \bibinfo {author}
  {\bibfnamefont {J.~A.}\ \bibnamefont {{Peacock}}}, \bibinfo {author}
  {\bibfnamefont {B.~A.}\ \bibnamefont {{Peterson}}}, \bibinfo {author}
  {\bibfnamefont {W.}~\bibnamefont {{Sutherland}}}, \bibinfo {author}
  {\bibfnamefont {K.}~\bibnamefont {{Taylor}}},\ }\emph {{The 2dF Galaxy
  Redshift Survey: the bias of galaxies and the density of the Universe}},\
  \href {\doibase 10.1046/j.1365-8711.2002.05620.x} {\bibfield  {journal}
  {\bibinfo  {journal} {\mnras}\ }\textbf {\bibinfo {volume} {335}},\ \bibinfo
  {pages} {432} (\bibinfo {year} {2002})},\ \Eprint
  {http://arxiv.org/abs/arXiv:astro-ph/0112161}
  {arXiv:astro-ph/0112161}\BibitemShut {NoStop}%
\bibitem [{{Weinberg}(1992)\citenamefont {{Weinberg}}}]{Weinberg1992}%
{(\PineGreen{{Weinberg}}, \PineGreen{1992})}  \BibitemOpen
  \bibfield  {author} {\bibinfo {author} {\bibfnamefont {D.~H.}\ \bibnamefont
  {{Weinberg}}},\ }\emph {{Reconstructing primordial density fluctuations. I -
  Method}},\ \href@noop {} {\bibfield  {journal} {\bibinfo  {journal} {\mnras}\
  }\textbf {\bibinfo {volume} {254}},\ \bibinfo {pages} {315} (\bibinfo {year}
  {1992})}\BibitemShut {NoStop}%
\bibitem [{{Yoshisato} {\textit{et~al}}\mbox{.}(2006)\citenamefont
  {{Yoshisato}}, \citenamefont {{Morikawa}}, \citenamefont {{Gouda}},\ \&\
  \citenamefont {{Mouri}}}]{Yoshisato2006}%
{(\PineGreen{{Yoshisato} {\textit{et~al}}\mbox{.}}, \PineGreen{2006})}
  \BibitemOpen
  \bibfield  {author} {\bibinfo {author} {\bibfnamefont {A.}~\bibnamefont
  {{Yoshisato}}}, \bibinfo {author} {\bibfnamefont {M.}~\bibnamefont
  {{Morikawa}}}, \bibinfo {author} {\bibfnamefont {N.}~\bibnamefont {{Gouda}}},
  \bibinfo {author} {\bibfnamefont {H.}~\bibnamefont {{Mouri}}},\ }\emph {{Why
  is the Zel'dovich Approximation So Accurate?}},\ \href {\doibase
  10.1086/498496} {\bibfield  {journal} {\bibinfo  {journal} {\apj}\ }\textbf
  {\bibinfo {volume} {637}},\ \bibinfo {pages} {555} (\bibinfo {year}
  {2006})},\ \Eprint {http://arxiv.org/abs/astro-ph/0510107}
  {astro-ph/0510107}\BibitemShut {NoStop}%
\bibitem [{{Yu} {\textit{et~al}}\mbox{.}(2011)\citenamefont {{Yu}},
  \citenamefont {{Zhang}}, \citenamefont {{Lin}}, \citenamefont {{Cui}},\ \&\
  \citenamefont {{Fry}}}]{Yu2011}%
{(\PineGreen{{Yu} {\textit{et~al}}\mbox{.}}, \PineGreen{2011})}  \BibitemOpen
  \bibfield  {author} {\bibinfo {author} {\bibfnamefont {Y.}~\bibnamefont
  {{Yu}}}, \bibinfo {author} {\bibfnamefont {P.}~\bibnamefont {{Zhang}}},
  \bibinfo {author} {\bibfnamefont {W.}~\bibnamefont {{Lin}}}, \bibinfo
  {author} {\bibfnamefont {W.}~\bibnamefont {{Cui}}}, \bibinfo {author}
  {\bibfnamefont {J.~N.}\ \bibnamefont {{Fry}}},\ }\emph {{Gaussianizing the
  non-Gaussian lensing convergence field: The performance of the
  Gaussianization}},\ \href {\doibase 10.1103/PhysRevD.84.023523} {\bibfield
  {journal} {\bibinfo  {journal} {\prd}\ }\textbf {\bibinfo {volume} {84}},\
  \bibinfo {eid} {023523} (\bibinfo {year} {2011})},\ \Eprint
  {http://arxiv.org/abs/1103.2858} {arXiv:1103.2858 [astro-ph.CO]}\BibitemShut
  {NoStop}%
\bibitem [{{Yu} {\textit{et~al}}\mbox{.}(2012)\citenamefont {{Yu}},
  \citenamefont {{Zhang}}, \citenamefont {{Lin}}, \citenamefont {{Cui}},\ \&\
  \citenamefont {{Fry}}}]{Yu2012}%
{(\PineGreen{{Yu} {\textit{et~al}}\mbox{.}}, \PineGreen{2012})}  \BibitemOpen
  \bibfield  {author} {\bibinfo {author} {\bibfnamefont {Y.}~\bibnamefont
  {{Yu}}}, \bibinfo {author} {\bibfnamefont {P.}~\bibnamefont {{Zhang}}},
  \bibinfo {author} {\bibfnamefont {W.}~\bibnamefont {{Lin}}}, \bibinfo
  {author} {\bibfnamefont {W.}~\bibnamefont {{Cui}}}, \bibinfo {author}
  {\bibfnamefont {J.~N.}\ \bibnamefont {{Fry}}},\ }\emph {{Gaussianizing the
  non-Gaussian lensing convergence field II. The applicability to noisy
  data}},\ \href {\doibase 10.1103/PhysRevD.86.023515} {\bibfield  {journal}
  {\bibinfo  {journal} {\prd}\ }\textbf {\bibinfo {volume} {86}},\ \bibinfo
  {eid} {023515} (\bibinfo {year} {2012})},\ \Eprint
  {http://arxiv.org/abs/1201.4527} {arXiv:1201.4527 [astro-ph.CO]}\BibitemShut
  {NoStop}%
\bibitem [{{Zel'dovich}(1970)\citenamefont {{Zel'dovich}}}]{Zeldovich1970}%
{(\PineGreen{{Zel'dovich}}, \PineGreen{1970})}  \BibitemOpen
  \bibfield  {author} {\bibinfo {author} {\bibfnamefont {Y.~B.}\ \bibnamefont
  {{Zel'dovich}}},\ }\emph {{Gravitational instability: An approximate theory
  for large density perturbations.}},\ \href@noop {} {\bibfield  {journal}
  {\bibinfo  {journal} {\aap}\ }\textbf {\bibinfo {volume} {5}},\ \bibinfo
  {pages} {84} (\bibinfo {year} {1970})}\BibitemShut {NoStop}%
\end{thebibliography}%

\end{document}